\documentclass[twocolumn,tighten]{aastex63}
\usepackage{amsmath}
\usepackage{threeparttable}
\usepackage{float}
\usepackage{graphicx}
\usepackage{CJK}
\usepackage{xspace}
\defcitealias{NaiduALT24}{Naidu \& Matthee et al.}

\shorttitle{The environments of broad H$\alpha$ line emitters at $z\sim5$}
\shortauthors{Matthee et al.}

\begin{document}
\begin{CJK*}{UTF8}{gbsn}

\title{Environmental Evidence for Overly Massive Black Holes in Low Mass Galaxies and a Black Hole - Halo Mass Relation at $z\sim5$}

\correspondingauthor{Jorryt Matthee}
\email{jorryt.matthee@ista.ac.at}
\author[0000-0003-2871-127X]{Jorryt Matthee}
\affiliation{Institute of Science and Technology Austria (ISTA), Am Campus 1, 3400 Klosterneuburg, Austria}

\author[0000-0003-3997-5705]{Rohan P. Naidu}
\thanks{NASA Hubble Fellow}
\affiliation{MIT Kavli Institute for Astrophysics and Space Research, 70 Vassar Street, Cambridge, 02139, Massachusetts, USA}


\author[0009-0007-1062-0253]{Gauri Kotiwale} 
\affiliation{Institute of Science and Technology Austria (ISTA), Am Campus 1, 3400 Klosterneuburg, Austria}

\author[0000-0001-6278-032X]{Lukas J. Furtak}
\affiliation{Department of Physics, Ben-Gurion University of the Negev, P.O. Box 653, Be'er-Sheva 84105, Israel}

\author[0000-0001-5346-6048]{Ivan Kramarenko}
\affiliation{Institute of Science and Technology Austria (ISTA), Am Campus 1, 3400 Klosterneuburg, Austria}

\author[0000-0003-0417-385X]{Ruari Mackenzie}
\affiliation{Department of Physics, ETH Z{\"u}rich, Wolfgang-Pauli-Strasse 27, Z{\"u}rich, 8093, Switzerland}

\author[0000-0002-5612-3427]{Jenny Greene}
\affiliation{Department of Astrophysical Sciences, Princeton University,Princeton, NJ 08544, USA}


\author[0000-0002-8192-8091]{Angela Adamo}
\affiliation{The Oskar Klein Centre, Department of Astronomy, Stockholm University, AlbaNova, SE-10691 Stockholm, Sweden}

\author[0000-0002-4989-2471]{Rychard J. Bouwens}
\affiliation{Leiden Observatory, Leiden University, NL-2300 RA Leiden, Netherlands}

\author[0000-0003-1408-7373]{Claudia Di Cesare}
\affiliation{Institute of Science and Technology Austria (ISTA), Am Campus 1, 3400 Klosterneuburg, Austria}

\author[0000-0003-2895-6218]{Anna-Christina Eilers}
\affiliation{MIT Kavli Institute for Astrophysics and Space Research, 77 Massachusetts Avenue, Cambridge, 02139, Massachusetts, USA}
\affiliation{Department of Physics, Massachusetts Institute of Technology, Cambridge, MA 02139, USA}
\author[0000-0002-2380-9801]{Anna de Graaff}
\affiliation{Max-Planck-Institut f\"ur Astronomie, K\"onigstuhl 17, D-69117, Heidelberg, Germany}

\author[0000-0002-9389-7413]{Kasper~E.~Heintz}
\affiliation{Department of Astronomy, University of Geneva, Chemin Pegasi 51, 1290 Versoix, Switzerland}
\affiliation{Cosmic Dawn Center (DAWN), Copenhagen, Denmark}
\affiliation{Niels Bohr Institute, University of Copenhagen, Jagtvej 128, K{\o}benhavn N, DK-2200, Denmark}

\author[0000-0001-9044-1747]{Daichi Kashino}
\affiliation{National Astronomical Observatory of Japan, 2-21-1 Osawa, Mitaka, Tokyo 181-8588, Japan}

\author[0000-0003-0695-4414]{Michael V. Maseda}
\affiliation{Department of Astronomy, University of Wisconsin-Madison, 475 N. Charter St., Madison, WI 53706 USA}

\author[0000-0002-8224-4505]{Sandro Tacchella}
\affiliation{Kavli Institute for Cosmology, University of Cambridge, Madingley Road, Cambridge, CB3 0HA, UK}
\affiliation{Cavendish Laboratory, University of Cambridge, 19 JJ Thomson Avenue, Cambridge, CB3 0HE, UK}

\author[0000-0001-5586-6950]{Alberto Torralba}
\affiliation{Observatori Astron\`omic de la Universitat de Val\`encia, Ed. Instituts d'Investigaci\'o, Parc Cient\'ific. C/ Catedr\'atico Jos\'e Beltr\'an, n2, 46980 Paterna, Valencia, Spain}
\affiliation{Departament d'Astronomia i Astrof\'isica, Universitat de Val\`encia, 46100 Burjassot, Spain}


\begin{abstract}  
{\it JWST} observations have unveiled faint active galactic nuclei (AGN) at high-redshift that provide insights on the formation of supermassive black holes (SMBHs) and their coevolution with galaxies. However, disentangling stellar from AGN light in these sources is challenging. Here, we use an empirical approach to infer the average stellar mass of 6 faint broad line (BL) H$\alpha$ emitters at $z=4-5$ with BH masses $\approx6\, (4-15)\times10^6$ M$_{\odot}$, with a method independent of their spectral energy distribution (SED). We use the deep {\it JWST}/NIRcam grism survey ALT to measure the over-densities around BL-H$\alpha$ emitters and around a spectroscopic reference sample of $\sim300$ galaxies. In our reference sample, we find that Mpc-scale over-density correlates with stellar mass, while pair counts are flat below $\approx50$ kpc due to satellites. Their large-scale environments suggest that BL-H$\alpha$ emitters are hosted by galaxies with stellar masses $\approx5\times10^7$ M$_{\odot}$, $\approx40$ times lower than those inferred from galaxy-only SED fits. Adding measurements around more luminous $z\approx6$ AGNs, we find tentative correlations between line width, BH mass and the over-density, suggestive of a steep BH to halo mass relation. The main implications are (1) when BH masses are taken at face value, we confirm extremely high BH to stellar mass ratios of $\approx10$ \%, (2) the low stellar mass galaxies hosting growing SMBHs are in tension with typical hydrodynamical simulations, except those without feedback, (3) a 1 \% duty cycle implied by the host mass hints at super-Eddington accretion, which may imply over-estimated SMBH masses, (4) the masses are at odds with a high stellar density interpretation of the line broadening, (5) our results imply a diversity of galaxy masses, environments and SEDs among AGN samples, depending on their luminosity. 

\end{abstract}

\keywords{galaxies: high-redshift, galaxies: formation,  galaxies: active galaxies, early Universe: reionization, supermassive black holes: quasars}

\section{Introduction}
\label{sec:introduction}
The rapid formation of supermassive black holes (SMBH) in the early Universe and the role that feedback from active galactic nuclei (AGN) feedback has on galaxy formation are among the key questions in present-day extragalactic astrophysics. How could black holes with masses $\sim10^{10}$ M$_{\odot}$ \citep{Wu15,Eilers22} emerge already by $z\approx6$? How can we explain the seemingly high black hole to stellar mass ratio inferred in many distant AGNs \citep[e.g.][]{Pacucci23}? Do these indicate SMBHs formation scenarios from non stellar origin such as direct collapse \citep{Haiman13,Natarajan24}, or highly super-Eddington accretion \citep{Bennett24,Husko24}? Is AGN feedback responsible for quenching the highest-redshift passive galaxies \citep{Carnall23,dEugenio24,degraaff24,Weibel24b}?

While the brightest AGNs, quasars, above $z\gtrsim4$ have been known for decades \citep[see][for a recent review]{Fan23}, {\it JWST} has significantly expanded the parameter space of known AGNs at high-redshifts towards lower luminosities and black hole masses \citep{Scholtz23,Adamo24,Treiber24}. {\it JWST} has also identified indications of AGN activity beyond redshifts $z>8$ \citep{Kokorev23,Goulding23,Larson23,Maiolino23,Napolitano24}. Therefore, these new samples hold the promise of improving our understanding of the earliest stages of SMBH formation and growth.

The population of broad Balmer-line selected AGNs at $z\sim3-8$ \citep{Matthee24,Lin24,Greene23} have been a key focus of attention. A significant subset of these are compact, with red rest-frame optical colors, in particular those where the broad line component is more dominant, leading \cite{Matthee24} to nickname them Little Red Dots (LRDs). The term LRDs has more widely been used to describe galaxies with a range of selection criteria, such as V-shaped spectral energy distributions (SEDs) with blue UV and red optical continua \citep[e.g.][]{Labbe23,Killi24,PerezGonzalez24} or compact red sources \citep{Akins24}. Broad-line selected samples contain both sources with blue UV colours as well as objects that are red in the full rest-UV to optical regime \citep{Matthee24}. In this paper we primarily focus on broad line (BL) H$\alpha$ emitters.

What are BL-H$\alpha$ emitters and what do they teach us about early SMBH formation? So far, most effort has been spent on understanding the SEDs of LRDs and BL-H$\alpha$ emitters in particular \citep[e.g.][]{PerezGonzalez24,YMa24,Setton24b,Volonteri24}. These SEDs include unusual features such as a flat rest-frame near-infrared SED that is suggesting the lack of the typical hot dust emission around AGNs \citep{Akins24,BWang24a,Williams24}, an extreme X-Ray faintness \citep{Kocevski24,Ananna24,Yue24,Maiolino24}, and the presence of Balmer absorption, suggesting very dense gas \citep{Juodzbalis24b,Matthee24} that is possibly Compton thick \citep{Maiolino24,Inayoshi24}. 

Among the key goals of the study of BL-H$\alpha$ emitters has been to measure the stellar mass of their host galaxies, as the relationship between BH mass and host galaxy mass is one of the fundamental relations that encode the physics of SMBH formation and growth \citep[e.g.][]{Harikane23,Maiolino23b,Furtak23,Marshall24,Onoue24,Yue24,Kokorev24,Juodzbalis24a}. These studies typically find very high SMBH mass to stellar mass ratios relative to the local $z=0$ scaling relation \citep[e.g.][]{Pacucci23}, partly due to the selection effect that AGN activity is easier to measure in galaxies with more massive BHs \citep{Li24}. However, accurately measuring their stellar masses is challenging \citep[e.g.][]{Leung2024}. Without accounting for AGN light, stellar masses can easily be over-estimated -- indeed, some seemingly over-massive galaxies \citep{Labbe22} turn out to have a significant AGN component (see e.g. \citealt{Kocevski23b,BWang24}). Reliably accounting for the AGN component is not straightforward, as the AGN SEDs are poorly understood \citep[e.g.][]{YMa24}. In particular, dense gas whose presence is indicated by Balmer absorption may also produce a Balmer break \citep{Inayoshi24}, a spectral feature typically associated with $\sim100$ Myr old stellar populations, but seemingly ubiquitous among LRDs \citep{Setton24b}. These challenges call for complementary approaches to characterize the galaxies with broad H$\alpha$. 

In our paradigm of cosmic structure formation, it is well understood that the large scale environments of galaxies are correlated with the properties of the dark matter halos hosting these galaxies, i.e. on average, more massive galaxies reside in larger over-densities as they are hosted by more massive dark matter halos \citep[e.g.][]{Blumenthal84}. Abundance-matching studies of high-redshift galaxies have confirmed the existence of a strong stellar mass - halo mass relation out to $z\sim10$ \citep[e.g.][]{Shuntov24} and clustering measurements based on photometric redshifts show indications of a luminosity-dependent bias \citep{Dalmasso24}. The environments of luminous quasars at $z\sim6$ suggest that they, on average, reside in halos with masses $\approx3\times10^{12}$ M$_{\odot}$ \citep[e.g.][]{Eilers24}, in line with their stellar masses of $\gtrsim10^{10}$ M$_{\odot}$ \citep{Yue23}. The much higher number densities of BL-H$\alpha$ emitters on the other hand suggests that they reside in less massive halos \citep{Pizzati24b}, in rough agreement with first clustering measurements using photometric redshifts \citep{Arita24}.

Here we present a new empirical approach to understand the host galaxies of BL-H$\alpha$ emitters based on studies of their environments, in comparison to the large-scale environments of star-forming galaxies. We use BL-H$\alpha$ emitters and neighboring galaxies identified in spectroscopic data from the `All the Little Things' (ALT) survey (see Section $\ref{sec:data}$; \citetalias{NaiduALT24} \citeyear{NaiduALT24}). ALT is the deepest NIRCam grism survey undertaken to date and covers H$\alpha$ emission at $z=3.8-5.0$ down to star formation rates of 0.1 M$_{\odot}$ yr$^{-1}$. The NIRCam grism is ideal to map galaxy over-densities due to its simple line-flux limited selection function over a wide field of view and the high precisions ($\approx100$ km s$^{-1}$) of the redshifts \citep[e.g.][]{Kashino23,Wang23,HerardDemanche24,Sun24}. The ALT field is covered by 27 band {\it JWST} NIRCam photometry \citep[e.g.][]{Bezanson22,Suess24}, as well as photometric data from {\it HST}, which yields nearly model-limited characterization of the SEDs. Our key assumptions are that the SED fits of galaxies without broad H$\alpha$ emission are very good and that the presence of broad H$\alpha$ emission is not impacting the detectability of galaxies in their environments.

We present the data that we use in Section $\ref{sec:data}$. In Section $\ref{sec:sample}$, we present our sample of BL-H$\alpha$ emitters at $z=4-5$, the AGN properties inferred from the H$\alpha$ line profile, and the reference sample of star-forming galaxies. We measure the environments of BL-H$\alpha$ emitters and the reference galaxies in Section $\ref{sec:environment}$. In Section $\ref{sec:impliedhost}$, we investigate the relation between over-density and stellar mass and use this to infer the stellar mass of BL-H$\alpha$ emitters. In Section $\ref{sec:coev}$ we investigate whether the over-density depends on BH mass. We discuss the implications of our results in Section $\ref{sec:implications}$ and summarize our results in Section $\ref{sec:summary}$.

Throughout the paper we assume a standard flat $\Lambda$CDM cosmology \citep{Planck18}. Magnitudes are reported in the AB system.

\section{Data} \label{sec:data}
The main aim of this paper is to study the environments of a well-defined sample of broad-line H$\alpha$ emitters at $z\sim5$ and compare them to the environments of star-forming galaxies with excellent stellar mass measurements. Given the possibility that BL-H$\alpha$ emitters are hosted by galaxies with masses as low as $\approx10^8$ M$_{\odot}$ \citep[e.g.][]{Pacucci23}, we require a very deep survey that includes large numbers of such low mass galaxies. To achieve this, we use galaxies and AGN identified in the Cycle 2 JWST/NIRCam Wide Field Slitless Spectroscopic (WFSS) survey `All the Little Things (ALT)' (PID 3516; PIs Matthee \& Naidu). ALT is the deepest NIRCam grism survey undertaken to date, yielding a sample of $\approx1600$ galaxies with spectroscopic redshifts at $z\sim0.3-8.5$ (\citetalias{NaiduALT24} \citeyear{NaiduALT24}), with a redshift error of $\approx100$ km s$^{-1}$ \citep{Bordoloi24,TorralbaTorregrosa24}. ALT targets galaxies in the $\approx 30$ arcmin$^2$ region in the background of the powerful lensing cluster Abell 2744, building upon the legacy from earlier surveys with {\it HST} and {\it JWST}, primarily the NIRCam imaging from the UNCOVER project \citep{Bezanson22}, but see \citetalias{NaiduALT24} \citeyear{NaiduALT24} for a full list of programs that contributed data. Additionally, the Cycle 2 medium-band program MegaScience \citep{Suess24} further completed the NIRCam medium-band coverage, yielding optimal characterisation of galaxy SEDs with 27 bands. The spectroscopic galaxy sample was constructed by identifying emission-lines in two dimensional grism spectra with a S/N criterion of $>5$. The typical 5$\sigma$ line-flux sensitivity ranges from $6-20\times10^{-19}$ erg s$^{-1}$ cm$^{-2}$ at $3.15-3.95$ micron. The identification of the lines used the detection of multiple emission-lines (such as the [OIII] doublet or HeI + Paschen-$\gamma$), and/or photometric data. The median magnification of the sample is relatively modest, $\mu=1.83$, and 83 \% of the sample has a magnification $\mu<3$.  For full details of the survey design, data reduction and galaxy selection we refer to the survey paper (\citetalias{NaiduALT24} \citeyear{NaiduALT24}).

\begin{table*}
    \centering
    \caption{{\bf General properties of the Broad-Line H$\alpha$ emitters used in this work.} Coordinates are in the J2000 reference frame. Objects with a * have already been identified in \cite{Greene23} (ALT-22928 = G23-10686; ALT-66543 = G23-45924; ALT-75753 = G23-38108). We list the line-width ($v_{\rm FWHM}$) of the broad H$\alpha$ component, the bolometric luminosity derived from the H$\alpha$ line and the BH mass. We also list the over-density within a 1 cMpc radius around the BL-H$\alpha$ emitters measured in this work. Errors on magnifications are dominated by systematics that we estimate to be on the 20 \% level. } 
    \begin{tabular}{lcccccccc}
    ALT-ID & R.A. & Dec. & $z_{\rm spec}$ & $\mu$ & $v_{\rm FWHM}$/km s$^{-1}$ & log$_{10}$(L$_{\rm bol}$/erg s$^{-1}$) & log$_{10}$(M$_{\rm BH}$/M$_{\odot}$) &  $(1+\delta)_{1 \rm cMpc}$  \\ \hline
    11345  & 3.570070 & $-30.432089$ & 4.782 & $1.50\pm0.30$ & $1380\pm220$ & $44.1\pm0.2$ & $6.7^{+0.2}_{-0.2}$ & $4.0\pm1.7$  \\ 
    16986 & 3.598798 & $-30.418733$ & 4.316 & $2.69\pm0.54$ & $1760\pm290$ & $44.3\pm0.2$ & $7.1^{+0.2}_{-0.3}$ & $7.4\pm3.6$ \\        
    22928*  & 3.550840 & $-30.406599$ & 5.051 & $1.61\pm0.32$ & $1100\pm110$ & $44.4\pm0.2$ & $6.7^{+0.2}_{-0.2}$  & $6.3\pm3.0$ \\  
    34016  & 3.604805 & $-30.369871$ & 4.702 & $1.71\pm0.34$ & $2150\pm580$ & $43.7\pm0.3$ & $6.9^{+0.3}_{-0.9}$ & $1.2\pm1.2$  \\ %
    66543*  & 3.584759 & $-30.343629$ & 4.464 & $1.81\pm0.36$ & $4540\pm50$ & $46.3\pm0.2$ & $8.8\pm0.2$  & $30.9\pm5.4$ \\ 
    69688  & 3.569437 & $-30.348231$ & 4.307 & $2.08\pm0.42$ & $1670\pm50$ & $44.7\pm0.2$ & $7.2\pm0.2$ & $10.1\pm5.2$  \\  
    75753*  & 3.530008 & $-30.358013$ & 4.966 & $2.24\pm0.44$ & $1240\pm90$ & $44.4\pm0.2$ & $6.6\pm0.2$  & $4.3\pm2.5$ \\  \hline

    \end{tabular}
    \label{tab:LRDsample}
\end{table*}

\section{Sample} \label{sec:sample}
\subsection{AGN sample}\label{sec:sample_AGN}
In the ALT data, H$\alpha$ is covered at $z=4-5$, which is therefore our prime redshift of interest. Broad H$\beta$ lines could be found to higher redshifts \citep[e.g.][]{Kokorev23}, however, in broad-line AGN identified with {\it JWST}, the broad component of H$\beta$ line typically is more than three times fainter than that of the H$\alpha$ line \citep{Brooks24}. Further our environment measurements cover a smaller dynamic range at $z\sim6$, simply because there are less massive galaxies at earlier times due to the buildup of the galaxy stellar mass function.

\begin{figure*}
    \centering
    \begin{tabular}{ccc}\hspace{-0.5cm}
    \includegraphics[width=6.1cm]{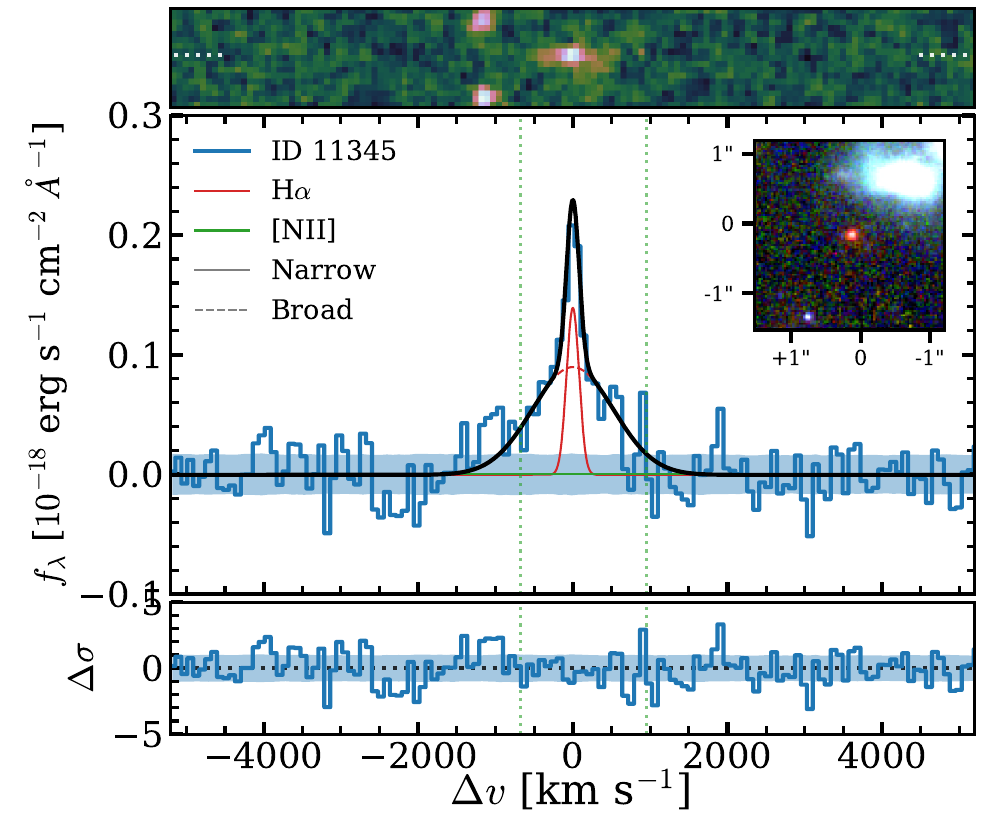} & \hspace{-0.55cm}
    \includegraphics[width=6.1cm]{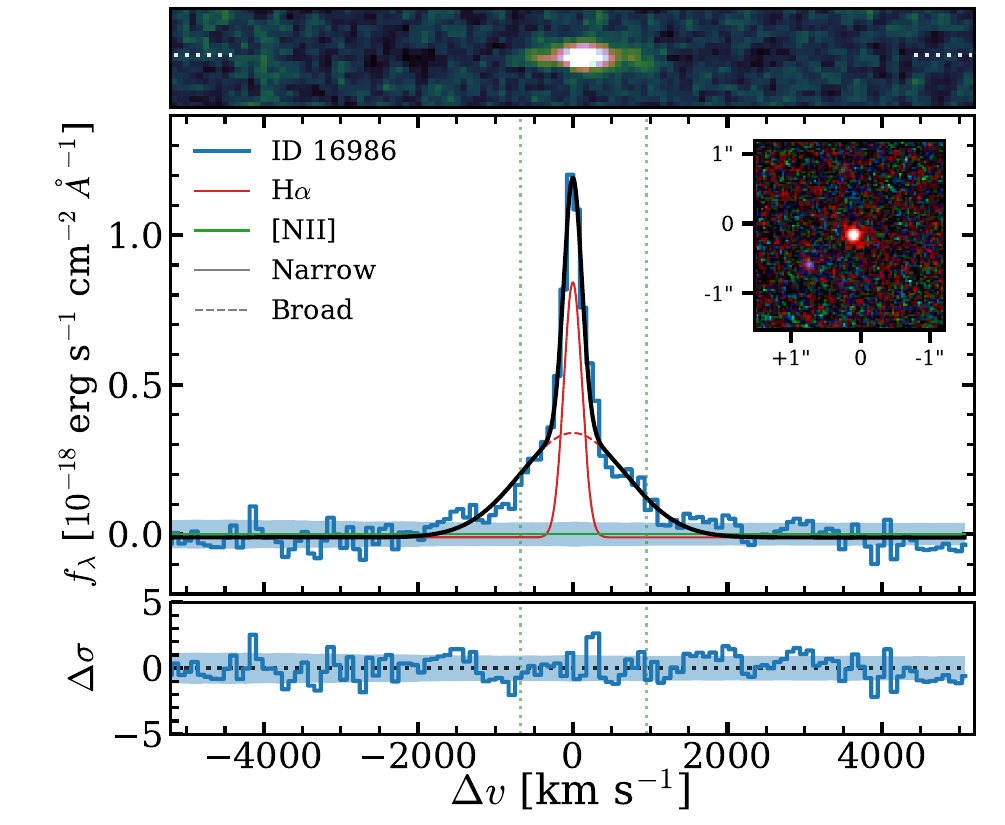} & \hspace{-0.5cm}
    \includegraphics[width=6.1cm]{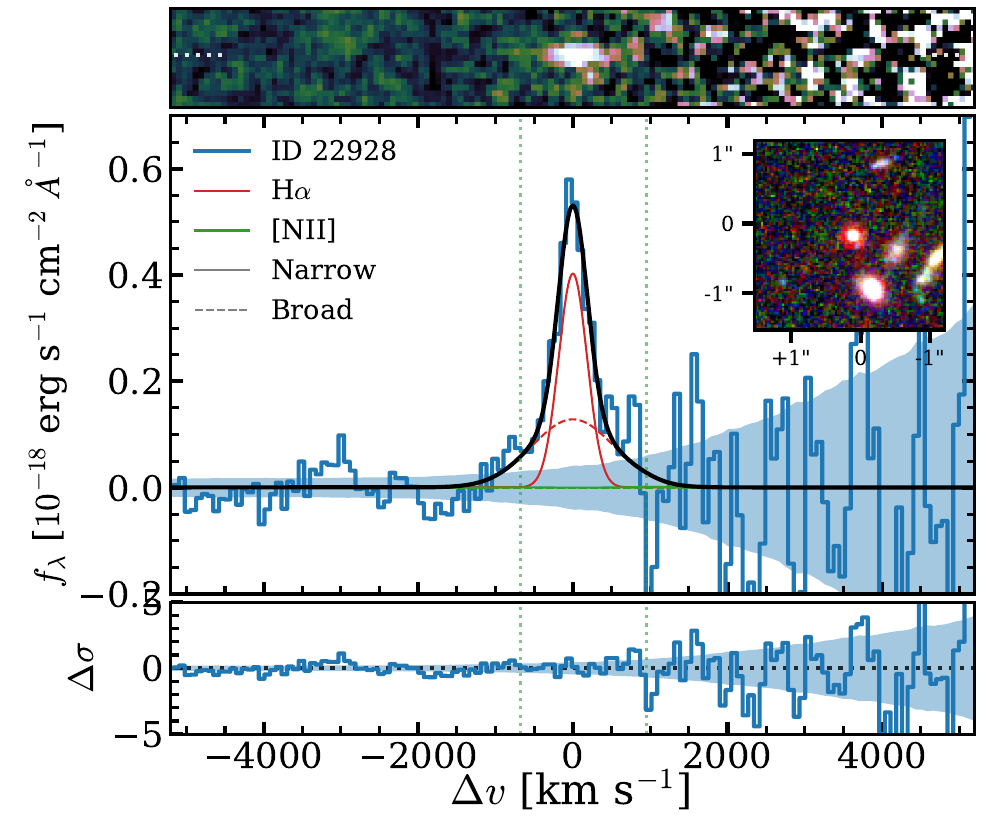} \\ \hspace{-0.5cm}
    \includegraphics[width=6.1cm]{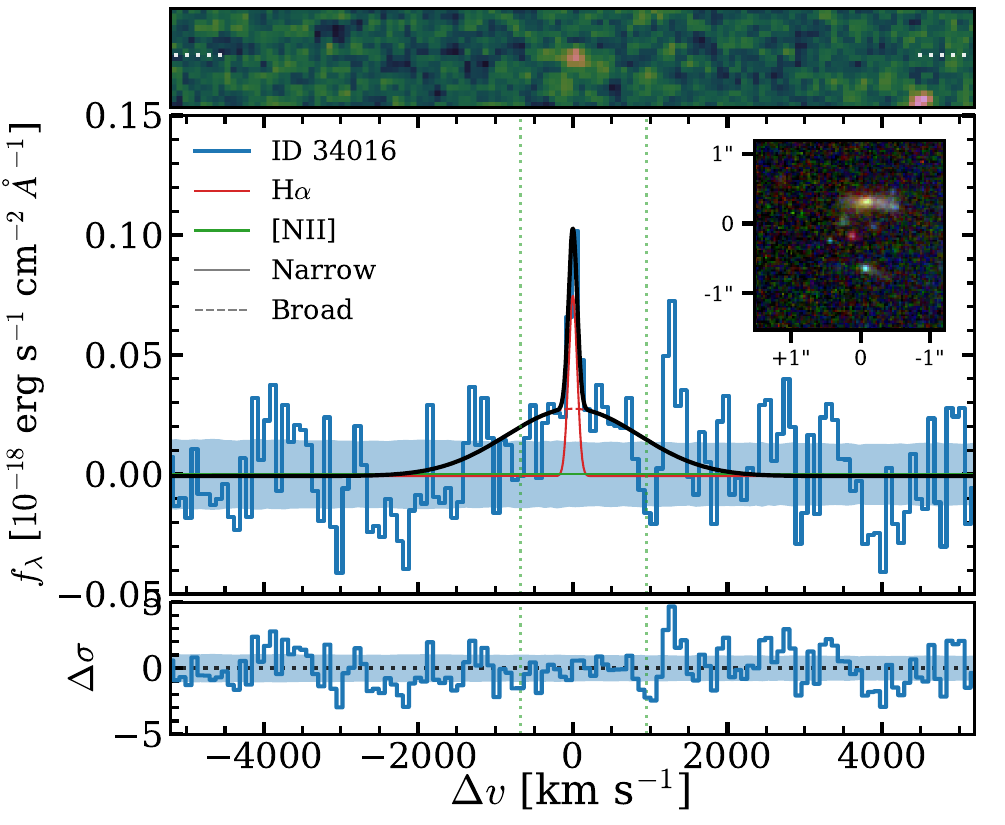} &\hspace{-0.55cm}
    \includegraphics[width=6.1cm]{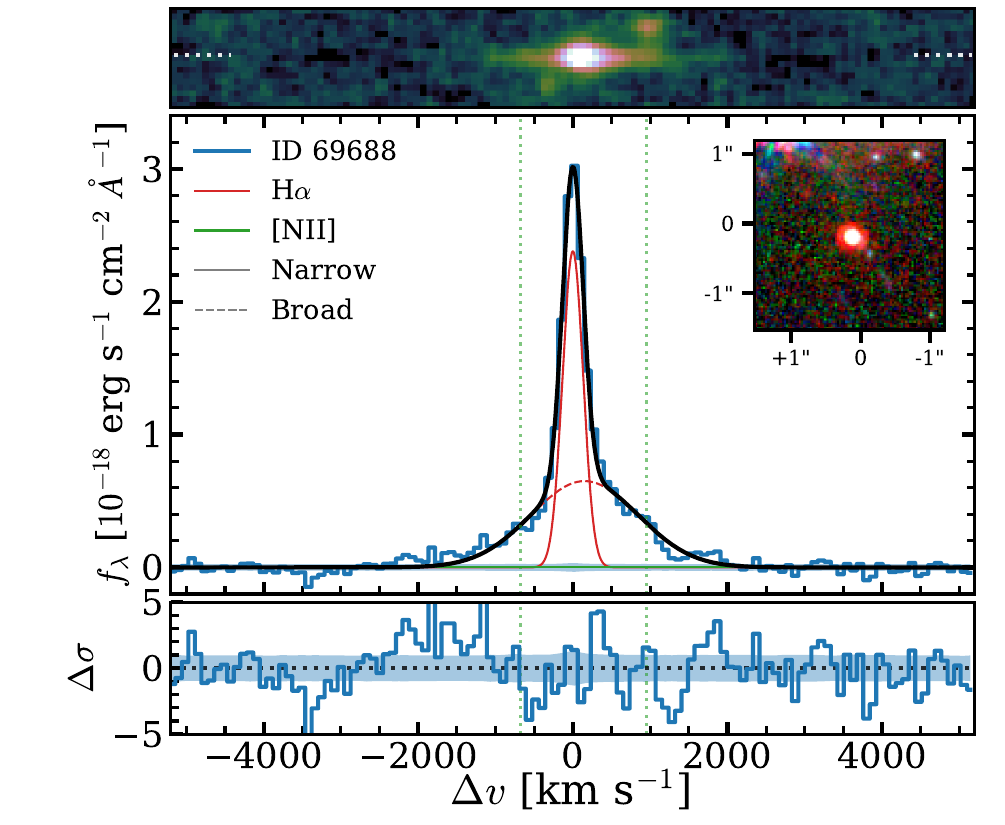} &\hspace{-0.5cm}
    \includegraphics[width=6.1cm]{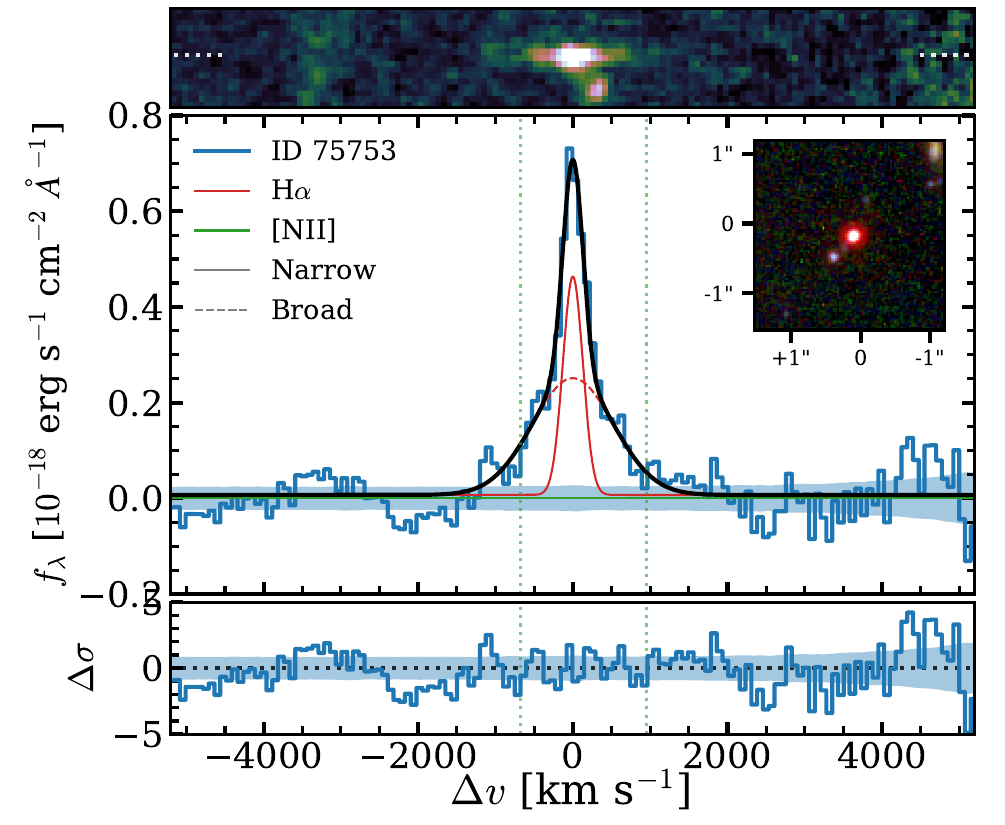} \\
    \end{tabular}
    \caption{The H$\alpha$ profiles of the main sample of BL-H$\alpha$ emitters studied in this paper measured in the NIRCam grism data. Top panels show the 2D continuum-subtracted grism spectra. The middle panels show the optimally extracted 1D spectra. Blue lines show the data, where shaded regions show the errors. The black line shows the combined fit that is composed of a narrow and a broad H$\alpha$ line and narrow [NII]. The red dashed and solid components show the broad and narrow H$\alpha$ component and green shows the best-fit [NII] line, whose wavelength we highlight with dotted green lines.  Bottom shows the residuals of the spectral fit. Inset panels show false-color rest-frame optical RGB images constructed from NIRCam F115W/F200W/F356W images, highlighting the point-source morphology of the objects. }
    \label{fig:Lineprofiles}
\end{figure*}

We select broad H$\alpha$ line emitters in the ALT data by inspecting all 628 H$\alpha$ emitters in the $z=3.8-5.05$ range. Following the methods outlined in \cite{Matthee24}, we optimise the continuum subtraction of their grism spectra by explicitly masking a large region around the H$\alpha$ line when estimating the continuum level, in order to prevent over-subtraction of broad components and then fit the H$\alpha$+[NII] complex with a combination of a narrow and a broad gaussian component. 

Seven broad H$\alpha$ line emitters are identified at $z=3.99-5.05$ with a robust detection (S/N$>5$) of a broad component with $v_{\rm FWHM}>1000$ km s$^{-1}$. Figure $\ref{fig:Lineprofiles}$ shows the fitted H$\alpha$ line profiles of 6/7 BL-H$\alpha$ emitters in our sample. The H$\alpha$ profile of the luminous ALT-66543 has been presented in Labbe et al. in prep. We also show false-color NIRCam images that illustrate their compact, red appearance.\footnote{The spectrum of the exceptionally luminous object ALT-66543 has such high signal to noise that we fit the broad component with a Lorentzian wing (see Labbe et al. in prep for details; which also discusses the detection of H$\alpha$ absorption in its spectrum). } Three of the seven broad H$\alpha$ line-emitters have been identified already in \cite{Greene23}. ALT-16986 was selected by \cite{Labbe23}, but not yet followed-up. The other three were not in their colour-selected parent sample, suggesting that not all BL-H$\alpha$ emitters display the so-called v-shaped spectrum, with a blue UV and a red optical continuum. Indeed, these are somewhat less red in the optical continuum (34016, 69688), or are relatively red in the UV (11345 with $\beta=-1.3$). No previously known broad H$\alpha$ emitter in our coverage has been missed by our selection.

After correcting for magnification, we find broad H$\alpha$ line luminosities ranging from $(1.7 - 1700)\times10^{41}$ erg s$^{-1}$, with a median of $10^{42}$ erg s$^{-1}$, with line-widths 1240 - 4540 km s$^{-1}$ (median 1350 km s$^{-1}$). Assuming that the broad lines originate from AGN activity (see for e.g., \citealt{Matthee24} for detailed arguments backing the AGN assumption, and also Section $\ref{sec:implication_AGNnature}$), we infer the SMBH mass and the bolometric luminosity from the H$\alpha$ line properties following \cite{Reines13,ReinesVolonteri15} and \cite{GreeneHo2005,Richards08}, respectively, assuming no attenuation corrections that could lead to a higher BH mass. The basic properties of our sample of broad-line H$\alpha$ emitters are listed in Table $\ref{tab:LRDsample}$. 

\begin{figure}
    \centering
    \includegraphics[width=8.6cm]{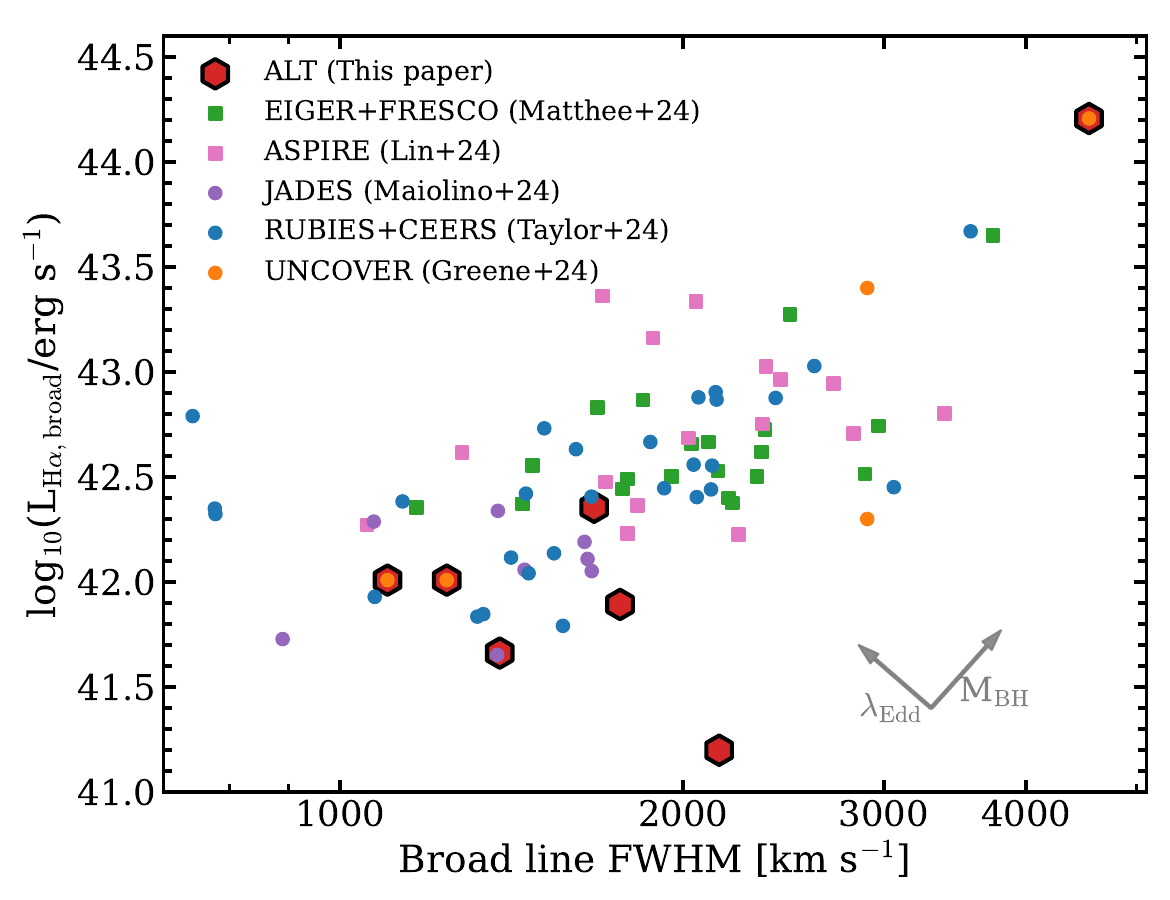}
    \caption{Compilation of H$\alpha$ broad lines measured with {\it JWST} at $z=4-6$ from NIRCam Grism (red hexagons this paper; green squares EIGER \& FRESCO and ASPIRE in pink) and NIRspec (points, purple JADES, RUBIES and CEERS in blue and UNCOVER in orange) spectroscopy. We highlight how the H$\alpha$ luminosity and width change with SMBH mass and Eddington ratio under the commonly adopted calibrations used. Three of the ALT sources are also in the UNCOVER sample. }
    \label{fig:lha_fwhm}
\end{figure}

In Fig. $\ref{fig:lha_fwhm}$, we show a compilation of broad H$\alpha$ line luminosities and line-widths at $z=4-5$ from various surveys. The NIRCam grism surveys include ASPIRE \citep{Lin24}, EIGER, FRESCO \citep{Matthee24} and ALT (This paper). The NIRspec surveys include JADES \citep{Maiolino23b}, a combination of the CEERS (which were included in e.g. \citealt{Harikane23} and \citealt{Kocevski23}) and RUBIES \citep{deGraaff24_survey} surveys presented in \cite{Taylor24} and from the UNCOVER program \citep{Greene23}. Due to the depth of the ALT data compared to shallower grism surveys and the lensing magnification, the majority of the BL-H$\alpha$ emitters in the ALT data probe the faint-end of the sample, overlapping with the faintest samples from JADES, UNCOVER and CEERS obtained through NIRspec spectroscopy. Fig. $\ref{fig:lha_fwhm}$ also highlights how unrepresentative the luminous ALT-66543 (Labbe et al. in prep) is for the overall BL-H$\alpha$ sample, warranting why we exclude it from calculations of sample averages in this paper.

\begin{figure*}
    \centering
    \begin{tabular}{ccc}
    \includegraphics[width=5.6cm]{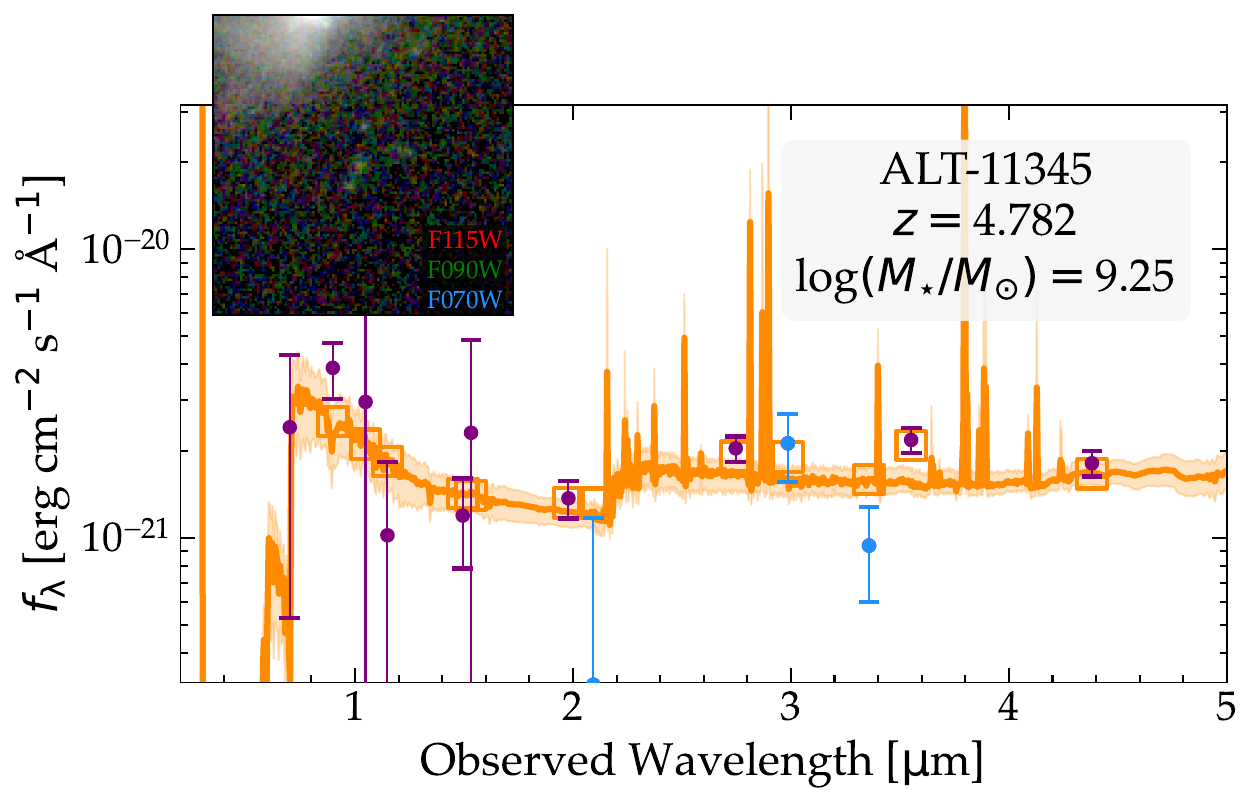}  &
    \includegraphics[width=5.6cm]{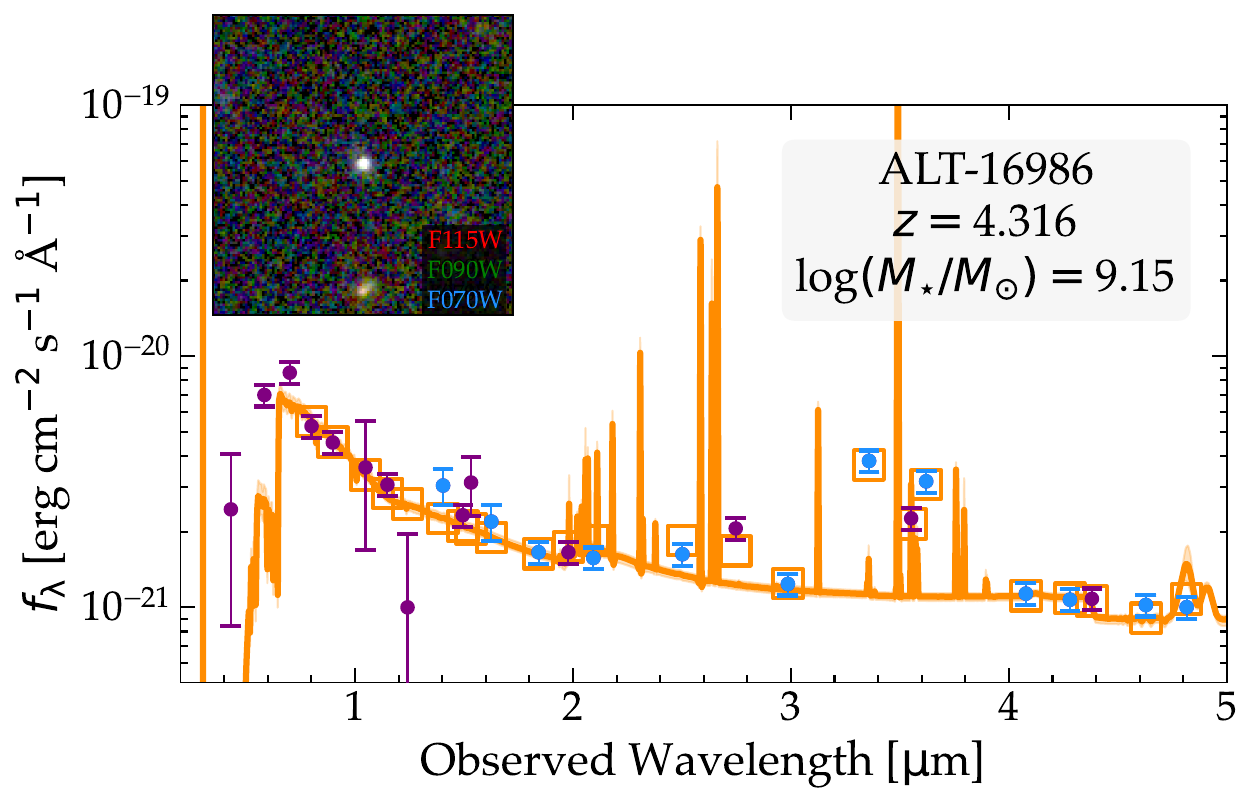}  & 
    \includegraphics[width=5.6cm]{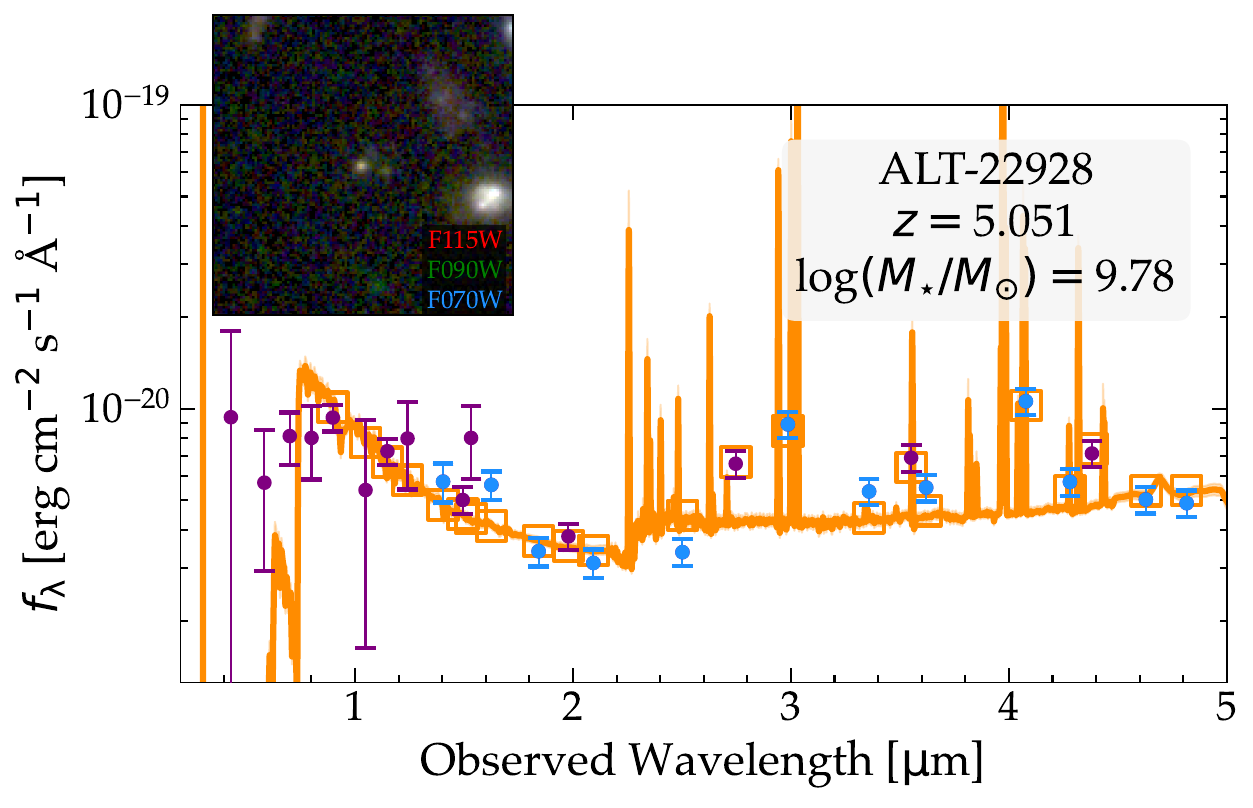}  \\  
    \includegraphics[width=5.6cm]{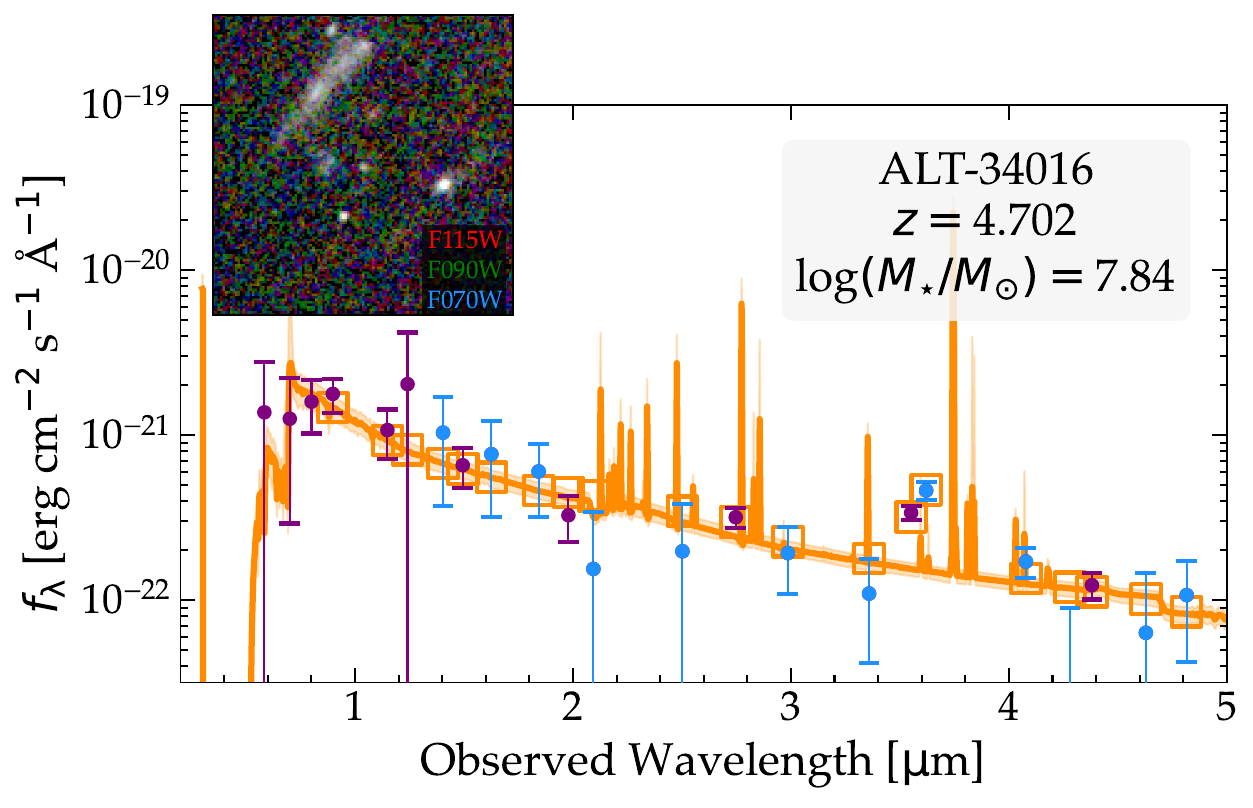}  &
    \includegraphics[width=5.6cm]{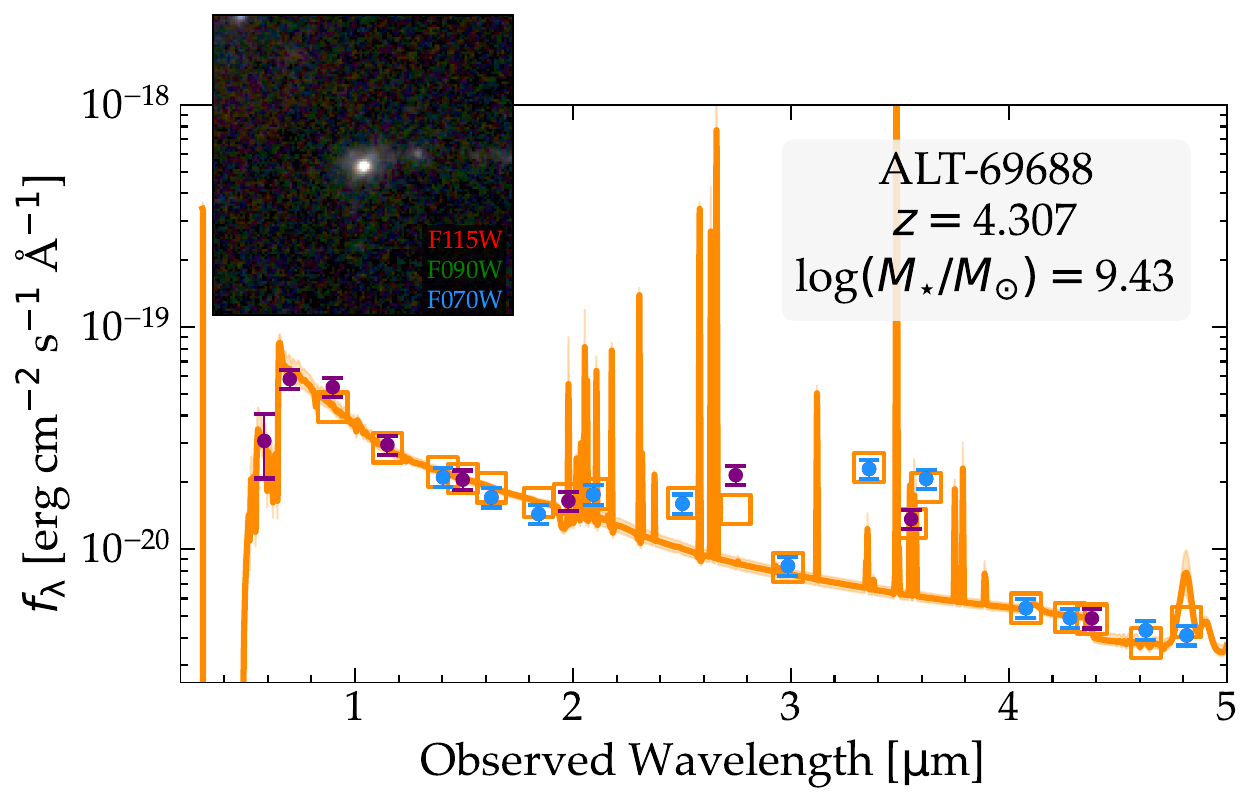}  & 
    \includegraphics[width=5.6cm]{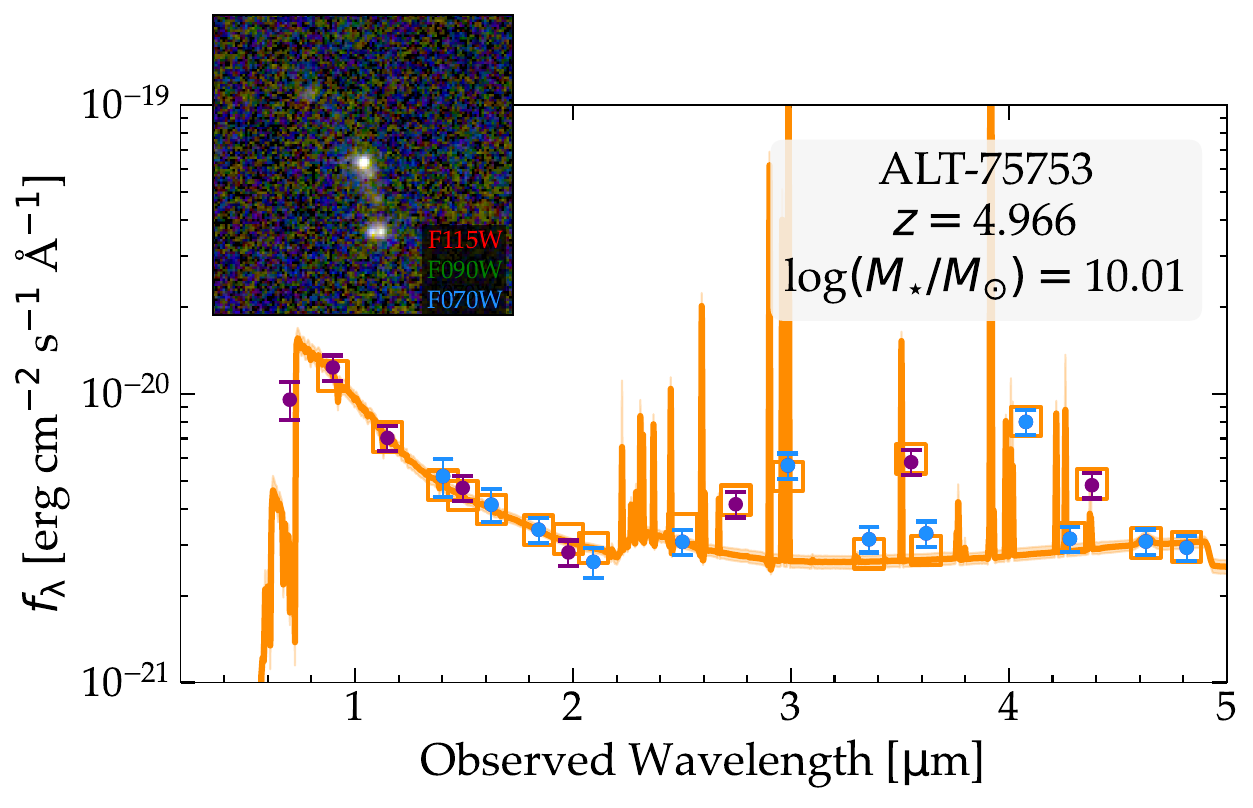}   \\
    \end{tabular}
    \caption{SED fits of the six BL-H$\alpha$ emitters that are the main focus of this paper. Orange curves and shaded regions show the best fit SEDs {\it assuming pure stellar and nebular emission} and their uncertainties that ignore an AGN contribution. Purple data points are measurements in broad-band filters, while blue data-points are medium-band filters. Inset stamps are false-color rest-frame UV RGB images of $2.4''\times2.4''$ based on the F070W, F090W and F115W NIRCam imaging data.} 
    \label{fig:LRDs_SEDs}
\end{figure*}

We used {\sc prospector} to fit the SEDs of all emission-line galaxies in the ALT survey with stellar population models that self-consistently include the nebular emission from the fitted stellar populations, as detailed in \citetalias{NaiduALT24} \citeyear{NaiduALT24}. All {\it JWST} photometry is used in these fits, excluding filters that include flux below 1240 {\AA} such that the fit is not impacted by variations in the Lyman-$\alpha$ emission line or the IGM transmission bluewards of the Ly$\alpha$ break. The emission-line fluxes measured from the grism data are not used in the fitting, but we note that various strong emission-lines typically strongly boost the photometry in medium-bands. Importantly, for almost all sources the available photometry also covers the optical continuum, free from emission-line contamination.  

In Figure $\ref{fig:LRDs_SEDs}$, we show the spectral energy distributions of the BL-H$\alpha$ emitters inferred from photometry. We stress that the SED fits for the BL-H$\alpha$ emitters are displayed primarily for illustrative purposes, as no AGN component is included in the fits. Generally, the {\sc prospector} models are flexible enough to yield reasonable fits to the data, with relatively high stellar masses M$_{\rm star} = 10^{7.7 - 10}$ M$_{\odot}$. However, as discussed in detail in for example \cite{BWang24}, the inclusion of a red AGN component could lead to drastically lower stellar masses (i.e. $\sim1$ dex). Inspecting the SEDs in Figure $\ref{fig:LRDs_SEDs}$ in detail, we remark somewhat poorer performance around the Balmer break region (observed wavelength $\approx2-2.5 \mu$m for most sources), in particular in ALT-22928, ALT-69688, and ALT-75753. Comparing the derived properties to the main galaxy sample at $z\sim4-5$, we notice that the BL-H$\alpha$ emitters relatively uniquely combine a high inferred stellar mass with a young age.

\subsection{Galaxy sample}\label{sec:sample_galaxy}
In order to accurately and uniformly map the environment of the BL-H$\alpha$ emitters, we only use galaxies with spectroscopic redshifts and with a conservative line luminosity threshold that effectively yields a volume-limited sample. While the photometric redshifts in our survey field are generally very accurate, this is not always true for specific redshifts (see \S5.5 of \citetalias{NaiduALT24} \citeyear{NaiduALT24} for a detailed discussion), as the photometric redshift accuracy may depend significantly on the accuracy of the (wings of the) filter transmission curves, and any redshift error could propagate strongly in the measured redshift difference between galaxy pairs. We impose a limiting H$\alpha$ luminosity of $2\times10^{41}$ erg s$^{-1}$, which corresponds to a typical S/N $>10$ and an unobscured star formation rate threshold of $\gtrsim0.6$ M$_{\odot}$ yr$^{-1}$ (see Di Cesare et al. in prep). Note that using a fixed luminosity threshold, rather than a flux threshold, also balances the wavelength-dependent sensitivity (that is somewhat better at higher wavelengths). We also ignore any region with a magnification $\mu>3$ such that our on-sky distributions are not strongly impacted by differential volumes due to changes in magnifications (we have verified that changing limits from $\mu=2-4$ does not impact the results we obtained). As a consequence, our final galaxy sample that we use to measure the environment consists of 308 galaxies with redshifts $z=3.8-5.05$.

\begin{figure}
    \centering
    \includegraphics[width=8cm]{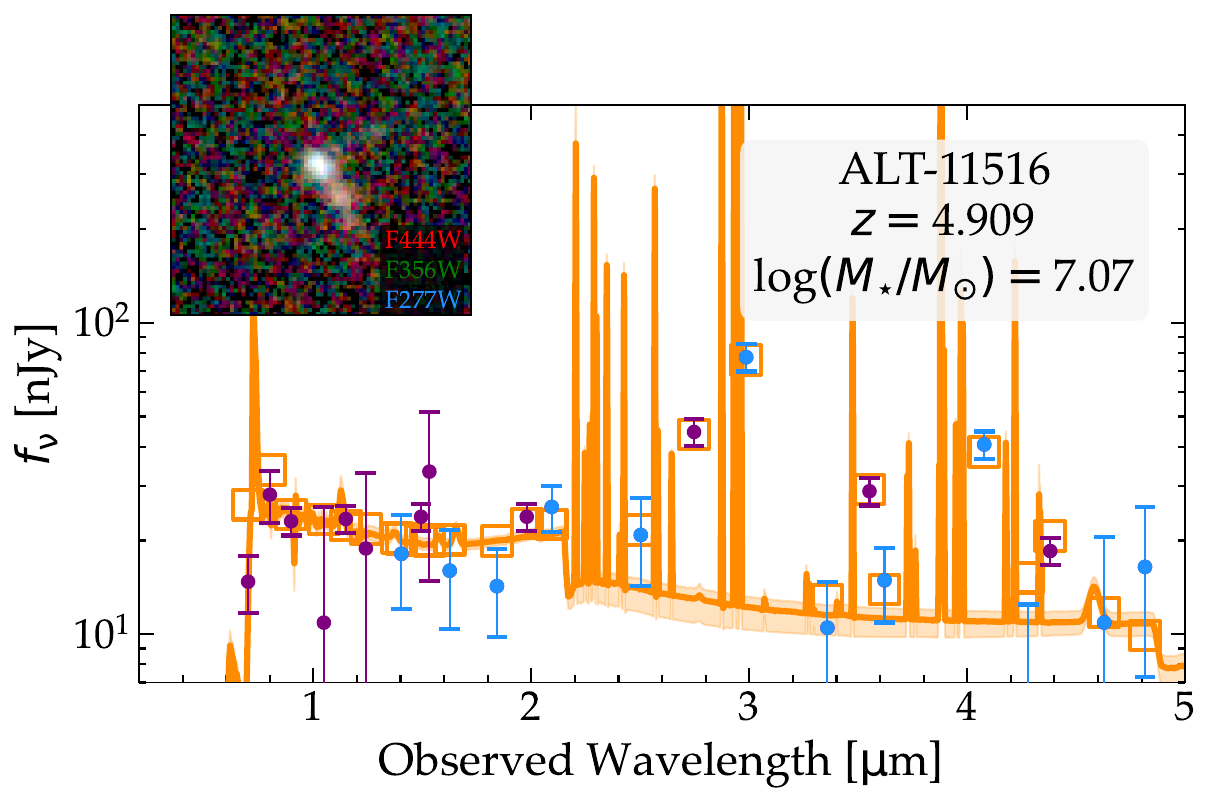}  \\ 
    \includegraphics[width=8cm]{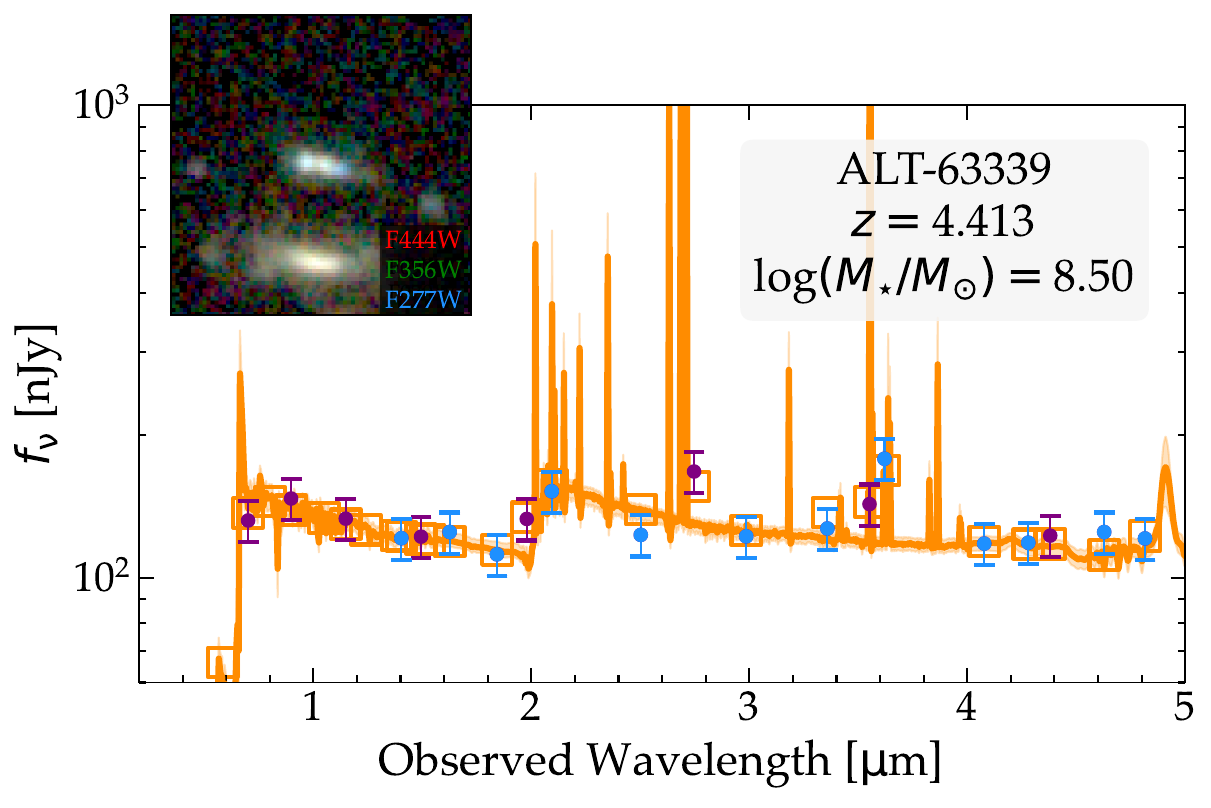}  \\ 
    \includegraphics[width=8cm]{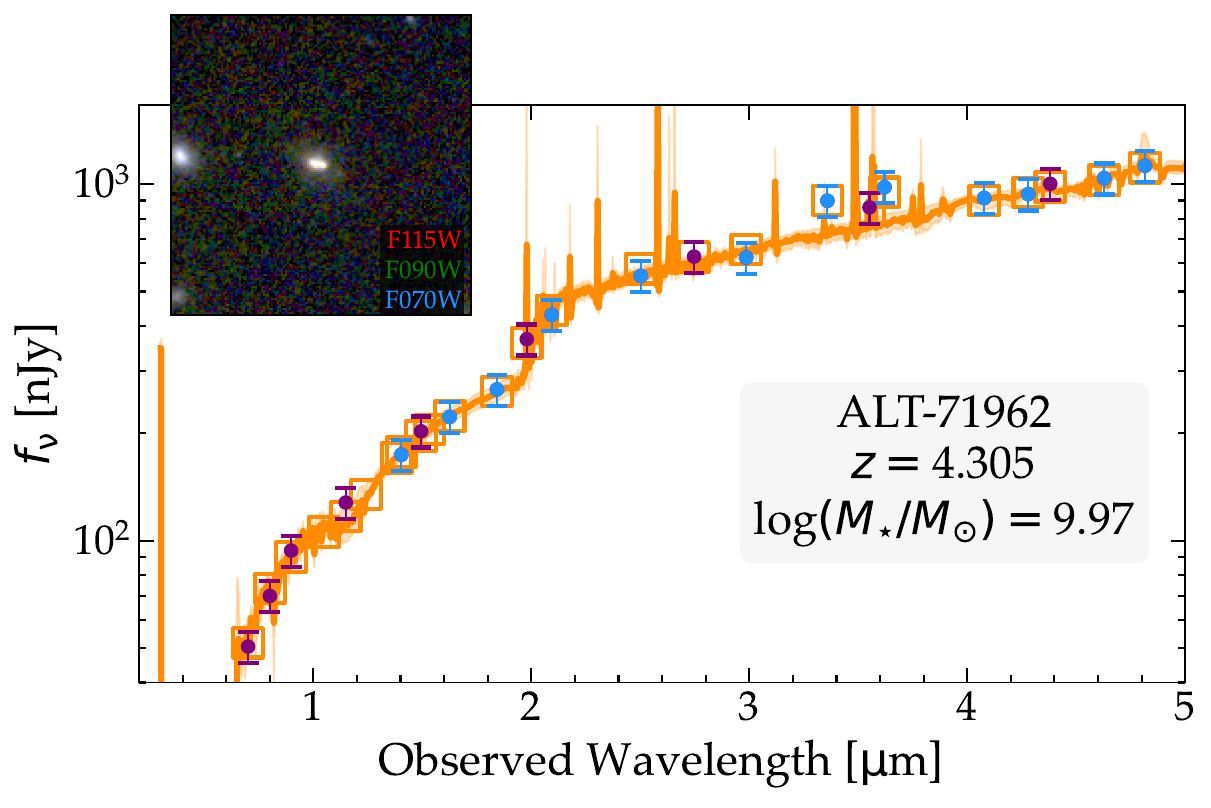}  \\     
    \caption{Example SED fits of galaxies in the reference sample of galaxies without broad H$\alpha$ lines in the ALT data. The masses increase from top to bottom -- $10^{7}$, $10^{8.5}$, $10^{10}$ M$_{\odot}$, respectively. Orange curves and shaded regions show the best fits and their uncertainties. Purple data points are measurements in broad-band filters, while blue data-points are medium-band filters. Inset stamps are false-color RGB images of $3''\times3''$ based on the F070W, F090W and F115W NIRCam imaging data, revealing the diverse and resolved morphologies of these galaxies.} 
    \label{fig:SEDs}
\end{figure}

Based on the {\sc prospector} fits to the photometry, we find that the typical galaxy in our sample has a mass of $10^8$ M$_{\odot}$ (ranging from $5\times10^6$ to $2\times10^{10}$ M$_{\odot}$), a UV luminosity M$_{\rm UV}=-18.8$ (ranging from $-14.5$ to $-21.5$) and an unobscured SFR$_{\rm H\alpha} =1.4$ M$_{\odot}$ yr$^{-1}$ (ranging from $0.6-24$ M$_{\odot}$ yr$^{-1}$). In Figure $\ref{fig:SEDs}$, we show example SED fits of galaxies in our sample, with stellar masses of $10^{7}$, $10^{8.5}$, $10^{10}$ M$_{\odot}$, from top to bottom, respectively. The models are very tightly constrained thanks to the spectroscopic redshift and the large number of filters, implying that stellar mass uncertainties are limited by systematics. We also note that the SED of ALT-71962, a normal galaxy with narrow H$\alpha$ emission and with a similar fitted mass as some BL-H$\alpha$ emitters, is characterized by a relatively old stellar population, with a red UV slope.

 \begin{figure}
     \centering
     \includegraphics[width=8.6cm]{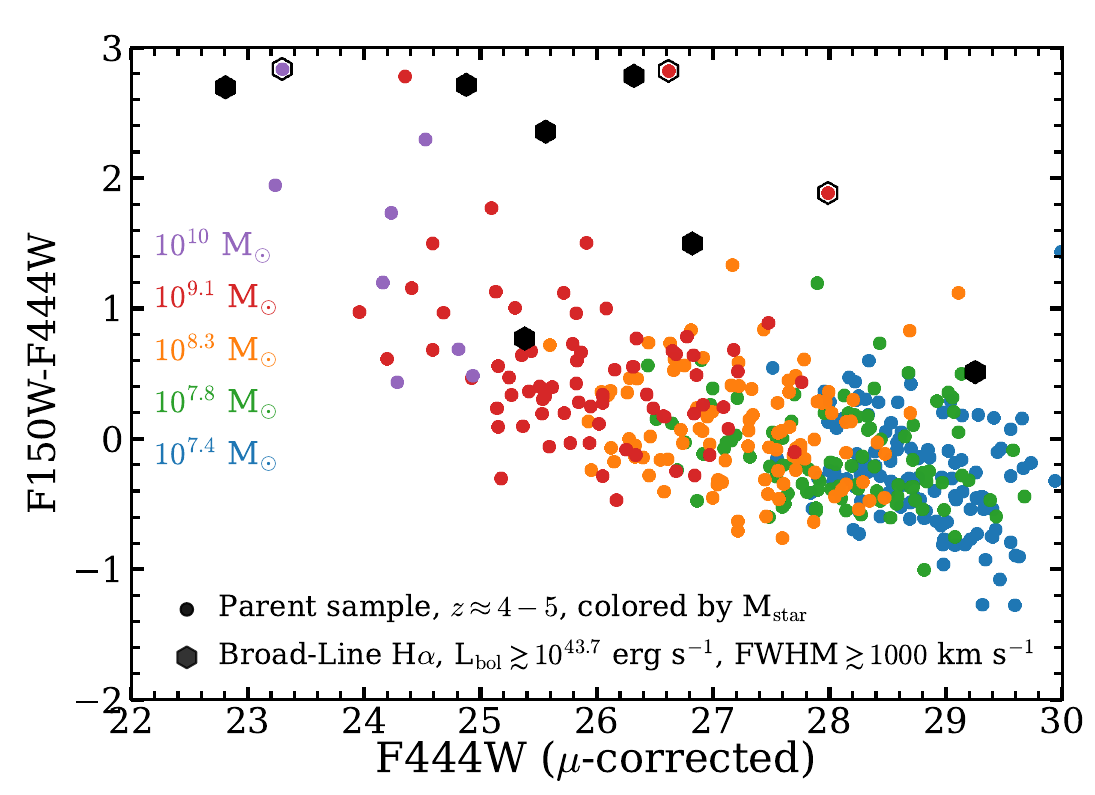}
     \caption{The observed F150W-F444W colors versus the (magnification corrected) F444W magnitude of our galaxy (points colored by their stellar mass) and Broad-Line H$\alpha$ samples (black hexagons). The F150W-F444W colors probe the difference in the continuum level from rest-frame wavelengths $\lambda_0=0.27-0.8 \mu$m, free from emission-line contamination. Open hexagons show sources with suspected broad lines (based on the colors and morphology). The stellar masses of the sample correlate with the F444W magnitude and range from $\approx2\times10^7$ M$_{\odot}$ (blue) to $\approx10^{10}$ M$_{\odot}$ (purple). }
     \label{fig:sample}
 \end{figure}

\subsection{The colours of normal and broad H$\alpha$ line emitters}
Figure $\ref{fig:sample}$ shows the distribution of the F150W-F444W colours from the galaxy sample versus the (magnification corrected) F444W magnitude, where galaxies are colored by their stellar mass. We chose to use observed magnitudes and colors for a model-independent comparison between the galaxies and the BL-H$\alpha$ emitters. At $z=4-5$, the F150W and F444W photometry is not sensitive to strong emission-lines and probes the difference in the continuum level from rest-frame wavelengths $\lambda_0=0.27-0.8 \mu$m and thus captures the Balmer discontinuity that is of particular interest \citep[e.g.][]{Setton24b, BWang24}. 

Galaxies with a fainter F444W magnitude are generally less massive and they have flat or even blue F150W-F444W colors, the latter indicative of Balmer jumps \citep[e.g.][]{Katz24}. Massive galaxies tend to have redder F150W-F444W colors due to Balmer breaks, but few as red as the colors of the most luminous BL-H$\alpha$ emitters. Generally, all BL-H$\alpha$ emitters have red F150W-F444W colors, suggesting that the AGN activity is reddening the rest-frame optical colours. Four BL-H$\alpha$ emitters have extremely red F150W-F444W colors ($>2$), that are particularly exceptional among the fainter magnitudes (F444W$>25$). Two of the BL-H$\alpha$ emitters (69688 and 34016) are significantly less red and have colours similar to galaxies without broad H$\alpha$ lines. Object 34016 has the weakest and least significant broad H$\alpha$ line of our sample, whereas 69688 has a relatively strong narrow component on top of a significant broad component. This is in agreement with the sample presented in \cite{Matthee24} that showed a correlation between optical redness and broad to total H$\alpha$ flux. There are two galaxies with a relatively high mass given their faint F444W magnitude (F444W$>26$; ALT IDs 16772 and 62404\footnote{Coordinates and redshifts can be found in the public ALT catalog (\citetalias{NaiduALT24} \citeyear{NaiduALT24}) available at: \url{https://zenodo.org/records/13871850}}). These have colours as the typical BL-H$\alpha$ emitters, but our data is likely not sensitive to identify their broad H$\alpha$ components. The luminous, red galaxy ALT-62975 is a possible AGN due to its compact appearance, but the H$\alpha$ profile analysis is complicated by an ongoing merger such that the two components are separated almost exactly in the direction of the (single available) dispersion angle, complicating the identification of a possible faint broad component, see Appendix $\ref{app:extra}$.

\section{The environments of galaxies and of Little Red Dots} \label{sec:environment}
\subsection{Redshift distribution}
\begin{figure}
    \centering
    \includegraphics[width=8.7cm]{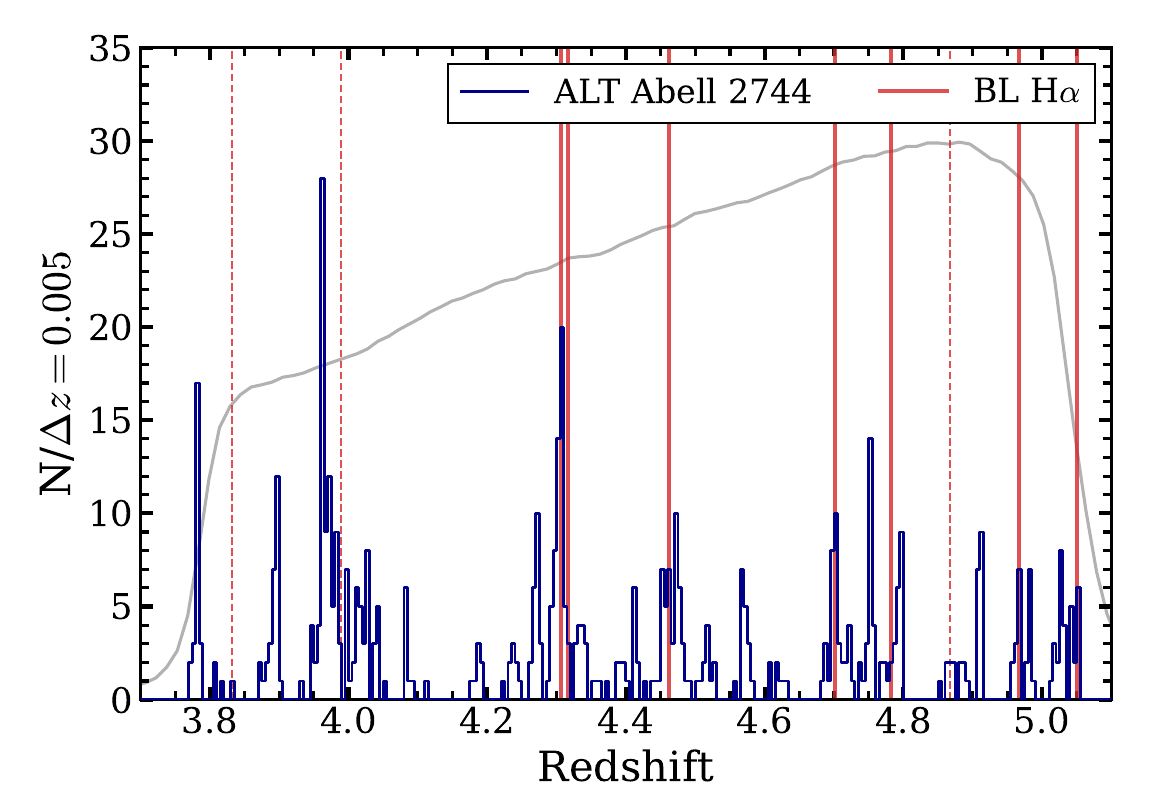}
    \caption{The redshift distribution of our galaxy and BL-H$\alpha$ sample behind the Abell 2744 lensing cluster. Red lines mark the redshifts of the BL-H$\alpha$ emitters. Dashed lines show the redshifts of the tentative broad line emitters discussed in the text and highlighted in Fig. $\ref{fig:sample}$. The grey curve shows the F356W filter curve that was combined with the grism observations that we use.}
    \label{fig:zdist}
\end{figure}

In Figure $\ref{fig:zdist}$ we show the redshift distribution of the galaxy sample and we highlight the redshifts of the BL-H$\alpha$ emitters. Most galaxies are found in about ten redshift spikes, i.e. $\approx50$ \% of the galaxies are found within 10 regions of $\Delta z<0.02$. The strongest over-densities are at redshifts $z\approx3.8$, $z\approx4.0$ (which is associated with the passive galaxy identified recently by \citealt{Setton24}, and the massive ALT galaxy ALT-62975 that we flag as suspected AGN), $z\approx4.3$ and $z\approx4.7$. Besides spikes, there are also various notable under-densities, such as the one at $z\approx4.1$. The most massive galaxies (without broad H$\alpha$ emission) are at redshifts $z=3.975, 4.305, 4.296, 4.462$ and $z=4.272$, ordered by stellar mass. Each of these thus corresponds to a large redshift spike, with $\gtrsim10$ galaxies within 1500 km s$^{-1}$. This suggests that there is a strong correlation between the over-density and stellar mass that we explore and utilize below. The BL-H$\alpha$ emitters tend to be found in somewhat over-dense regions, but none of them is found in the most over-dense region in the field. This is also the case for the BL-H$\alpha$ emitters identified in the FRESCO survey \citep{Oesch23,Helton24,CoveloPaz24}. Whether these results are simply because over-dense regions have more galaxies, thus boosting the likelihood of catching a galaxy with broad-line AGN activity, or whether there is also relation between AGN activity and halo mass, will be explored in the next section.

\begin{figure*}
    \centering
    \begin{tabular}{cc}\vspace{-0.3cm}
    \includegraphics[width=7.8cm]{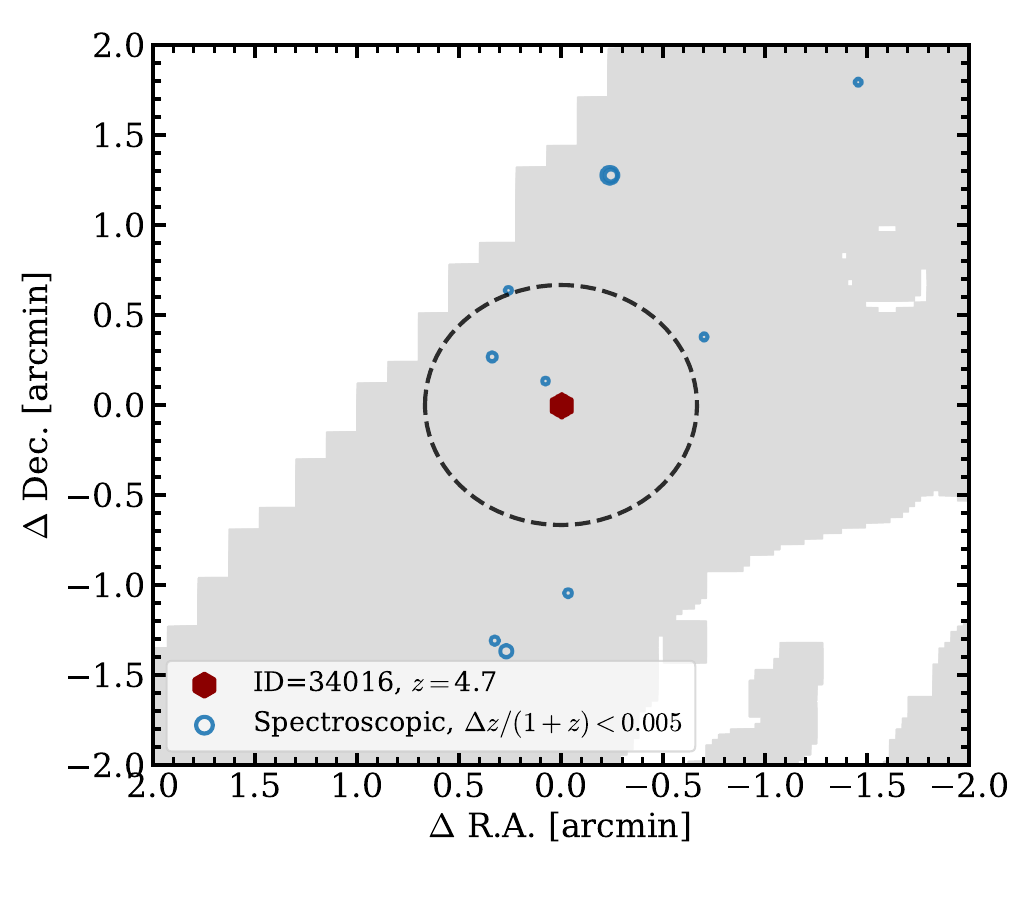} &
    \includegraphics[width=7.8cm]{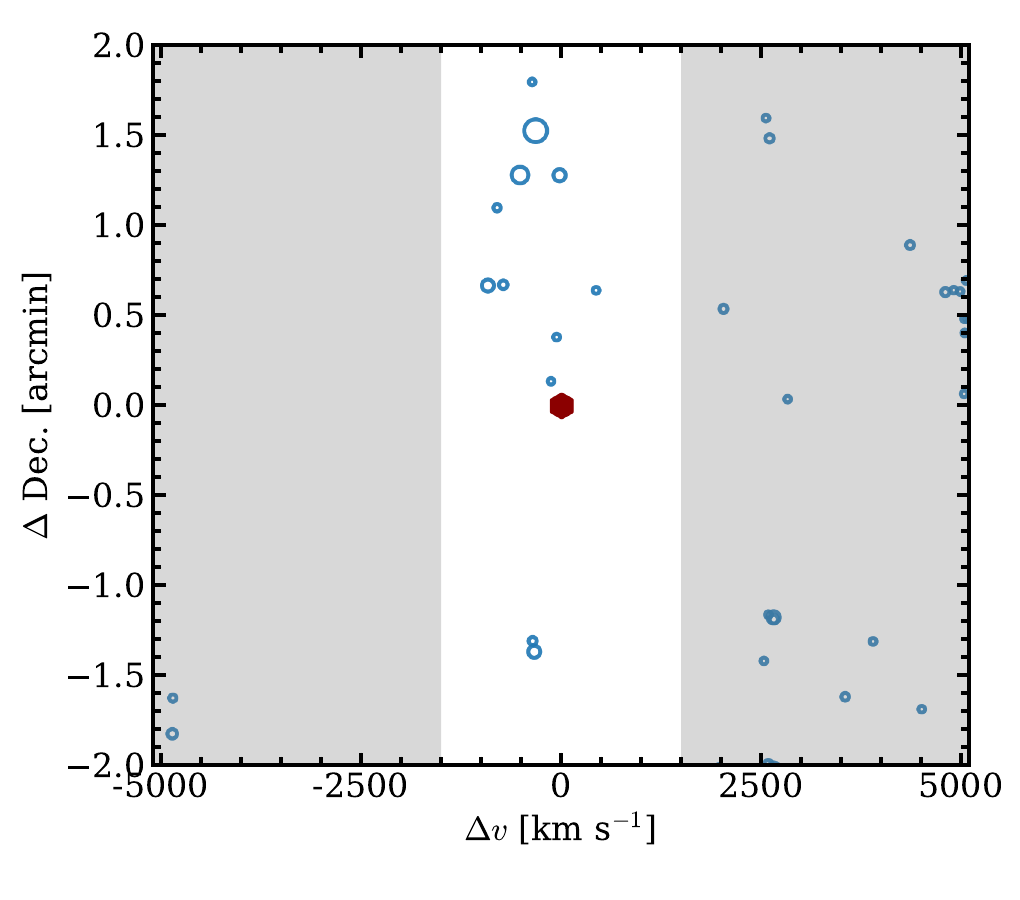} \\ \vspace{-0.3cm}
    \includegraphics[width=7.8cm]{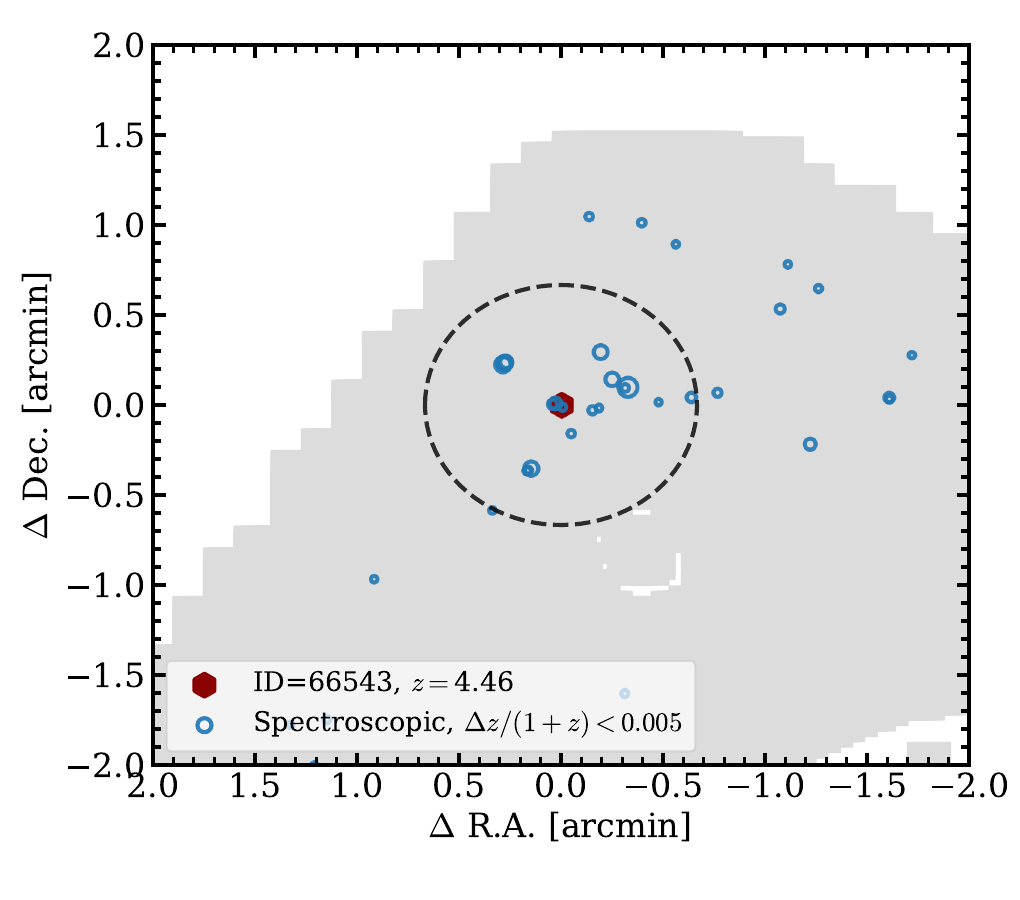} &
    \includegraphics[width=7.8cm]{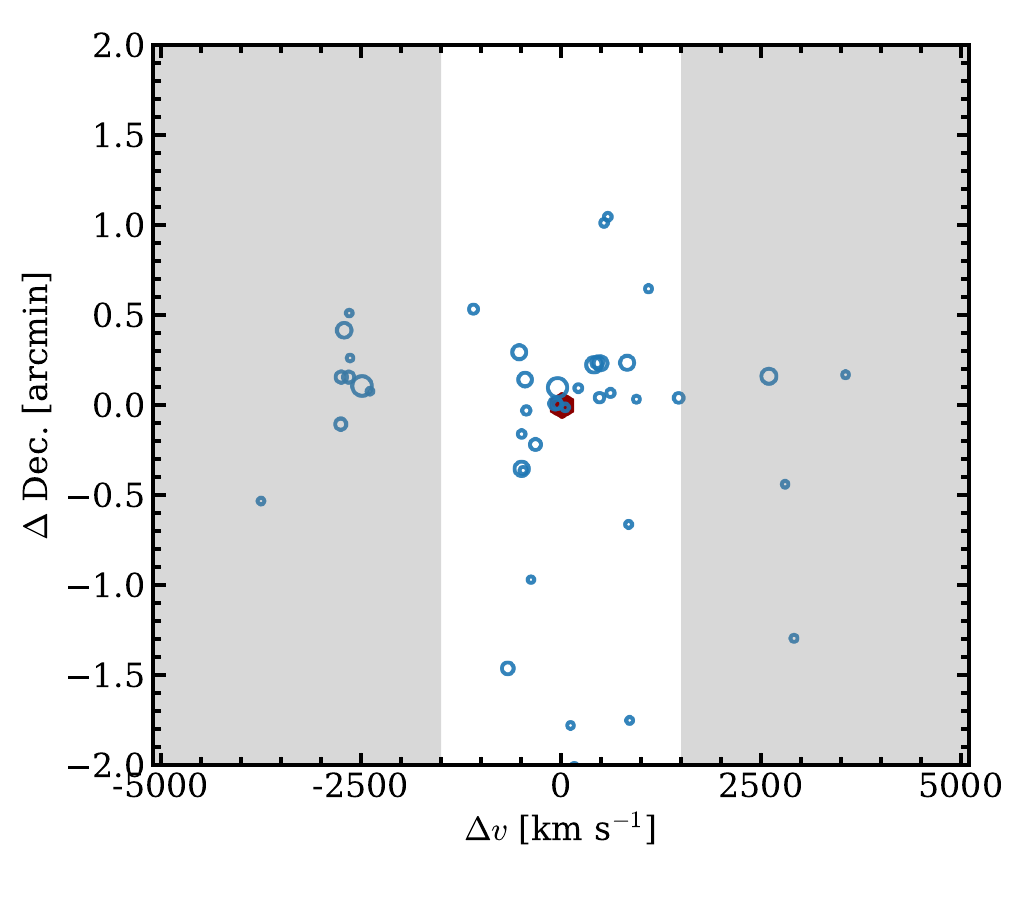} \\
     \includegraphics[width=7.8cm]{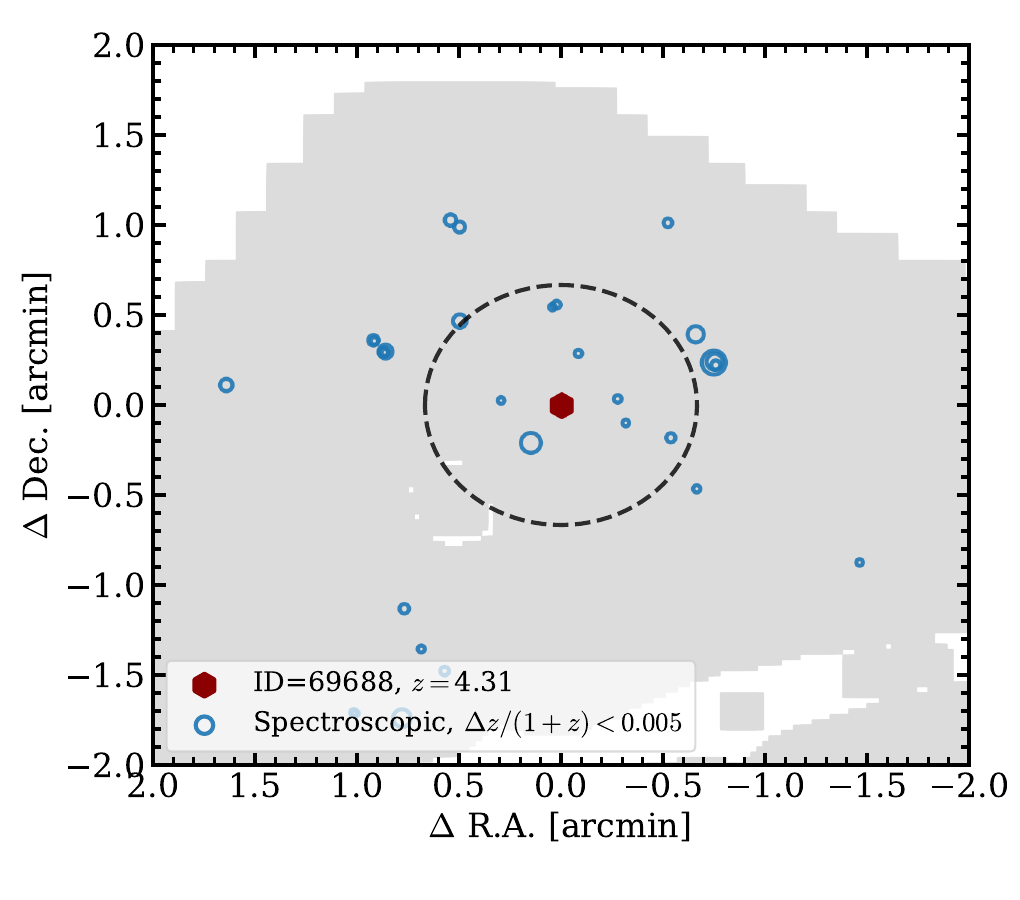} & 
    \includegraphics[width=7.8cm]{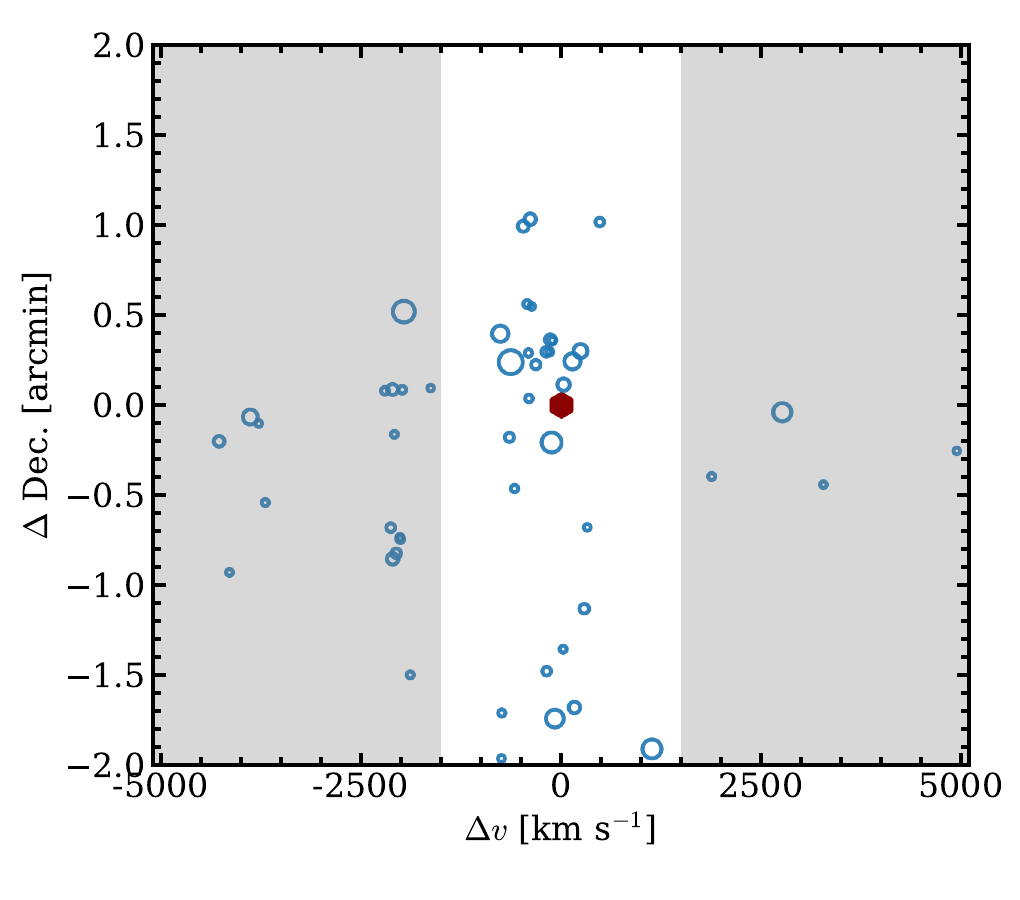} \\
    \end{tabular}
    \caption{The environments of BL-H$\alpha$ emitters (red hexagons), ordered from low density (top) to intermediate density (bottom). Left panels show the sky-plane distribution of the galaxy sample, where the grey region highlights the coverage (with $\mu<3$). Blue circles are our spectroscopic galaxy sample (with a redshift difference $\Delta z/(1+z)<0.005$ to the BL-H$\alpha$ emitter) and their size scales with stellar mass. The dashed circle roughly corresponds to a radius of 1 cMpc in the source-plane. Right panels show the redshifts of the galaxies. Environments of the other BL-H$\alpha$ emitters in our sample are shown in the Appendix.} 
    \label{fig:RADEC_1}
\end{figure*}

\subsection{The projected pair count distributions}  \label{sec:LRDenvs}
In this section we characterize the environment of the BL-H$\alpha$ emitters using the galaxy sample described above. We focus on the projected separations among galaxies in the plane of the sky. We compute source-plane positions based on the spectroscopic redshifts and the lens model from \cite{Furtak23Lens} (updated as described in \citealt{Price24}). We find that the distribution of velocity differences between all pairs of galaxies with separations $(0,2]$ arcmin, has a very tight peak centered around $\Delta v=0$ km s$^{-1}$, with a FWHM of 500 km s$^{-1}$. Therefore, we assume that galaxies with velocity differences $\Delta v < 1500$ km s$^{-1}$ ($\Delta z/(1+z)<0.005$) are associated to each-other. Our results are stable when changing this cut from 500 to 3000 km s$^{-1}$. If redshift differences were purely be due to the Hubble flow, $\Delta v\approx1500$ km s$^{-1}$ corresponds to $\approx3$ cMpc at $z=4.5$.

In Fig. $\ref{fig:RADEC_1}$, we illustrate the environments of three of our BL-H$\alpha$ emitters that display a range in over-densities. The environments of the others are shown in Appendix $\ref{app:A}$. The right panels illustrate that the typical over-densities are confined within $\Delta v < 10000$ km s$^{-1}$, justifying our choice of associating galaxies in a redshift window as narrow as 1500 km s$^{-1}$. A range of environments can be seen among the BL-H$\alpha$ emitters. ALT-34016 has a very typical environment with a few neighbors within a radius of 1 cMpc (that we use as a reference scale, but our results are robust to changes from 0.5 to 4 cMpc), while ALT-66543 and ALT-69688 both show several companions. Noteworthy is that ALT-69688 is at a redshift with a large over-density, but the object itself is not in the spatial center of the over-density (which instead is close to one of the most massive star-forming galaxies in our sample). The extremely luminous ALT-66543, on the other hand, appears in the center of an extreme over-density -- the largest on 1 cMpc scales in our full survey. Generally, these illustrations suggest that BL-H$\alpha$ emitters sample a range in environments, and do not appear strongly correlated with themselves. 

We quantify the over-density of each object by normalising the number of neighbors within a cylinder with radius 1 cMpc and redshift difference $\Delta z/(1+z)<0.005$ to the random expected number counts, i.e. $1+\delta_{<R} = N(<R)/\langle N(<R) \rangle$, where $\delta$ is the over-density within radius $R$. At 1 cMpc, the two-halo clustering term dominates over the one-halo term in low mass galaxies at high-redshift such as Lyman-$\alpha$ emitters \citep[e.g.][]{Herrero23}. The random expectation is empirically measured using the average of the number counts measured around a uniform grid of source-plane coordinates (RA, DEC, $z$), correcting for the fraction of the volume within radius $R$ that is covered by our survey and has a $\mu<3$. At our reference scale of 1 cMpc, this number is 0.8 per $\Delta z/(1+z)=0.005$. Table $\ref{tab:LRDsample}$ lists the over-densities for the BL-H$\alpha$ emitters, which range from $1+\delta=1.2$ to $30.9$, with a median of $5.3\pm1.4$ (mean $5.6\pm1.2$) when excluding ALT-66543 which has an exceptional luminosity and line-width and is therefore not representative (Fig. $\ref{fig:lha_fwhm}$).

\subsection{Comparison to SFGs} \label{sec:comparison_sfgs}
In order to interpret the over-densities measured around the BL-H$\alpha$ emitters, we here perform an empirical comparison to over-densities measured around the other galaxies in our sample that do not show broad H$\alpha$ line-emission. The SEDs of these galaxies are well described with stellar population models (see Fig. $\ref{fig:SEDs}$) and we can therefore infer their stellar masses, and subsequently investigate the relationship between stellar mass and over-density factor. The main motivation for this approach is that this reference sample is subject to similar systematic effects in their over-density measurements.

\begin{figure}
    \centering
    \includegraphics[width=8.7cm]{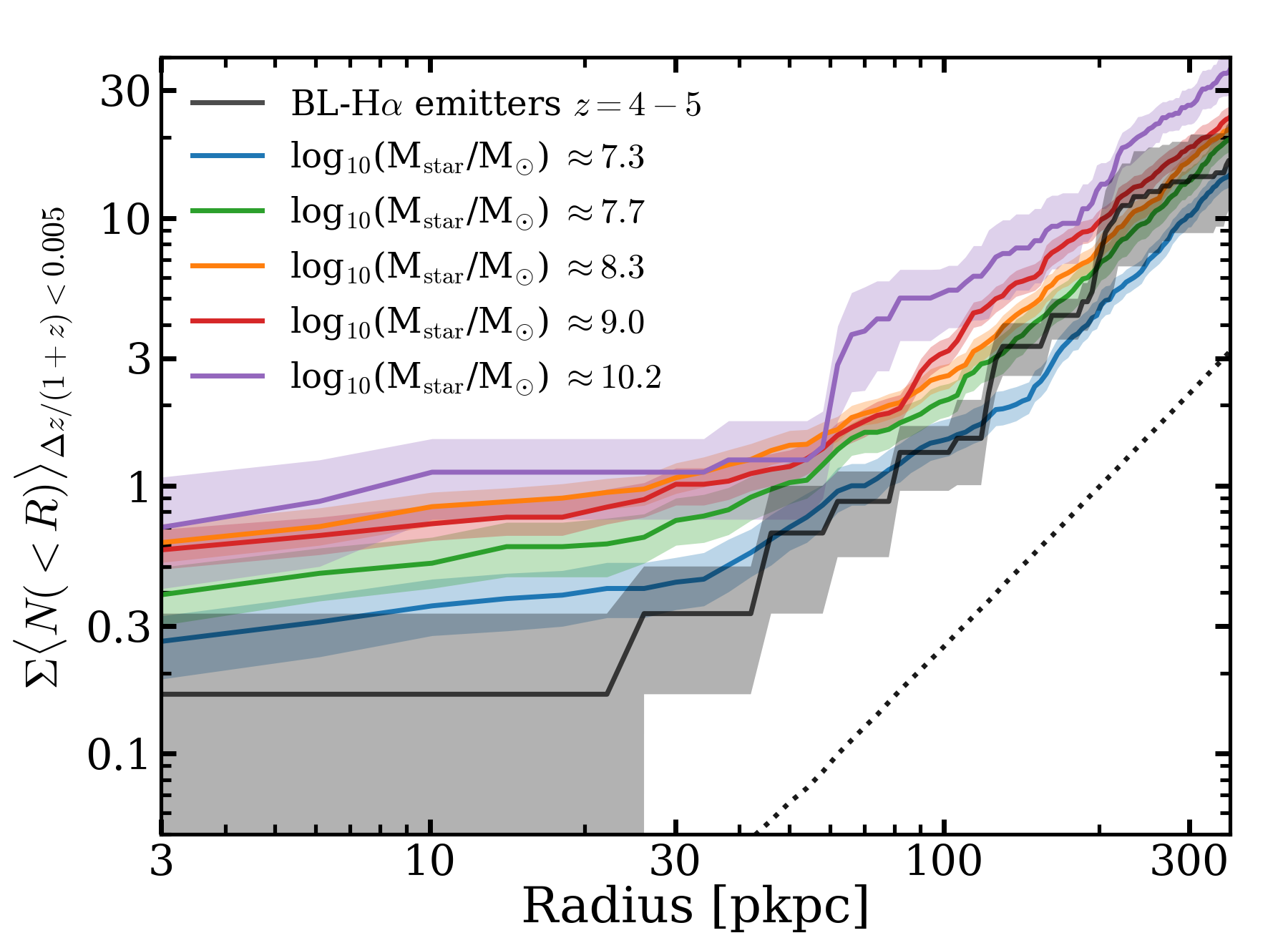} 
    \caption{The average number of neighbouring galaxies within a radius $R$ and $\Delta z/(1+z)<0.005$ for BL-H$\alpha$ emitters (black; excluding ALT-66543 due to its exceptional luminosity, see Fig. $\ref{fig:lha_fwhm}$) and for star-forming galaxies with various masses at $z\approx4-5$ (colored lines). The dotted black line shows the expectation for random regions in our coverage. Shaded regions show the 16-84 percentiles from bootstrap resampling the various subsets.  }
    \label{fig:z45_ALT}
\end{figure}

In Figure $\ref{fig:z45_ALT}$, we show the average number of galaxy pairs as a function of projected radius (in the source-plane) within $\Delta z/(1+z)<0.005$, taking into account the effective area around each galaxy that  is determined by our field of view and the area that has a magnification $\mu<3$. Our average curve for BL-H$\alpha$ emitters excludes ALT-66543 given its exceptional BH mass and luminosity that are more than an order of magnitude higher than all others in the sample (see Table $\ref{tab:LRDsample}$, and Section $\ref{sec:coev}$). We also show the average pair counts around galaxies in bins of stellar mass. The first four stellar mass bins are chosen to have a similar number of galaxies ($\approx60$), whereas the highest stellar mass bin is chosen to have (at least) the same of number galaxies (7) as our BL-H$\alpha$ sample. We illustrate the uncertainty due to variation within the samples with the shaded regions, which show the 16-84 percentiles of the pair counts when bootstrap resampling the subsets 1000 times (with replacement). These uncertainties are mostly important for small samples, i.e. the most massive galaxies and BL-H$\alpha$ emitters. Compared to this random expectation, an excess number of pairs is detected at all radii for all masses. The measured slopes are also somewhat shallower than the random pair counts, indicative of a significant clustering signal \citep[e.g.,][]{Pizzati24a}. At large radii ($R>50$ pkpc; $\gtrsim0.25$ cMpc), the average pair counts uniformly increase with stellar mass, showing that more massive galaxies are located in more over-dense regions. This is expected when galaxy stellar mass is correlated with halo mass \citep[e.g.][]{Shuntov22}. The average pair counts around BL-H$\alpha$ emitters suggest that their environments are similar to those of galaxies with stellar masses $\approx10^{8}$ M$_{\odot}$. This is explored in more detail in Section $\ref{sec:impliedhost}$.

At radii below $\sim100$ pkpc, the pair counts are relatively flat (independent of radius), especially below 30 pkpc, yet they are still dependent on mass. This flat slope may not be surprising as these separations are in the one-halo regime, as the virial radius of a $\approx10^{12(11)}$ M$_{\odot}$ NFW halo is $\approx60 (30)$ pkpc at $z\approx4.5$. It appears that BL-H$\alpha$ emitters typically are surrounded by a smaller number of neighbours, although the uncertainties are large due to the small number statistics. In any case, these measurements do not suggest that BL-H$\alpha$ have an excessively large number of nearby pairs, as may be expected in case their AGN activity were triggered by galaxy mergers.

\subsection{Controlling for the impact of satellite galaxies} \label{sec:satellites}
The relatively flat pair counts within radii that roughly correspond to the virial radii of halos with masses $\approx10^{11}$ M$_{\odot}$ strongly suggests that we are identifying multiple galaxies belonging to the same halo. This could significantly impact studies of the larger scale environment, in particular investigating a mass dependence. Low mass satellite galaxies likely reside in a different environment than low mass galaxies that are not satellites, as their large scale environments are more biased \citep[e.g.][]{Diemer08,OrtegaMartinez24}. In order to investigate and control for this effect, we attempt to identify which galaxies are satellites, and remove them when calculating the average pair count distributions as a function of stellar mass. 

\begin{figure}
    \centering
    \includegraphics[width=8.6cm]{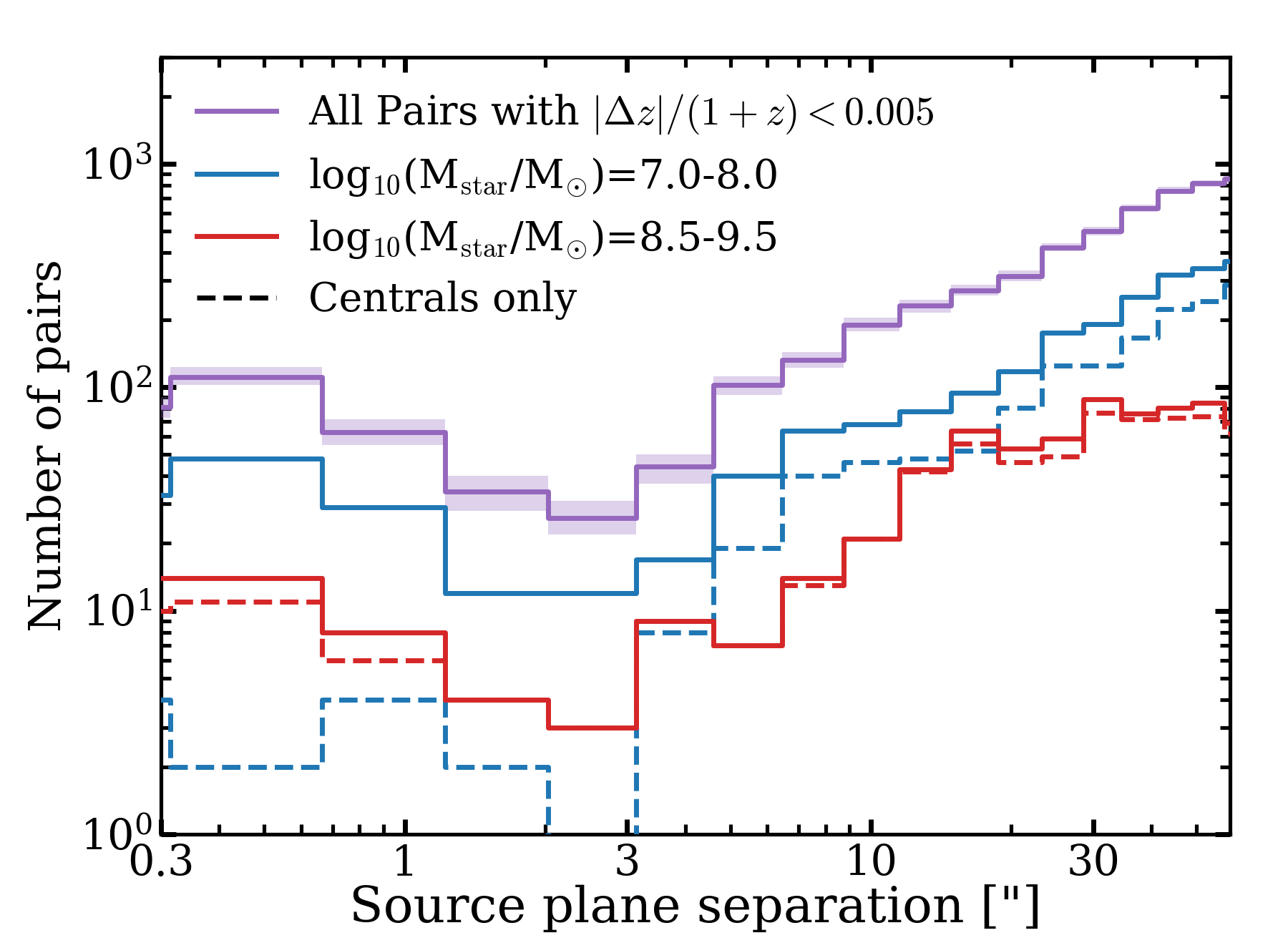} 
    \caption{The distribution of the projected source-plane separations between pairs of galaxies (with $\Delta z/(1+z)<0.005$). Blue shows all pairs within our galaxy sample (where the shaded regions display the variation from bootstrap resamples). Blue and red solid lines show the distribution centering on galaxies with masses between log$_{10}$(M$_{\rm star}$/M$_{\odot}$)=7.0-8.0 and 8.5-9.5, respectively. Dashed lines show the distribution around `centrals' only, where centrals are defined as galaxies that have no companion within 3$''$ that is more massive than itself.} 
    \label{fig:sat_motivation}
\end{figure}

To develop an empirically motivated definition of a central - satellite distinction, we investigate the distribution of projected separations between galaxy pairs (with $\Delta z/(1+z)<0.005$). As can be clearly seen in Fig. $\ref{fig:sat_motivation}$, the pair count distribution displays a minimum at around 2-3$''$, and increases both to larger and smaller separations (similar to $z\sim6$ galaxies, e.g. \citealt{Matthee23}). Motivated by this, we define a satellite as a galaxy that has a more massive companion within a projected separation of $3''$. This yields a satellite fraction ranging from about 30 \% at masses below $10^8$ M$_{\odot}$ to 15 \% at $10^9$ M$_{\odot}$ and virtually zero beyond that. Figure $\ref{fig:sat_motivation}$ illustrates that the pair count distribution around central galaxies (i.e. those that remain after removing satellites with this definition) drops at small separations, most prominently for low mass galaxies. 

\begin{figure}
    \centering
    \includegraphics[width=8.7cm]{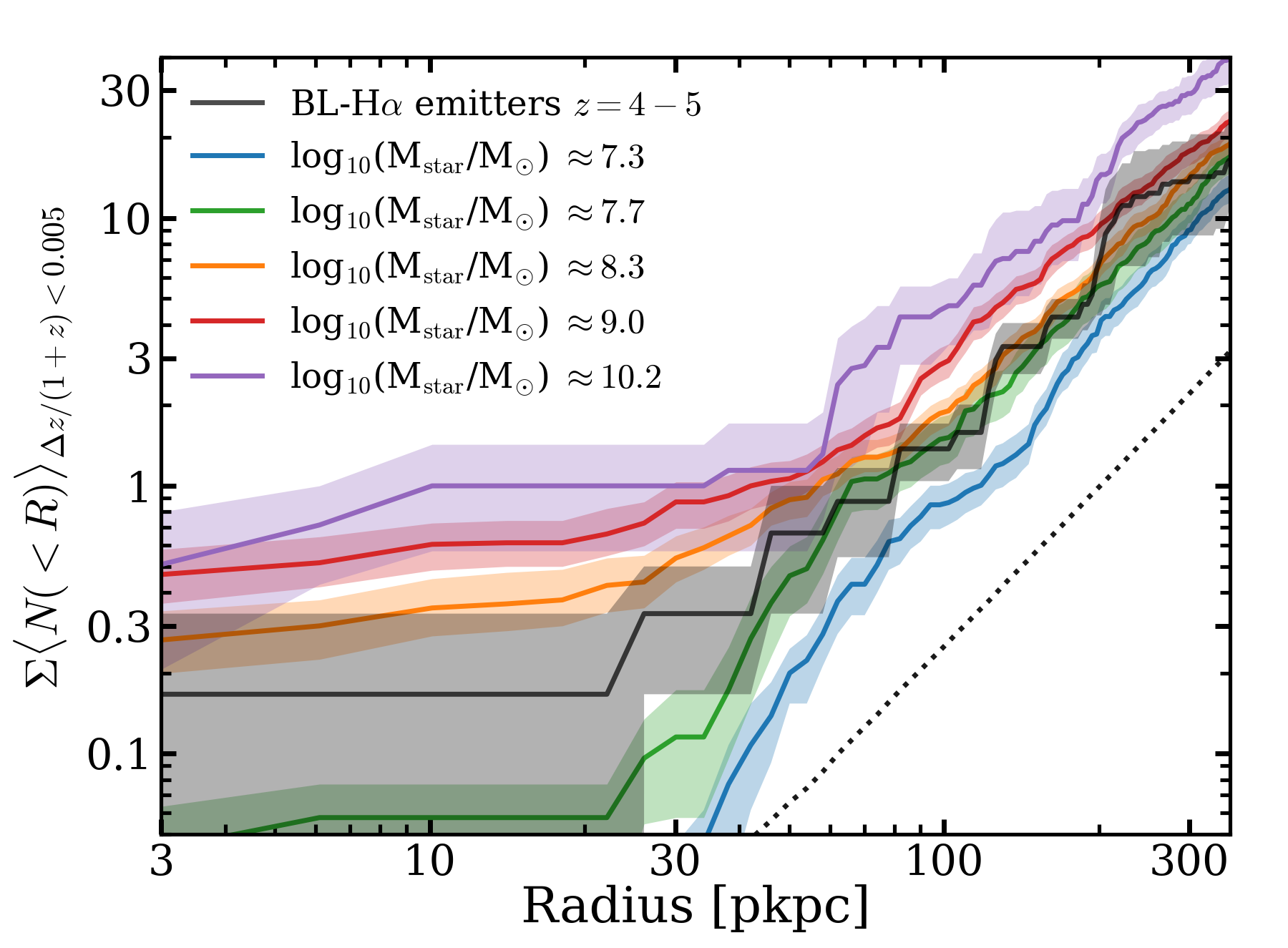}
    \caption{The average number of neighboring galaxies as a function of projected radius (as Fig. $\ref{fig:z45_ALT}$), but now centering on centrals only. Here we implicitly assume that BL-H$\alpha$ emitters are centrals.}
    \label{fig:z45_ALT_centrals}
\end{figure}

Fig. $\ref{fig:z45_ALT_centrals}$ shows the pair count distribution for galaxies that we identified as centrals. Compared to Fig. $\ref{fig:z45_ALT}$, we identify an even clearer distinction between the average number of neighbours around galaxies with various masses, especially at smaller radii. This is because for low mass galaxies, we now no longer include galaxies that are satellites to more massive galaxies (and which would therefore inherit their larger bias). With our sensitivity thresholds we detect one companion to ALT-75753 within $3''$ (see Fig. $\ref{fig:Lineprofiles}$) which has a stellar mass of $2\times10^7$ M$_{\odot}$, while we detect three companions to the luminous ALT-66543 with masses ranges from $6-40\times10^7$ M$_{\odot}$ (see also Labbe et al. in prep). The low stellar mass of the companions supports our assumption that BL-H$\alpha$ emitters are typically centrals.

\subsection{Controlling for biases due to the use of H$\alpha$ as environment tracer} \label{sec:control_density_bias}
To infer the physical properties of galaxies hosting broad H$\alpha$ lines based on their large-scale environments, it is crucial to assume that our measurement of the large-scale environment is independent of the properties of the galaxies around which the environment is measured. This assumption could be invalid in case the H$\alpha$ emission-line selection may lead to biased estimates of the over-density, for example when the star formation rates in the large scale environments around (passive) massive galaxies would systematically be lower (this is typically called galactic conformity), such that a H$\alpha$-line luminosity limited selection may primarily miss galaxies in such environments. Indications of galactic conformity at high-redshift have recently been identified in proto-cluster environments at $z\approx3$ that show a relatively high passive fraction \citep{McConachie24}.

\begin{figure}
    \includegraphics[width=8.6cm]{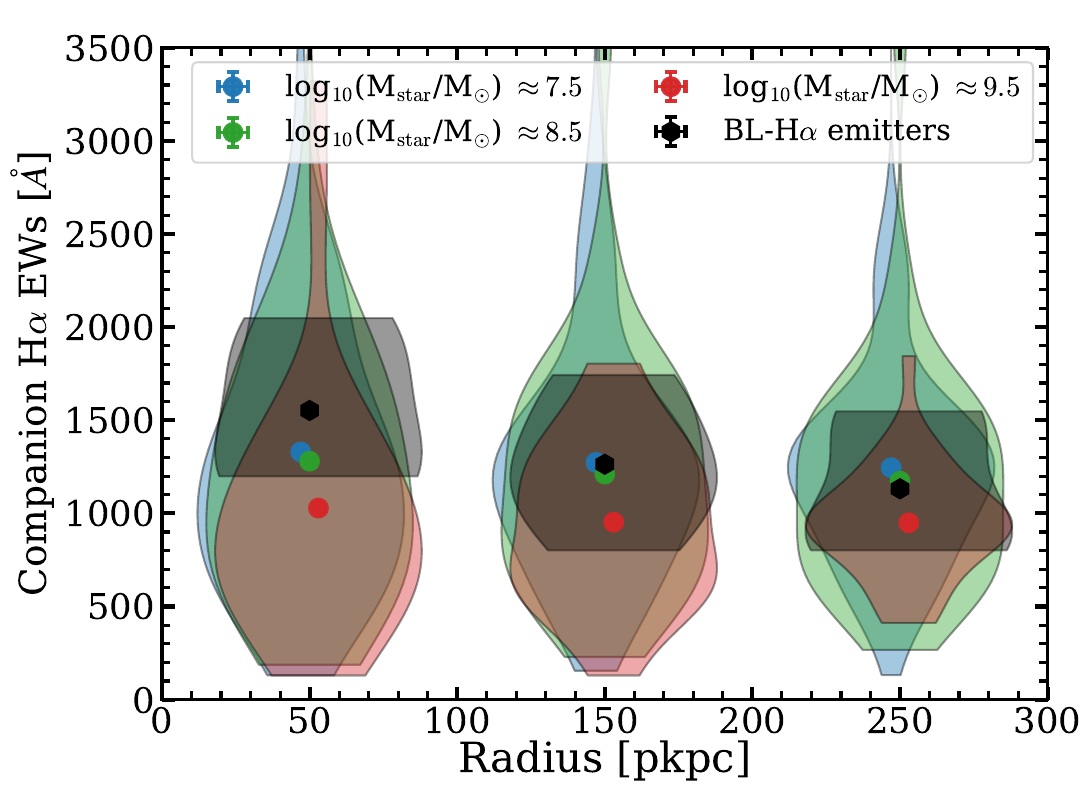} 
    \caption{The distribution of rest-frame H$\alpha$ EW of the companion galaxies as a function of projected radius around galaxies with masses $\approx10^{7.5}$ M$_{\odot}$ (blue), $\approx10^{8.5}$ M$_{\odot}$ (green) and $\approx10^{9.5}$ M$_{\odot}$ (red) and BL-H$\alpha$ emitters (black), demonstrating that there is no significant bias due to the use of H$\alpha$. Violins are slightly shifted horizontally randomly for illustrative purposes.  } \label{fig:bbreakcheck}
\end{figure}

In order to check whether the properties of galaxies in the large-scale environments depend on stellar mass, or whether they systematically differ around BL-H$\alpha$ emitters, we compare the H$\alpha$ equivalent width (EW) distributions of the companion galaxies around galaxies as a function of mass or presence of BL-H$\alpha$, split by distance. These distributions are illustrated in Figure $\ref{fig:bbreakcheck}$. We find that the companions around the most massive galaxies in our sample systematically tend to have a somewhat lower EWs, although the variation among galaxies is large. A similar trend is seen if we use tracers of the Balmer break strength instead of H$\alpha$ EW. These trends are a sign of galactic conformity, i.e. galaxies around the most massive galaxies are somewhat more evolved themselves as well (see also other evidence for accelerated evolution in over-dense regions in \citealt{Morishita24,NaiduALT24}), leading to relatively lower H$\alpha$ luminosities. Nevertheless, while the effect is relatively small, the typical EWs are also significantly higher than the lowest EWs in our sample ($\approx200$ {\AA}), suggesting that the possible underestimation of galaxy over-densities around massive galaxies are likely small.

On scales beyond 100 pkpc, the H$\alpha$ EWs of galaxies around BL-H$\alpha$ emitters are similar to those of the companion galaxies around galaxies with masses $\sim10^{7.5-8.5}$ M$_{\odot}$, suggesting that the large-scale environments of BL-H$\alpha$ emitters are not particularly biased compared to normal galaxies. There is an indication of a skew towards larger H$\alpha$ EWs at smaller distances, which could be indicative of a slight excess star formation activity in the nearby environment of BL-H$\alpha$ emitters.

\begin{table}
    \centering
    \caption{{\bf Over-densities as a function of stellar mass.} We list the edges and average of the mass bins.  Average over-densities around galaxies with these masses are calculated within 1 cMpc, both for all galaxies in that mass range and only for galaxies identified as central galaxies. Errors are estimated by bootstrapping the galaxy sample in each bin. } 
    \begin{tabular}{ccc}
    log$_{10}$(M$_{\rm star}$/M$_{\odot}$) & $(1+\delta)_{1 \rm cMpc, All}$ & $(1+\delta)_{1 \rm cMpc, Centrals}$ \\ \hline 
7.0-7.5 (7.3) & $4.5\pm0.6$ & $3.8\pm0.7$  \\ 
7.5-8.0 (7.7)  & $6.8\pm0.9$ &  $5.8\pm1.1$\\
8.0-8.6 (8.3) & $7.9\pm0.7$ & $6.6\pm0.8$  \\  
8.6-9.6 (9.0)  & $10.2\pm1.1$ & $9.7\pm1.1$ \\
9.6-11.0 (10.2)  & $12.6\pm3.0$ & $13.2\pm3.5$ \\ 
    \end{tabular}
    \label{tab:massbins}
\end{table}

\section{The implied host galaxies of BL-H$\alpha$ emitters} \label{sec:impliedhost}
\begin{figure}
    \centering
    \includegraphics[width=8.7cm]{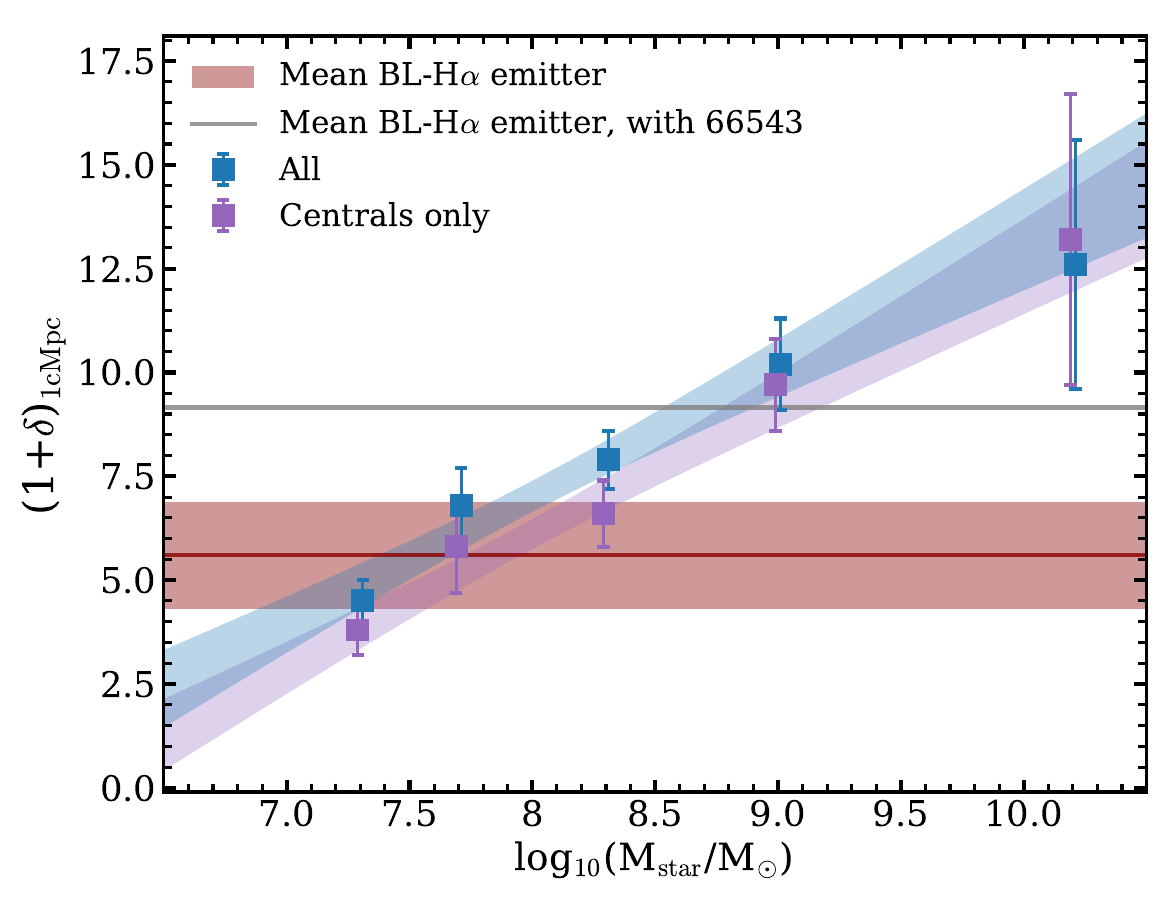}
    \caption{The relation between the over-density (at a radius of 1 cMpc) and stellar mass of star-forming galaxies, either centering on all galaxies with a certain mass (blue), or centrals only (purple). The red shaded region shows the mean over-density around BL-H$\alpha$ emitters, and its uncertainty based on bootstrapping, excluding ALT-66543. The grey line shows the mean when including ALT-66543.}
    \label{fig:delta_Mstar}
\end{figure}

Figure $\ref{fig:delta_Mstar}$ shows that the over-density measured on 1 cMpc scales strongly correlates with stellar mass. We illustrate results for a 1 cMpc radius as a reference, but note our results do not strongly depend on this choice. The measured over-densities are listed in Table $\ref{tab:massbins}$. We also illustrate the mean over-density around the BL-H$\alpha$ emitters (excluding the exceptional ALT-66543) and its uncertainty, which is $1+\delta_{\rm 1 cMpc}=5.6\pm1.2$. We fit a linear relation between over-density and stellar mass: 

\begin{equation}
    1+\delta_{\rm 1 cMpc} = a+ b \, {\rm log}_{10}({\rm M}_{\rm star}/10^9 {\rm M_{\odot}}),
\end{equation}

For all galaxies, we find $a=10.15\pm0.46$ and $b=3.19\pm0.34$, while for central galaxies alone we find $a=9.33\pm0.42$ and $b=3.22\pm0.31$. When measuring over-densities on 2 cMpc scales, over-densities are generally slightly lower ($1+\delta_{\rm 2 cMpc}=5.3\pm0.6$ for BL-H$\alpha$ emitters), with a resulting $a=8.04\pm0.40$ and $b=1.9\pm0.31$ for all galaxies and $a=7.63\pm0.37$ and $b=2.11\pm0.30$ for centrals only. In all cases, a positive correlation between stellar mass and over-density is measured at $\gtrsim6 \sigma$ significance ($>10 \sigma$ on 1 cMpc scales). We caution against extrapolating this relation towards higher over-densities as the relation may not be linear or changing, for example due to a flattening in the stellar to halo mass relation \citep[e.g.][]{Shuntov24}. As discussed in Section $\ref{sec:control_density_bias}$, there is a slight indication that galaxies around the most massive galaxies have lower H$\alpha$ EWs, such that we could possibly under-estimate their galaxy over-density, which could lead to a somewhat steeper relation, especially at the highest masses.

Using these correlations, we can infer the implied stellar masses of the galaxies hosting broad H$\alpha$ lines, based on their over-densities. Here, we assume that BL-H$\alpha$ emitters follow the same overdensity - stellar mass relation as galaxies without broad H$\alpha$. This method implies a stellar mass of log$_{10}$(M$_{\rm star}$/M$_{\odot}$)$ = 7.5\pm0.5\, (7.8\pm0.4)$ when comparing the 1 cMpc over-densities for all (central) galaxies. At 2 cMpc, the implied masses would be log$_{10}$(M$_{\rm star}$/M$_{\odot}$)$ = 7.6\pm0.3 \, (7.9\pm0.3)$, respectively. Therefore, averaging these masses, our indirect inference of the stellar mass based on their environments suggests that the typical BL-H$\alpha$ emitter in our sample, which has a BH mass of log$_{10}$(M$_{\rm BH}$/M$_{\odot}$)$ = 6.8\pm0.2$, is hosted by a galaxy with a stellar mass of log$_{10}$(M$_{\rm star}$/M$_{\odot}$)$ = 7.7\pm0.2$. 

\begin{figure}
    \centering
    \includegraphics[width=8.7cm]{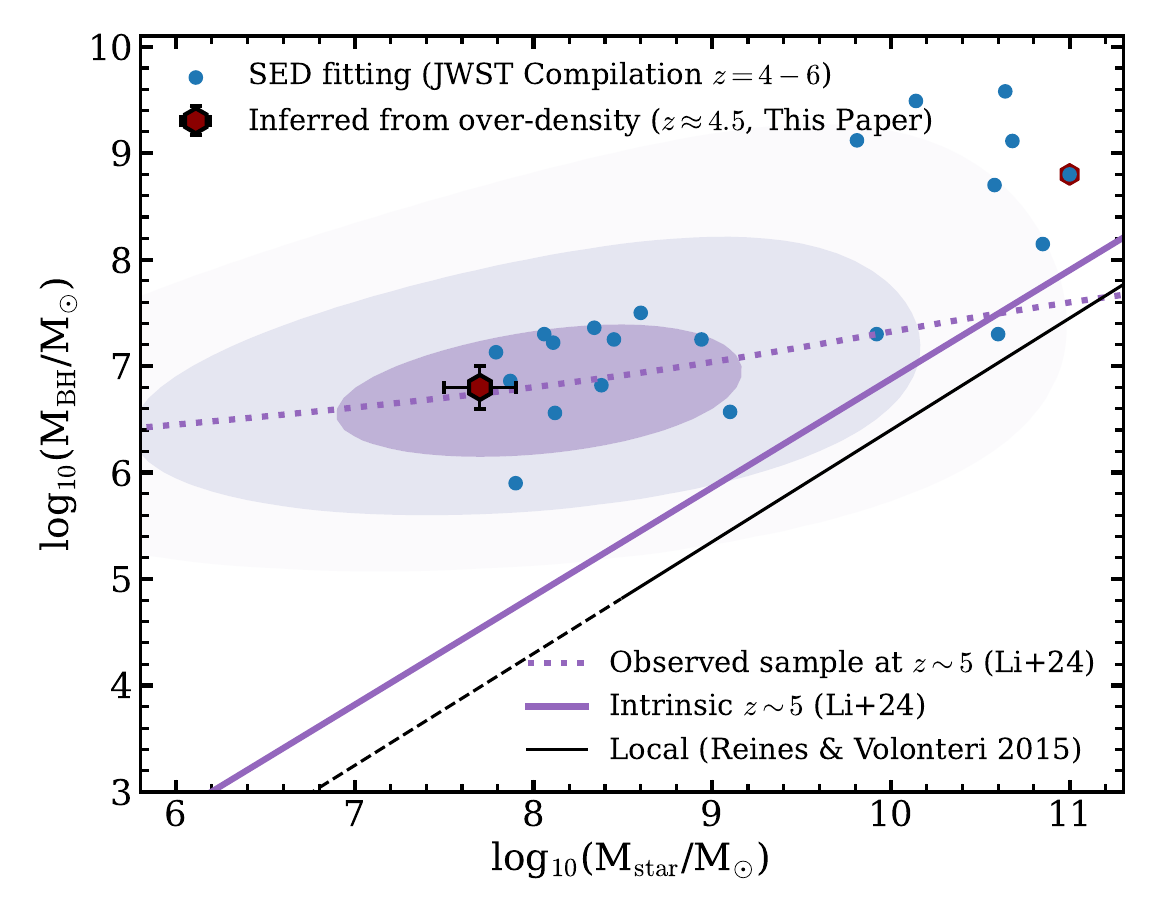}
    \caption{The relation between the stellar mass and SMBH mass for our BL-H$\alpha$ emitter sample (dark red hexagon, where stellar mass is indirectly inferred from their environments), a {\it JWST} literature compilation at $z\approx4-6$ (blue points; where stellar mass is inferred from SED fitting, from \citealt{Carnall23,Maiolino23b, Harikane23,Yue23,Kokorev24,Onoue24}). We highlight the luminous ALT-66543 (Labbe et al in prep) also with a red hexagon. For reference, we show the local relation measured by \cite{ReinesVolonteri15} in black (showing the extrapolated range with a dashed line). The luminosity-bias corrected fit to {\it JWST} data from \cite{Li24} is shown in purple, purple contours show the expected observed distribution from this modeling work. }
    \label{fig:Mstar_MBH}
\end{figure}

In Figure $\ref{fig:Mstar_MBH}$, we compare our BH and stellar mass estimates with other estimates based on recent {\it JWST} measurements at $z=4-6$ (combined from \citealt{Carnall23,Maiolino23b, Harikane23,Kokorev24}). Similar to our work, BH masses are estimated from the single epoch virial estimate using the broad Balmer lines (in most cases H$\alpha$). In these works, the stellar masses are estimated through SED fitting, using an image decomposition of point and extended sources \citep[e.g.][]{Harikane23,Yue23}, including AGN model components \citep{Maiolino23b} or by fitting stellar absorption features to spectra \citep{Carnall23,Kokorev24,Onoue24}. The mean stellar mass of our sample, that we indirectly inferred based on their environments, is in good agreement with these more direct methods, as well as recent results based on PSF-modeling approaches \citep[e.g.][]{Chen24}. The stellar mass is on average a factor 40 lower than the mass inferred from fitting galaxy-only SEDs (Fig. $\ref{fig:LRDs_SEDs}$). The typical stellar mass of $10^{7.7}$ M$_{\odot}$ suggests that these BL-H$\alpha$ emitters have a high BH to stellar mass ratio of $\approx12.5$ \%. While this ratio is much higher than typically found for galaxies in the local Universe \citep[e.g.][]{Pacucci23}, we caution that this is partly due to biases resulting from the sensitivity limits at which broad H$\alpha$ emission can be detected \citep[e.g.][]{Zhang23,Li24}. Indeed, in Fig. $\ref{fig:Mstar_MBH}$, we show the contour levels highlighting the expected distribution of an AGN sample with our H$\alpha$ luminosity limit at $z\sim4.5$ \citep{Li24}, here assuming the \cite{Weibel24} galaxy stellar mass function and the intrinsic relation shown in the solid purple line. 

It is interesting to compare our inferred stellar mass to the SEDs of the BL-H$\alpha$ emitters (shown in Fig. $\ref{fig:LRDs_SEDs}$). Based on correlations between the optical colours, the compactness and the broad H$\alpha$ line profile, \cite{Matthee24} argued that the AGN SEDs are (strongly) obscured and mainly outshine the host galaxy in the rest-frame optical, whereas the rest-frame UV emission is dominated by a (young) host galaxy \citep[see also][]{Kocevski23,Akins24b,Killi24,YMa24}. However, it is also possible that the rest-frame UV emission originates from a small fraction of scattered AGN light \citep[e.g.][]{Greene23,Stepney24}, in particular as some BL-H$\alpha$ emitters appear compact at virtually all wavelengths. We investigate the origin of the UV emission of the BL-H$\alpha$ emitters by comparing their UV luminosity to the typical UV luminosity that is expected for a normal galaxy. Using the typical UV mass-to-light ratio in our reference sample, we find that their UV emission implies a stellar mass of $10^{7.7 \pm 0.2 (\rm stat) \pm 0.5 (\rm sys)}$ M$_{\odot}$, where the statistical errors reflect the variation among UV luminosities of the BL-H$\alpha$ emitters and the systematical errors reflects the variation in mass to light ratios of the reference sample. The similarity in the stellar masses inferred from large-scale environments and the UV emission suggests that the UV emission of these BL-H$\alpha$ emitters indeed mainly originates from the star-forming host galaxy. This is in line with their very blue UV slopes $\beta\approx-2$, which is bluer than typical quasars (see the compilation in e.g. \citealt{Fujimoto22}). In the rest-frame optical (i.e. $\lambda_0=0.6-0.8$ micron), the typical mass-to-light ratio would imply that BL-H$\alpha$ emitters have much higher stellar masses of $\approx10^{9.3}$ M$_{\odot}$, indicating a significant AGN contribution to the light.

\begin{table*}
    \centering
    \caption{The AGN and over-density measurements for a compilation of high-redshift AGN. For the UV-luminous quasars, EIGER measured their H$\beta$ line-width and derived the BH properties as detailed in \cite{Yue24}. The bolometric luminosity is derived from L$_{5100}$. The errors on the BH mass and bolometric luminosity of the EIGER quasars are dominated by systematic errors. For the $z\sim6$ LRDs from the UNCOVER survey, their BH properties are derived from the H$\alpha$ line as detailed in \cite{Greene23}. G23-13821 corresponds to ALT-26902 and G23-41225 to ALT-73104. We measure the AGN properties of J1148+5253 based on it H$\beta$ line in the EIGER data. Over-densities within 1 cMpc are measured in this paper. Upper limits are at the 2 $\sigma$ level.  }
    \begin{tabular}{ccccccccc}
    ID & $z_{\rm spec}$ & $\mu$ & v$_{\rm FWHM}$/km s$^{-1}$ & log$_{10}$(M$_{\rm BH}$/M$_{\odot}$) & log$_{10}$(L$_{\rm bol}$/erg s$^{-1}$)  &  $(1+\delta)_{1 \rm cMpc}$ & Survey  \\ \hline
  G23-13821 & 6.34 & $1.62\pm0.32$ & $3100\pm710$ & $8.1\pm0.2$ & $45.4\pm0.2$ & $23.8\pm9.5$ & UNCOVER+ALT \\
  G23-41225 &  6.77 & $1.88\pm0.38$  & $2000\pm600$ & $7.7\pm0.4$ & $45.3\pm0.5$ & $<9.5$ & UNCOVER+ALT \\ \hline
  J0100+2802 & 6.33 & 1. & $6045\pm20$ & 10.06 & 47.18& $54\pm27$ & EIGER \\ 
  J0148+0600 & 5.98 & 1. & $7828\pm106$ & 9.89 & 46.39 & $106\pm38$ & EIGER \\
   J1030+0524 & 6.30 & 1. & $3670\pm15$ & 9.19 & 46.30 & $<25$  &EIGER \\
  J1148+5251 & 6.42 & 1. & $5370\pm80$ & 9.64 & 46.54 & $12\pm12$ &EIGER\\
  J159-02 & 6.38  & 1. & $3493\pm30$ & 9.10 & 46.20 & $<26$ &  EIGER\\
  J1148+5253 & 5.69 & 1. & $2910\pm450$ & $8.3\pm0.1$ & $45.8\pm0.1$ & $14\pm14$ & EIGER \\ \hline
    \end{tabular}
    \label{tab:literature_sample}
\end{table*} 

\section{The variation among high-$z$ AGN environments}  \label{sec:coev}
As illustrated in Figure $\ref{fig:Mstar_MBH}$, current samples of AGN discovered and observed by {\it JWST} span three orders of magnitude in BH mass. This implies that we can expect similar variations in their host galaxies. Indeed, detailed fitting of the spectra and photometry of AGNs with SMBHs as massive as $\sim10^9$ M$_{\odot}$ \citep[e.g.][]{Ding23,Yue23,Juodzbalis24a,Marshall24,Onoue24,BWang24a} suggests that their host galaxies are also significantly more massive (M$_{\rm star}\sim10^{10-11}$ M$_{\odot}$) than those around lower mass AGNs. In our analysis, we have excluded ALT-66543 from our sample averages due to its much higher luminosity and BH mass. Its deep spectrum suggests a much higher stellar mass $\sim10^{11}$ M$_{\odot}$ as well as AGN emission (Labbe et al. in prep). This is consistent with its very large over-density $1+\delta\approx30$ (Table $\ref{tab:LRDsample}$), which is significantly higher than the typical over-density we measure for normal galaxies with masses $2\times10^{10}$ M$_{\odot}$, $1+\delta\approx13$ (Table $\ref{tab:massbins}$). Therefore, there is substantial evidence that there is significant variation among the host galaxies of high-redshift AGN that may correlate with the BH mass.

We now directly investigate the relation between SMBH mass and the environment. Early clustering studies of BL-H$\alpha$ emitters and the comparison of their number densities to those of quasars suggest that BL-H$\alpha$ emitters reside in lower mass halos than quasars \citep{Arita24,Pizzati24b}. Here, we try to connect galaxy over-density measurements of these different samples directly by extending our dynamic range. First we add two broad line selected AGN at $z\sim6.5$ whose AGN properties are based on NIRspec spectroscopy \citep{Greene23}. G23-13821 at $z=6.34$ has a relatively high BH mass of $10^{8.1\pm0.2}$ M$_{\odot}$, while G23-41225's BH mass appears lower ($10^{7.7\pm0.4}$ M$_{\odot}$). We use the ALT data to measure their environments, similar to the methodology detailed in Section $\ref{sec:environment}$. Here, we use [OIII] emitters as a probe of their environment rather than H$\alpha$ selected sources. As above, we only include galaxies with $\mu<3$ and with a luminosity threshold of L$_{\rm [OIII]5008}>1.5\times10^{41}$ erg s$^{-1}$. We detect four galaxies within a radius of 1 cMpc and $\Delta z/(1+z)<0.005$ around G23-13821, but none around G23-41225, despite similar coverage and sensitivity. We measure over-densities on 1 cMpc scales of $1+\delta=23.8\pm9.5$ and $1+\delta<9.5$ (at $2\sigma$), respectively, where errors are poissonian. On 2 cMpc scales, their over-densities are $58.7\pm8.2$ and $7.6\pm7.6$. These measurements are listed in Table $\ref{tab:literature_sample}$. We note that these over-density are similar to those measured around a $z\sim7$ broad-line AGN by \cite{Schindler24} that has comparable H$\beta$-based AGN properties as these two sources.

\begin{figure}
    \centering
    \includegraphics[width=8.7cm]{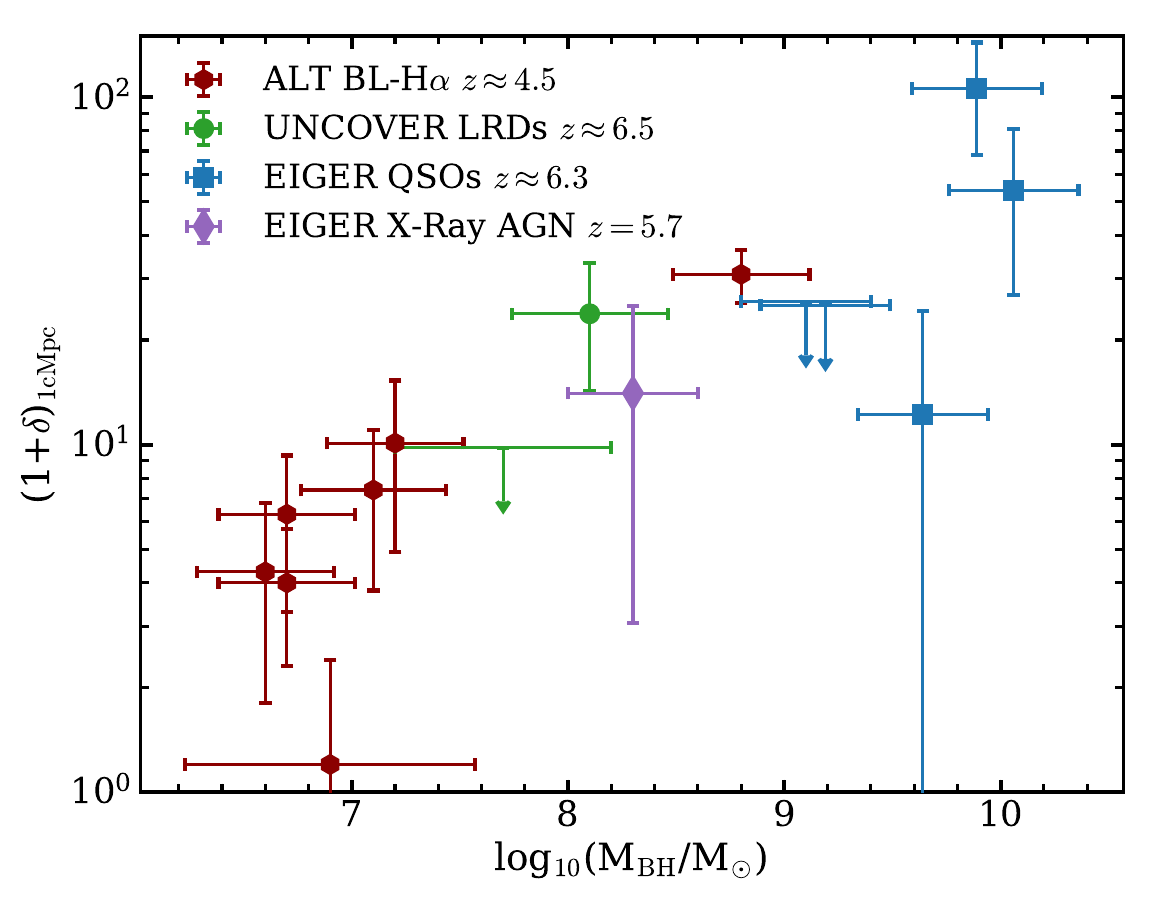} \\
    \includegraphics[width=8.7cm]{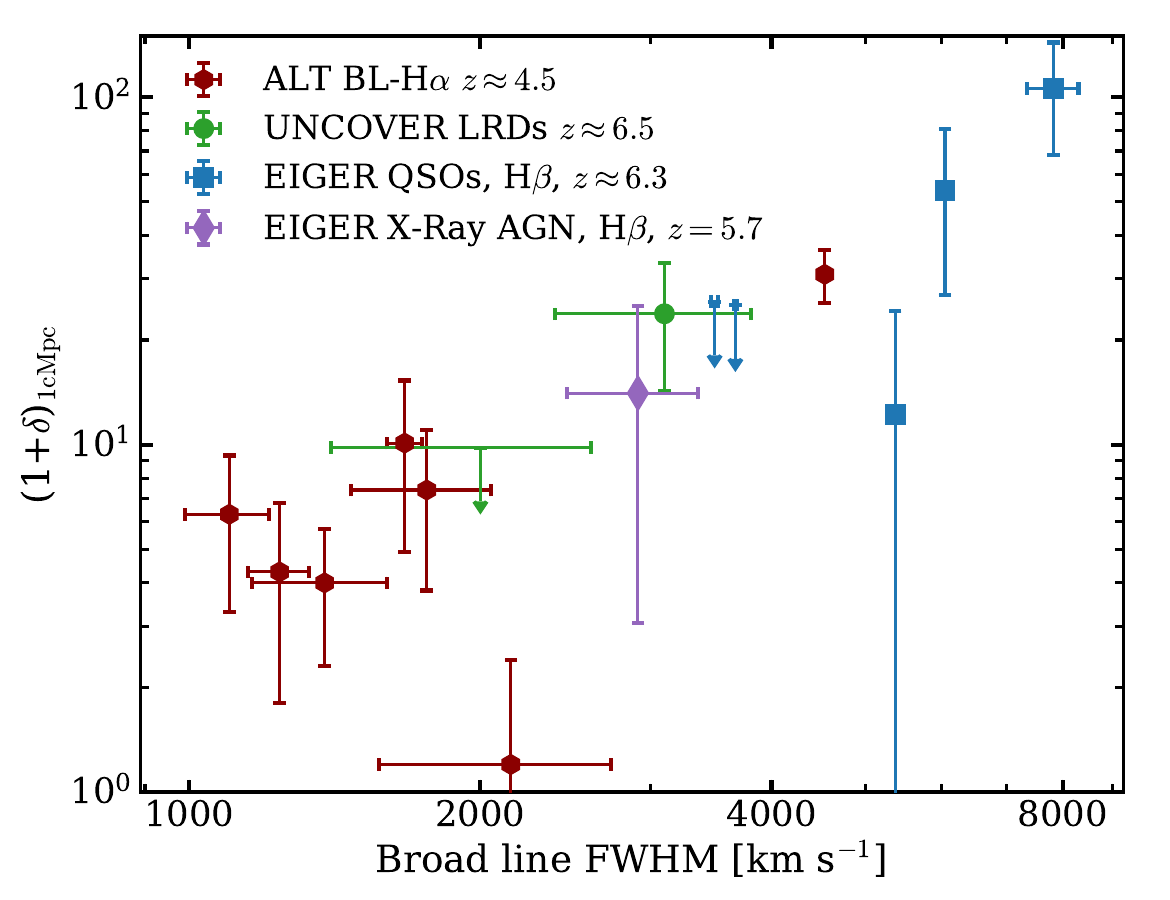} \\
    \caption{The relation between the over-density (within 1 cMpc cylinders centered on the AGN) and BH mass (top) and broad Balmer line FWHM (bottom). Dark red hexagons show our results for the BL-H$\alpha$ emitters at $z\approx4.5$ from ALT, green points are ALT over-density measurements around NIRSpec confirmed broad-line AGN, blue squares show the results for the luminous quasars from the EIGER survey at $z\approx6.3$ and the purple diamond is the over-density around a UV faint, X-Ray selected AGN at $z=5.7$ covered by the EIGER survey. }  
    \label{fig:delta_BH} 
\end{figure}

Second, we add over-density measurements around five UV luminous quasars at $z=6.0-6.4$ measured using data from the EIGER survey \citep{Kashino23}, see Mackenzie et al. (in prep). The quasar properties are based on NIRCam grism spectra \citep{Yue23}. The statistical errors are very low due to the very high signal-to-noise spectra, thus, in Fig. $\ref{fig:delta_BH}$ we show the 0.3 dex systematic errors for their BH mass. We also add the X-ray detected AGN J1148+5253 at $z=5.69$ \citep{Mahabal05} that is covered by the EIGER survey. We derive its AGN properties by fitting the H$\beta$ and [OIII] complex similar to the method detailed in \cite{Yue23}, which yields a BH mass of M$_{\rm BH}=10^{8.3\pm0.1}$ M$_{\odot}$. To measure the over-density around these quasars, we use the same method as \cite{Eilers24}, thus only including galaxies with [OIII]$_{5008}$ luminosities above $10^{42}$ erg s$^{-1}$. The full details on the identification of [OIII] emitters in the EIGER data will be presented in Kashino et al. in prep, but see also \cite{Kashino23}. Apart from variations in the sensitivity and the lack of cluster lensing magnification, the EIGER and ALT methodologies are very similar. As listed in Table $\ref{tab:literature_sample}$, we measure very high over-densities for some quasars, while we can only report upper limits for other quasars indicating significant scatter (see also \citealt{Wang23}). For objects without companions within 1 cMpc, we list upper limits that would correspond to $N=2$ detected galaxies.

Combining these measurements with those presented in Section $\ref{sec:environment}$, we find correlations between the galaxy over-density, broad Balmer line-width and BH mass, see Figure $\ref{fig:delta_BH}$. As Balmer line-width is used to derive BH mass, the two results are not independent of each other, but the relation appears slightly stronger showing the direct observable quantity. We show the over-density measured on 1 cMpc scales, but note that the trends are qualitatively similar on 2 cMpc scales. This correlation is in line with the result that more massive SMBHs reside in more massive galaxies (Fig. $\ref{fig:Mstar_MBH}$), but it is also suggestive of a BH mass - halo mass relation, as more massive halos reside in larger over-densities.

Previous clustering measurements have yielded halo masses $\approx2.5\times10^{12}$ M$_{\odot}$ for the EIGER quasars (with BH mass $\sim3\times10^{9}$ M$_{\odot}$) using spectroscopic redshifts \citep{Eilers24}, while \cite{Arita24} derived halo masses $\sim6\times10^{11}$ M$_{\odot}$ for broad-line selected AGN (with BH mass $\sim3\times10^7$ M$_{\odot}$) using photometric redshifts. Assuming that our sample of BL-H$\alpha$ emitters (with BH mass $\sim6\times10^6$ M$_{\odot}$)  resides in typical dark matter halos given their stellar mass, their halo masses would be $\approx5\times10^{10}$ M$_{\odot}$ \citep[e.g.][]{Behroozi19,Shuntov22}, but we caution that more detailed studies are required. Quantitatively establishing the shape of the BH mass - halo mass relation at redshifts $z\sim5$ directly would provide new constraints on models of SMBH formation growth and AGN feedback \citep[e.g.][]{Bower17,Dayal24,Li24_models}. 

Various complications are involved in accurately quantifying such a BH mass - halo mass relation. The scatter in over-densities among individual halos at a given halo mass is substantial (e.g. Fig. 12 in \citealt{TorralbaTorregrosa24}). This challenges accurate individual halo mass inferences. Further, the galaxy samples that are used for over-density measurements have different sensitivity limits and are at different redshifts, which could mean that their bias with respect to the matter density field varies \citep[e.g.][]{Dalmasso24}, which impacts halo mass inferences \citep[e.g.][]{Pizzati24a}. Further, we note that \cite{Shen07} find much larger bias and halo masses for quasars at $z\sim3-4$ compared to the halo masses of quasars at $z\sim6$ \citep{Eilers24}. While this could suggest strong redshift evolution, we note that the $z\sim3-4$ halo mass measurements are much higher than both the $z\sim2.5$ and $z\sim6$ measurements, which is challenging to reconcile (see \citealt{Eftekharzadeh2015} for a detailed discussion). Therefore, a future robust clustering analysis to quantify a possible BH mass - halo mass relation at $z\approx5-6$ requires larger statistical samples of AGNs with consistent redshifts and over-density probes.

\section{Implications} \label{sec:implications} 
\subsection{The AGN nature of BL-H$\alpha$ emitters} \label{sec:implication_AGNnature}
Due to the unusual spectrum of BL-H$\alpha$ emitters, and especially due to their X-Ray faintness \citep[e.g.][]{Akins24,Yue24}, some studies have questioned their AGN nature. The most common key indicator of AGN emission is currently broad Balmer line-emission (but see also Labbe et al. in prep for the detection of strong FeII lines that unambiguously prove AGN activity). Theoretically, it is possible that broad Balmer lines may not originate from a broad line region around an AGN, but rather due to a high velocity dispersion of gas in compact massive galaxies with high stellar densities \citep[e.g.][]{Baggen24}. In this scenario, the typical dispersion of $\approx600$ km s$^{-1}$ of our sample would imply a stellar+gas mass of $\approx3\times10^{10}$ M$_{\odot}$ assuming a size of 100 pc, following the same methodology as described in \cite{Baggen24} based on \cite{vanDokkum15}. As shown in Section $\ref{sec:environment}$, such a mass for the typical BL-H$\alpha$ emitters in our sample would be at strong odds with the measured over-densities (which would need to be about three times higher), unless the gas fractions would be extremely high. Therefore, their over-densities rather point towards an AGN explanation for the broad Balmer line emission in the {\it typical} BL-H$\alpha$ emitter. Note however, as discussed above in Section $\ref{sec:coev}$, that some BL-H$\alpha$ emitters could still reside in massive galaxies, but more evolved stellar populations are not the only cause of their red appearance.

\subsection{A luminosity-dependent diversity in AGN hosts} \label{sec:implication_diversity}
The trends between stellar mass, BH mass and over-densities shown in Figures $\ref{fig:Mstar_MBH}$ and $\ref{fig:delta_BH}$ imply that more luminous BL-H$\alpha$ emitters reside in more massive galaxies, such that their SED will differ from the typical BL-H$\alpha$ emitter from our sample. Indeed, we notice that ALT-66543, the most luminous and massive object in the sample with a strong Balmer break, has the reddest UV slope of $\beta=-0.7$, while the UV slopes of the others are much bluer, $\beta\approx-2$, implying an older or more obscured host galaxy and/or a stronger AGN contribution to the UV. UV-luminous quasars at $z\approx6$ with similar BH masses as ALT-66543 have also recently been shown to display Balmer breaks suggestive of older and massive galaxy populations \citep{Onoue24}. However, in case Balmer breaks may also arise from the AGN phenomena themselves \citep{Inayoshi24}, the relative stellar and AGN contributions may be challenging to disentangle. A natural consequence of such a luminosity-dependent diversity in AGN host galaxies is that different surveys may arrive at seemingly discrepant conclusions on the nature of BL-H$\alpha$ emitters depending on their sensitivity and covered volumes. For example, BL-H$\alpha$ emitters identified in deeper spectroscopic surveys will more likely reside in relatively low mass galaxies, with blue UV slopes and without clear over-densities. Rarer and more luminous BL-H$\alpha$ emitters, on the other hand, are more likely reside in more massive galaxies with redder colors \citep[e.g.][]{BWang24}, larger over-densities, stronger Balmer breaks and, possibly, a higher AGN contribution to the rest-frame UV emission.

\begin{figure}
    \centering
    \includegraphics[width=8.7cm]{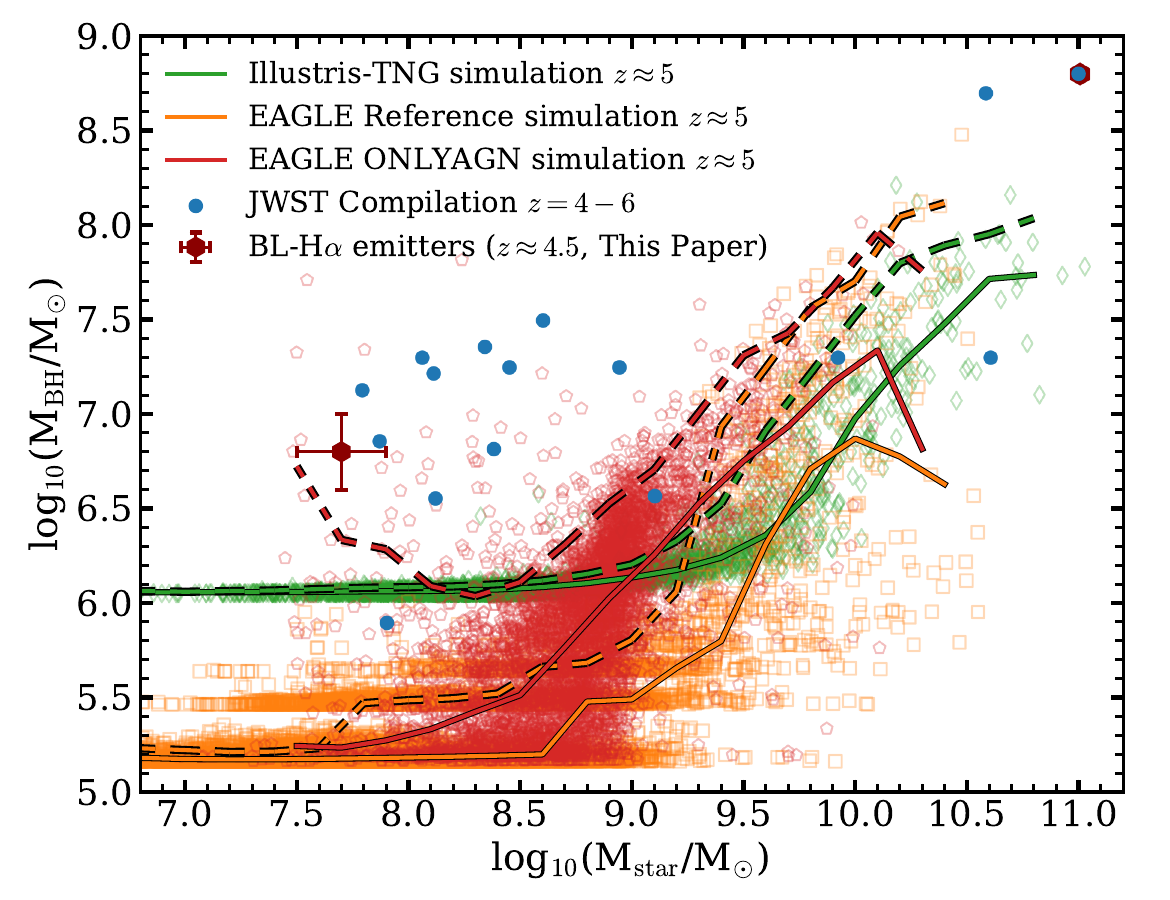}
    \caption{The relation between the stellar mass and supermassive black hole mass at $z\approx5$, comparing measurements at $z\approx5$ (dark red and blue points from data as in Fig. $\ref{fig:Mstar_MBH}$) with the 100 cMpc box EAGLE (\citealt{Schaye15,Crain15}) and Illustris-TNG (\citealt{Weinberger17,Pillepich18}) hydrodynamical simulations (orange and green, respectively). Red shows an EAGLE model without stellar feedback (ONLYAGN; with a 50 cMpc box size). Data-points show individual galaxies. Solid lines show the running median. Dashed lines show the running 95th percentile (where we note the low and high mass ends are impacted by low number statistics). The seed SMBH masses are $\approx10^5$ M$_{\odot}$ and $\approx10^6$ M$_{\odot}$ in EAGLE and Illustris-TNG, respectively.} 
    \label{fig:Mstar_MBH_sims} 
\end{figure}

\subsection{Hydrodynamical simulations need to grow SMBHs in lower mass galaxies} \label{sec:implication_sims}
State-of-the-art cosmological models of galaxy formation virtually all invoke the presence of AGN feedback associated with the growth of SMBHs to simulate realistic populations of galaxies  \cite[e.g.][]{Somerville08,Crain15,Weinberger17,Dave19}. These simulations are mostly tuned to reproduce properties of galaxy and cluster populations in the low-redshift Universe. However, the majority of SMBH seeding and growth occurs at high-redshift, because models typically seed in relatively low mass halos \citep[e.g.][]{Rosas2016} and massive galaxies form rapidly. In Figure $\ref{fig:Mstar_MBH_sims}$, we show how the measured SMBH to stellar mass relation of the observed galaxies at $z\approx5$ compares to the EAGLE \citep{Schaye15,Crain15} and Illustris-TNG \citep{Weinberger17,Pillepich18} hydrodynamical simulations.\footnote{See \cite{Habouzit22,Habouzit24} for a detailed comparison of EAGLE, Illustris-TNG and other cosmological simulations in the context of high-redshift SMBHs.} We highlight that the stellar masses of AGN with SMBHs with mass $\sim10^7$ M$_{\odot}$ are a factor 30-100 lower than the typical stellar masses that host such SMBHs in the Illustris-TNG and EAGLE Reference simulations (which are $\sim5\times10^9$ M$_{\odot}$, see the solid lines in Fig. $\ref{fig:Mstar_MBH_sims}$). In both simulations virtually all SMBHs in galaxies with masses $\approx10^8$ M$_{\odot}$ are still at the seed mass from the simulation (see the dashed lines in Fig. $\ref{fig:Mstar_MBH_sims}$ that show the rolling 95th percentiles), meaning that the simulations have not yet enabled their growth.

What would be needed to reconcile these differences? A key concern is the validity of the calibrations used to derive SMBH mass from broad H$\alpha$ line-width and luminosity. Given the X-Ray faintness of {\it JWST}'s broad H$\alpha$ line emitters, it has been argued that they are experiencing super-Eddington accretion \citep[e.g.][]{Lambrides24,PacucciNarayan24}. In this case, SMBH masses could be over-estimated by an order of magnitude \citep{Lupi24}, mitigating the tension. Super-Eddington accretion is likely linked to a low AGN duty cycle due to the short-duration of such inefficient bursts of SMBH growth. As discussed in \cite{Pizzati24b}, the observed number densities of broad H$\alpha$ line emitters at $z\sim5$ can only be reconciled with a low ($\sim1$ \%) duty cycle if their halo masses are as low as $\sim10^{11}$ M$_{\odot}$. The number density of broad-line H$\alpha$ emitters with implied BH masses $\sim6\times10^6$ M$_{\odot}$ (similar to our sources) is $\approx2 \times10^{-4}$ cMpc$^{-3}$ \citep{Taylor24}, while the galaxy stellar mass function yields number densities $\approx2\times10^{-2}$ cMpc$^{-3}$ for galaxies with stellar masses $\approx5\times10^7$ M$_{\odot}$ at these redshifts \citep[e.g.][]{Lovell21,Weibel24}. This implies a duty cycle of $0.5-1$ \%. Therefore, the Mpc-scale over-densities of the BL-H$\alpha$ emitters are in agreement with this scenario where short bursts of super-Eddington accretion characterize the first stages of SMBH growth in low mass halos.

However, given the large corrections of BH masses that would be required to reconcile the results shown in Fig. $\ref{fig:Mstar_MBH_sims}$, it is also justified to discuss possible changes in the models. \cite{Crain15} and \cite{Bower17} have shown how various model variations (in particular the seed mass, the feedback associated to star formation, and the parameters controlling the SMBH accretion rate) impact the BH to stellar mass relation in the EAGLE simulation. While they find that changes in the seed mass have a negligible impact, the strength of the stellar feedback controls the lowest mass at which SMBHs start their rapid growth \citep{Bower17,McAlpine18,Trebitsch18,Li24_models}. Additionally, allowing for super-Eddington accretion in simulations -- not varied in EAGLE -- likely leads to stronger BH growth compared to stellar mass growth \citep{Schneider23,Shi23,Bennett24,Husko24}. Focusing on EAGLE, \cite{Bower17} showed that rapid SMBH growth occurs as soon as they are seeded in the absence of stellar feedback. As we show in Fig. $\ref{fig:Mstar_MBH_sims}$, this ONLYAGN model without stellar feedback indeed leads to more efficient SMBH growth compared to stellar assembly, generally shifting the median BH - stellar mass relation to about 0.5 dex lower stellar masses.The ONLYAGN model variation produces a non negligible number of low mass galaxies with BH to stellar mass ratios comparable to the observations, which indeed likely probe such a biased sample \citep{Li24}. Therefore, a negligible efficiency of stellar feedback (at high-redshift) would enable the formation of low-mass galaxies that host SMBHs about 10 \% of their stellar mass, although such a model does not reproduce realistic galaxies at low-redshift.

Finally, we remark that scenarios for feedback-free star formation in the early Universe have recently been explored in the context of the high abundance of UV-luminous galaxies at $z>10$ \citep{Naidu22,Castellano22,Harikane24,Casey24,Carniani24}. As discussed in e.g. \cite{Dekel23,Mayer24,Renzini25}, the high gas densities and low metallicities of halos in the early Universe could lead to a significantly higher star formation efficiency due to the decreased impact of feedback, and simultaneously promote the formation of intermediate mass black holes and their efficient mergers into SMBHs \citep[e.g.][]{Mayer24,Dekel24}. Thus, exploring such models in the context of a cosmological simulation that also reproduces galaxies at later times is highly warranted.

\section{Summary} \label{sec:summary}
The population of faint AGN at high-redshift discovered in {\it JWST} data primarily through broad H$\alpha$-line emission (a subset of this population presenting red colors and compact shapes has colloquially been come to known as the "Little Red Dots") promises to unveil new insights into the formation and growth of supermassive black holes (SMBHs). A key measurement that is required to place these AGN in context is their host stellar mass. Directly measuring stellar masses through fitting their spectral energy distribution (SED) is challenged by the fact that the AGN SEDs are complex and uncertain, and their contribution to the continuum is difficult to disentangle from stellar emission. Here, we perform an independent and empirical approach to infer the typical stellar mass of galaxies hosting broad H$\alpha$ lines (BL-H$\alpha$ emitters) at $z\approx4-5$, based on comparing their Mpc-scale environments to the environments around star-forming galaxies. We also explore correlations between BH mass, line-width and their large-scale environments. We primarily use BL-H$\alpha$ emitters and galaxies identified using sensitive {\it JWST}/NIRCam grism data from the ALT survey behind the Abell 2744 lensing cluster (\citetalias{NaiduALT24} \citeyear{NaiduALT24}), which benefits from a high spectroscopic completeness down to galaxy masses $\approx10^7$ M$_{\odot}$, accurate redshifts, and excellent imaging data to inform SED fits of the star-forming galaxies. Our main results are:

\begin{itemize} 
\item We identify a sample of 7 broad H$\alpha$ line-emitters at $z\approx4-5$ with FWHM $>1000$ km s$^{-1}$ and a luminosity L$_{\rm bol}>10^{43.7}$ erg s$^{-1}$. Three of these were already confirmed as broad-line emitters from NIRSpec spectroscopy \citep{Greene23}, including the exceptional ALT-66543 that is the most luminous Little Red Dot known (Labbe et al. in prep). The BL-H$\alpha$ emitters have SMBHs with masses $10^{6.6-8.8}$ M$_{\odot}$, typically M$_{\rm BH}=10^{6.8}$ M$_{\odot}$, and overlap in parameter space with typical {\it JWST}-identified AGNs. [Section $\ref{sec:sample_AGN}$, Fig. $\ref{fig:Lineprofiles}$, Fig. $\ref{fig:lha_fwhm}$, Table $\ref{tab:LRDsample}$]

\item The ALT data shows a large range in galaxy densities over the full $\approx30$ arcmin$^2$ field. BL-H$\alpha$ emitters trace moderate galaxy over-densities and avoid the lowest density regimes, but they avoid the most prominent over-density. On 1 cMpc scales, the over-densities around the BL-H$\alpha$ emitters range from $(1+\delta)_{\rm 1 cMpc}=1 - 30$, typically $(1+\delta)_{\rm 1 cMpc}\approx5$, excluding the exceptional ALT-66543. [Section $\ref{sec:environment}$, Figure $\ref{fig:RADEC_1}$, Table $\ref{tab:LRDsample}$]

\item In the reference sample, we demonstrate that more massive galaxies are surrounded by more galaxies on $3-300$ pkpc scales along the plane of the sky (and likely beyond). Within cylinders of $\Delta z/(1+z)=0.005$ (1500 km s$^{-1}$), we find that the over-density increases from $(1+\delta)_{\rm 1 cMpc}\approx4$ for galaxies with stellar masses of $2\times10^{7}$ M$_{\odot}$ to $(1+\delta)_{\rm 1 cMpc}\approx10 (13)$ at a mass of $\approx10^{9 (10)}$ M$_{\odot}$. [Section $\ref{sec:environment}$, Figures $\ref{fig:z45_ALT}$, $\ref{fig:z45_ALT_centrals}$, $\ref{fig:delta_Mstar}$; Table $\ref{tab:massbins}$]   

\item We detect a clear and flat excess number of galaxy pairs below separations of $\approx50$ pkpc, which is due to satellites that cluster strongly within the typical virial radius of our galaxy sample. Based on a data-driven definition, we identify galaxies that are satellites to more massive galaxies and find satellite fractions of $\approx30 (15)$ \% at M$_{\rm star}\approx10^{8(9)}$ M$_{\odot}$, and virtually zero above that.  [Section $\ref{sec:satellites}$, Figure $\ref{fig:sat_motivation}$]

\item Our estimates of the \textit{relative} over-density around galaxies of varying stellar mass are robust to our choice of H$\alpha$ as a tracer. We demonstrate this by comparing H$\alpha$ EW distributions around BL-H$\alpha$ emitters and the reference sample galaxies, finding no difference on 1 cMpc scales (however, the AGN show some signs of increased H$\alpha$ EW at smaller $\lesssim0.3$ cMpc scales). The H$\alpha$ EW distributions also show that the most massive galaxies are surrounded by galaxies with lower H$\alpha$ EWs and stronger Balmer breaks suggestive of accelerated evolution, possibly depressing their over-density estimates. [Section $\ref{sec:control_density_bias}$, Figure $\ref{fig:bbreakcheck}$]
 
\item We use the correlation between over-density and stellar mass to infer that the typical BL-H$\alpha$ emitter in our sample has a stellar mass of $10^{7.7\pm0.2}$ M$_{\odot}$, $\sim1.5$ dex lower than the median stellar mass inferred from galaxy-only SED fits. Taking the BH mass estimates based on single-epoch virial calibrations at face value, this implies a BH to stellar mass ratio as high as 12.5 \%, in line with earlier measurements, and in line with a slight $0.2$ dex increase in the normalisation of the local BH to stellar mass relation given our luminosity-limited selection effects. Interestingly, this stellar mass is similar to the mass one would infer from the UV luminosity for a typical mass-to-light ratio in the reference sample, but is lower by an order of magnitude in the rest-frame optical, indicating significant AGN contribution to the rest-optical SED. [Section $\ref{sec:impliedhost}$, Figure $\ref{fig:Mstar_MBH}$]

\item By extending our over-density measurements to other samples of AGN at $z\approx6$ (from the EIGER and ALT surveys), we find an indication of a correlation between the over-density and BH mass, suggestive of a BH to halo mass relation. However, larger statistics of sensitive over-density measurements are required to quantify this relation with uniform tracers. [Section $\ref{sec:coev}$, Figure $\ref{fig:delta_BH}$, Table $\ref{tab:literature_sample}$]
\end{itemize}

The main implications of our results are:
\begin{itemize}
    \item As the typical BL-H$\alpha$ emitter in our sample resides in a moderate over-density that is smaller than typical over-densities for galaxies with masses $\sim10^{10}$ M$_{\odot}$, our results disfavor alternative explanations of broad H$\alpha$ lines that arise purely due to virial broadening of kinematics in very high stellar densities, at leat for this studies sample that covers the lower-luminosity half of the literature samples. [Section $\ref{sec:implication_AGNnature}$, Fig. $\ref{fig:lha_fwhm}$]
    
    \item The indicative correlation between BH mass and galaxy over-density implies that we can expect a luminosity-dependent diversity among AGN hosts, with more massive and older host galaxies in more luminous quasars with heavier SMBHs. Statistical analyses of BL-H$\alpha$ emitters, such as average clustering measurements, host galaxy mass or the shape of their SED, could thus be sensitive to the parameter space that AGNs occupy, which typically depend on survey characteristics.  [Section $\ref{sec:implication_diversity}$]
    
    \item Although current samples of BL-H$\alpha$ emitters probe only a luminosity-limited subset of the SMBH population, the presence of AGNs in galaxies with stellar mass as low as $\approx5\times10^7$ M$_{\odot}$, and the high accompanying BH to stellar mass ratios are not seen in hydrodynamical models of galaxy formation. From the observational side, this could imply the BH masses are significantly over-estimated, possibly due to short-lived super-Eddington accretion events, that would be in line with their low duty cycle and low masses. From the modeling side, the mechanisms controlling SMBH growth, such as the strength of stellar feedback, the maximum accretion rate or the seeding, may need to be revised, especially at early times. Interestingly, this highlights a possible common explanation for the presence of overly massive black holes in low mass, high-redshift galaxies, and the high abundance of UV-luminous galaxies beyond $z>10$.  [Section $\ref{sec:implication_sims}$, Figure $\ref{fig:Mstar_MBH_sims}$]

\end{itemize}


\facilities{\textit{JWST}}

\software{
    \texttt{Python},
    \texttt{matplotlib} \citep{matplotlib},
    \texttt{numpy} \citep{numpy}, 
    \texttt{scipy} \citep{scipy},
    \texttt{Astropy}
    \citep{astropy1, astropy2} }
    
\acknowledgments{
We thank Junyao Li for sharing model-output shown in Fig. $\ref{fig:Mstar_MBH}$ and Rob Crain for sharing results from the ONLYAGN EAGLE model shown in Fig. $\ref{fig:Mstar_MBH_sims}$ and Adi Zitrin for comments.
This work is based on observations made with the NASA/ESA/CSA James Webb Space Telescope. The data were obtained from the Mikulski Archive for Space Telescopes at the Space Telescope Science Institute, which is operated by the Association of Universities for Research in Astronomy, Inc., under NASA contract NAS 5-03127 for \textit{JWST}. These observations are associated with programs \# 3516. 
Funded by the European Union (ERC, AGENTS,  101076224). Views and opinions expressed are however those of the author(s) only and do not necessarily reflect those of the European Union or the European Research Council. Neither the European Union nor the granting authority can be held responsible for them. We acknowledge funding from {\it JWST} program GO-3516. Support for this work was provided by NASA through the NASA Hubble Fellowship grant HST-HF2-51515.001-A awarded by the Space Telescope Science Institute, which is operated by the Association of Universities for Research in Astronomy, Incorporated, under NASA contract NAS5-26555. AA acknowledges support by the Swedish research council Vetenskapsr{\aa}det (2021-05559).

}

\bibliography{Biblio}
\bibliographystyle{apj}

\appendix

\section{Over-densities around all ALT BL-H$\alpha$ emitters} \label{app:A}
\begin{figure*}
    \centering
    \begin{tabular}{cc}
    \includegraphics[width=7.8cm]{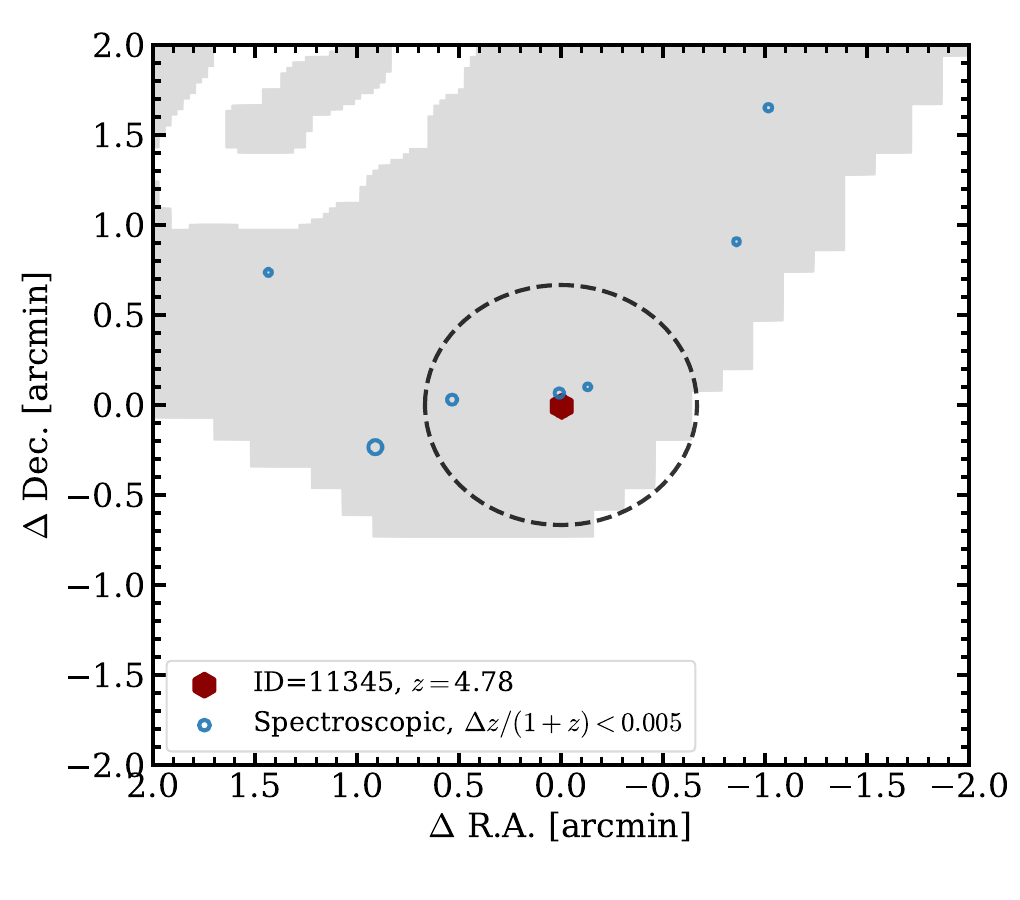} & 
    \includegraphics[width=7.8cm]{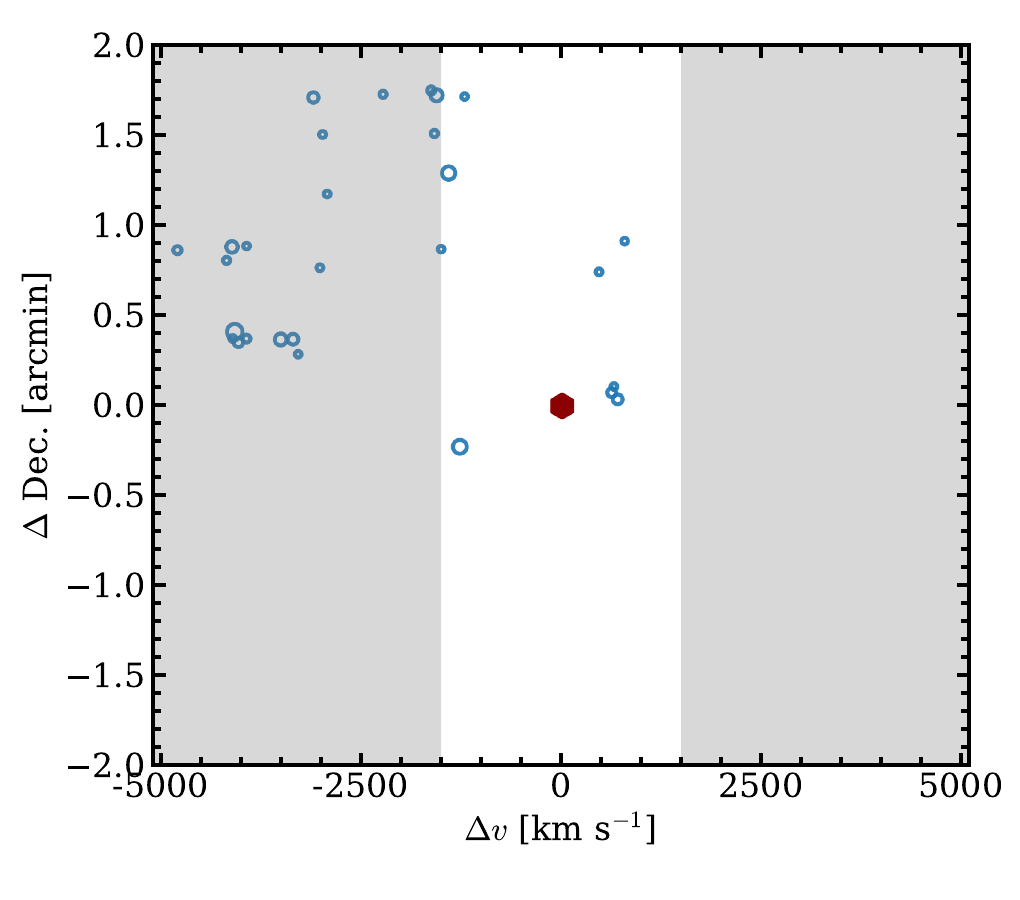} \\   \vspace{-0.3cm}
    \includegraphics[width=7.8cm]{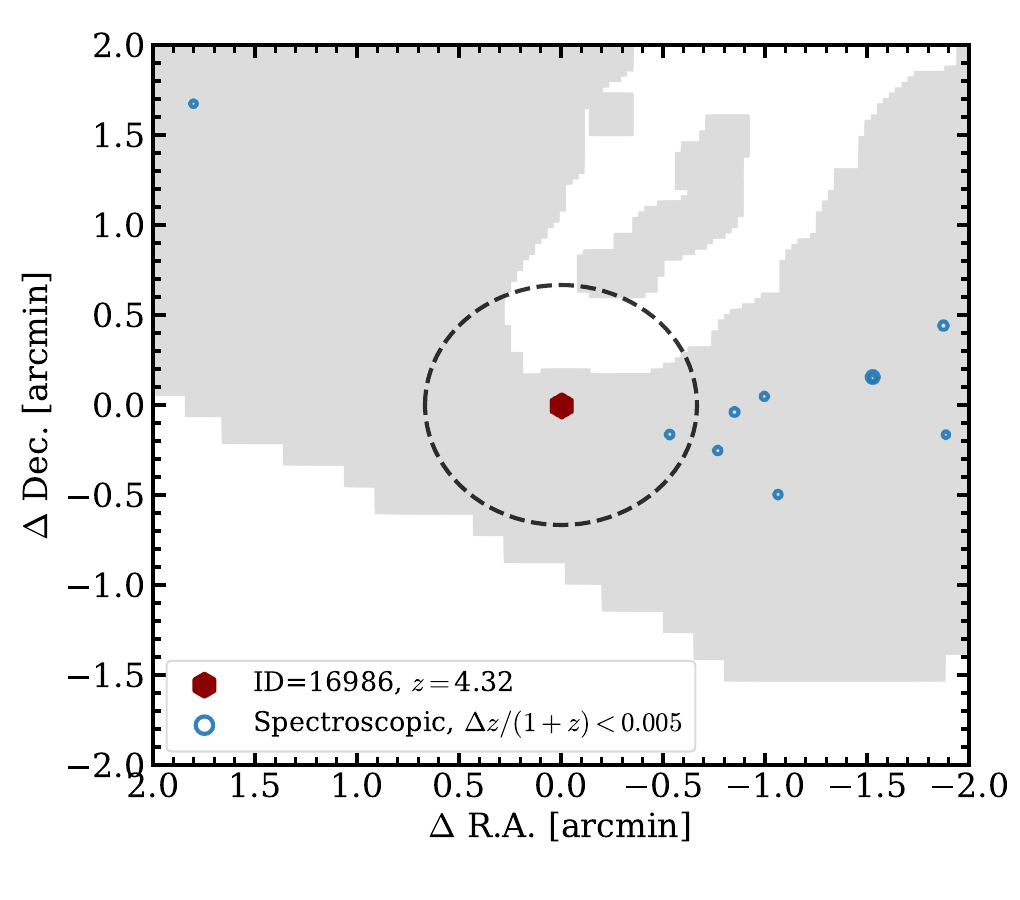} & 
    \includegraphics[width=7.8cm]{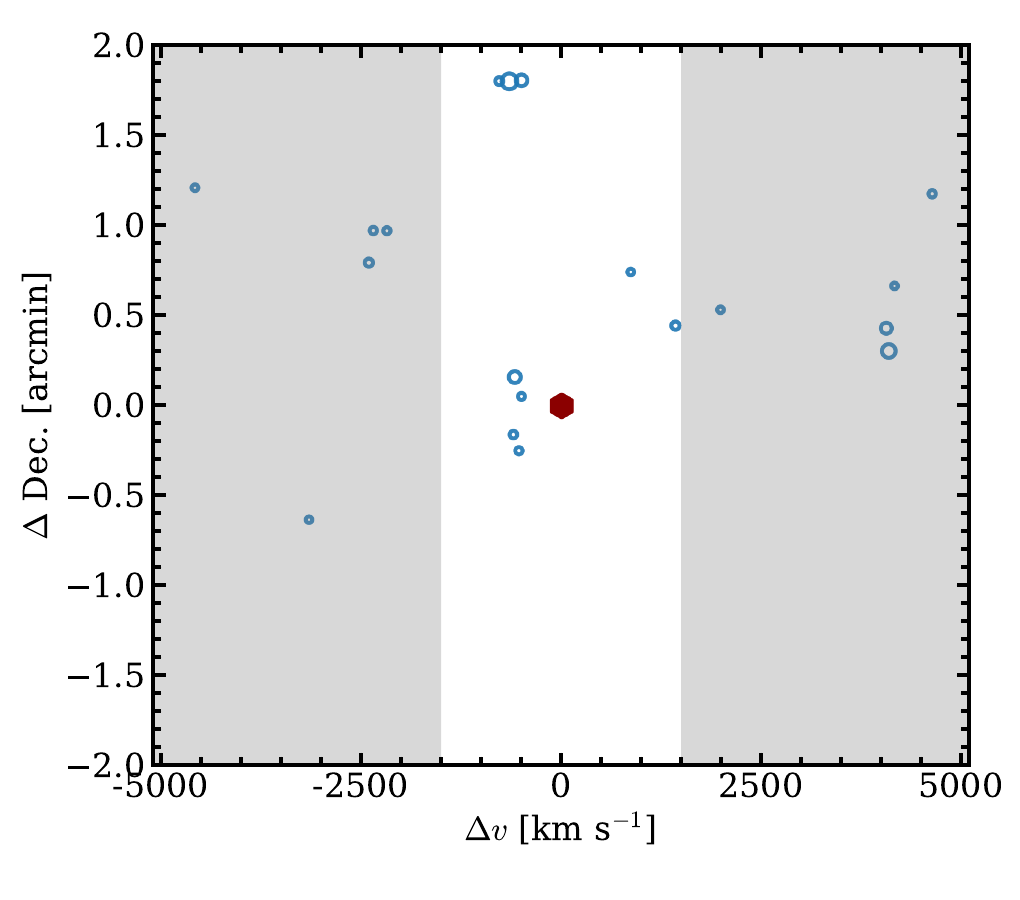} \\ \vspace{-0.3cm}    

    \end{tabular}
    \caption{The environments of Broad-line H$\alpha$ emitters (as Fig. $\ref{fig:RADEC_1}$).}
    \label{fig:RADEC_2}
\end{figure*}

\begin{figure*}
    \centering
    \begin{tabular}{cc}   
\includegraphics[width=7.8cm]{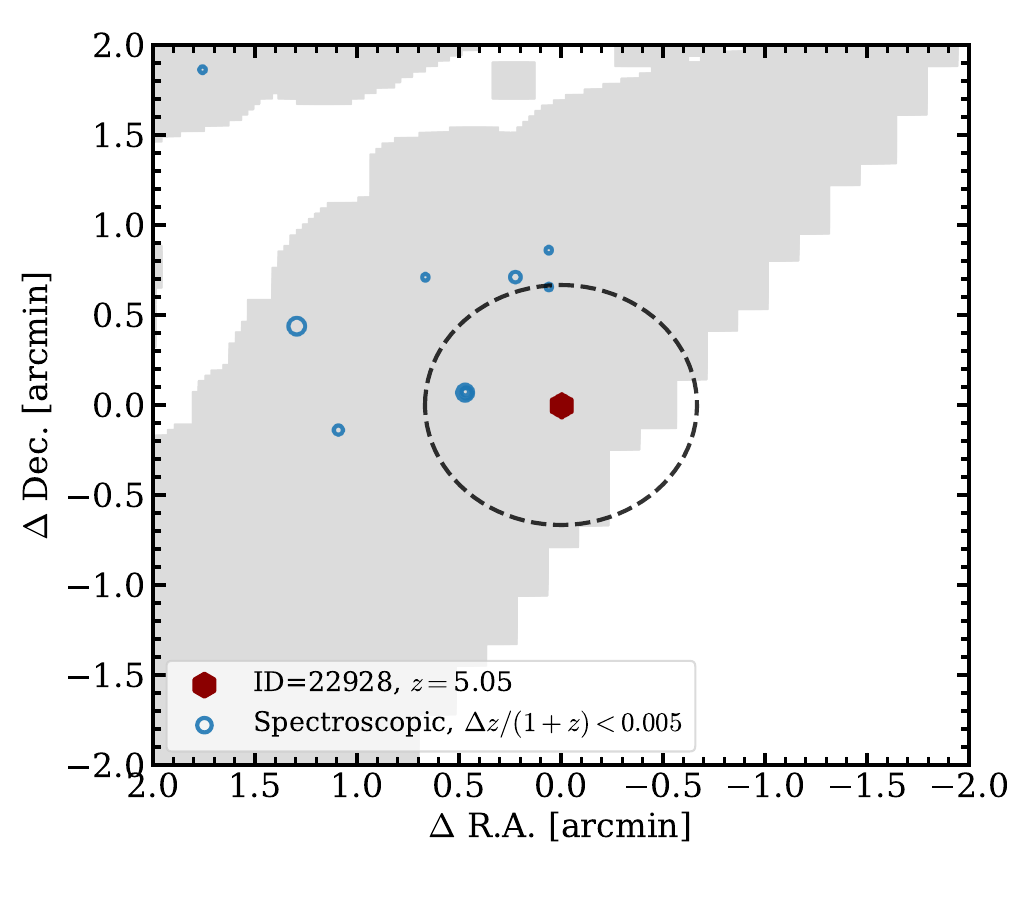} &     
\includegraphics[width=7.8cm]{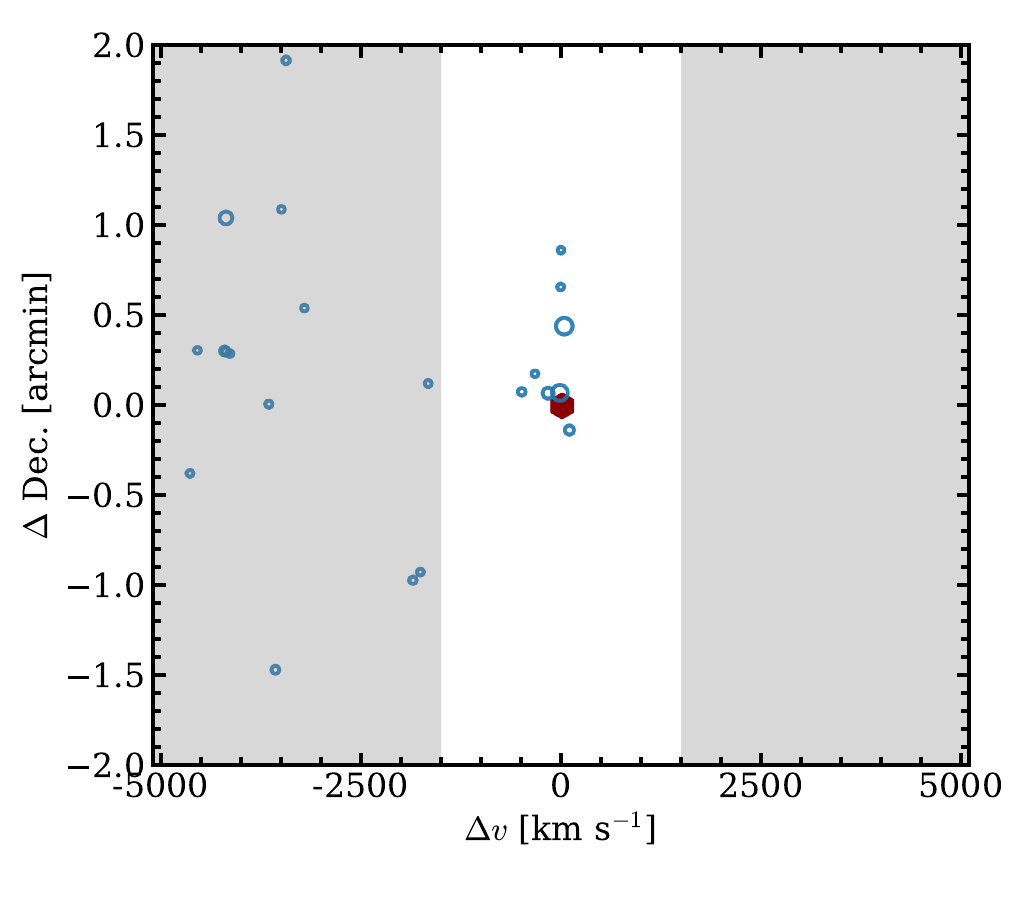} \\    \vspace{-0.3cm}
      \includegraphics[width=7.8cm]{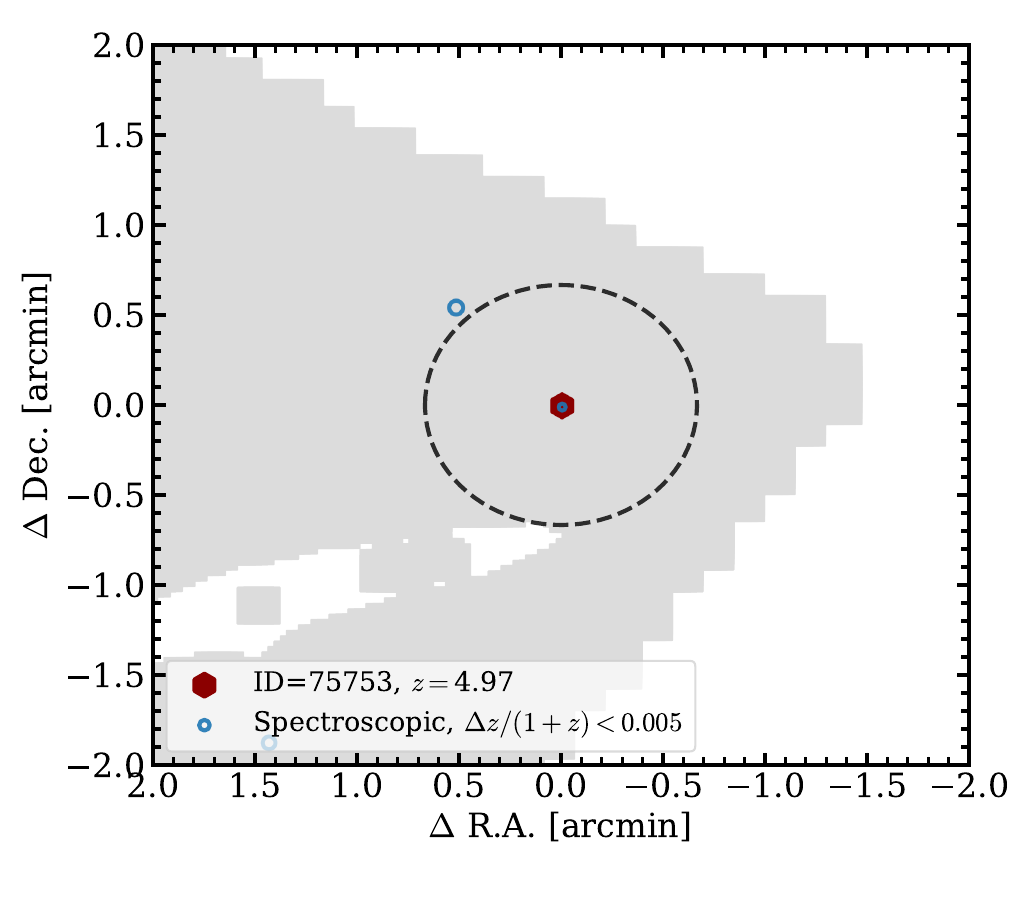} &
    \includegraphics[width=7.7cm]{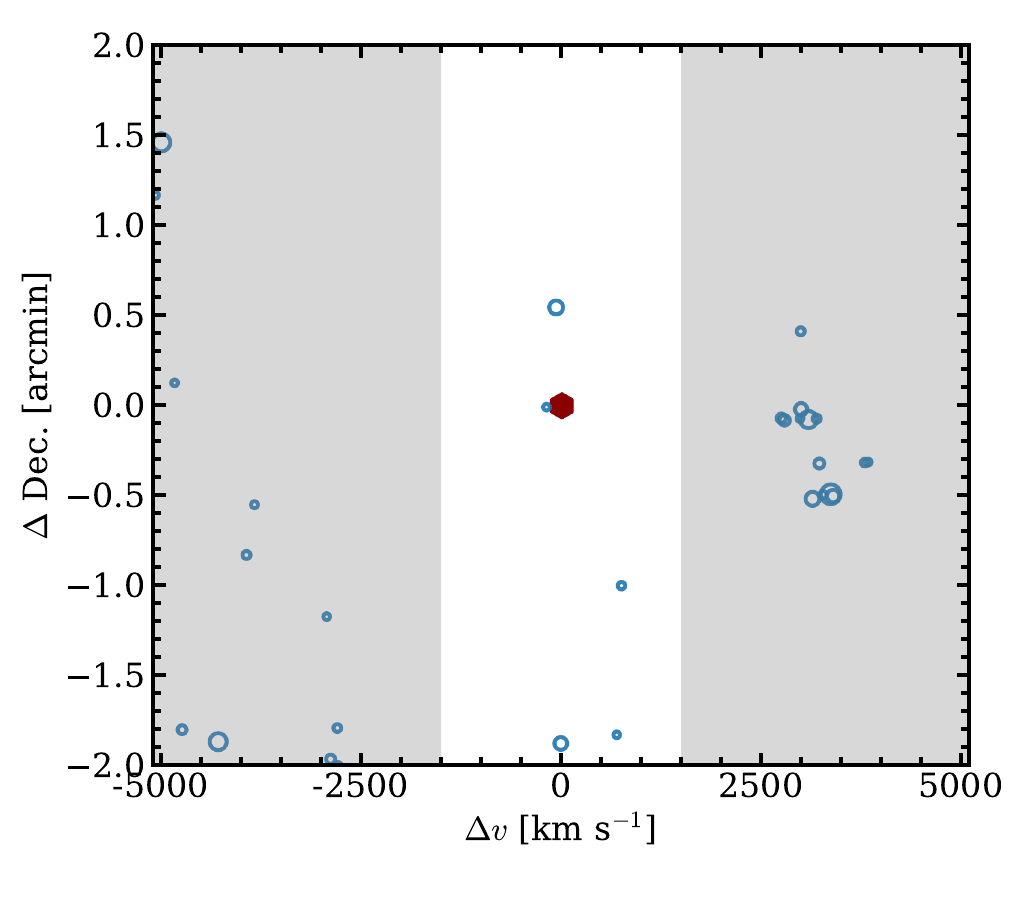} \\
    \end{tabular}
    \caption{The environments of Broad-line H$\alpha$ emitters (as Fig. $\ref{fig:RADEC_1}$).}
    \label{fig:RADEC_3}
\end{figure*}

Figures $\ref{fig:RADEC_2}$ and $\ref{fig:RADEC_3}$ show the large-scale over-densities around the BL-H$\alpha$ emitters that were not shown in the main text (Fig. $\ref{fig:RADEC_1}$).

\section{H$\alpha$ profiles and SEDs of suspected broad line emitters}\label{app:extra}

\begin{figure*}
    \centering
    \begin{tabular}{ccc}\hspace{-0.45cm}
    \includegraphics[width=6cm]{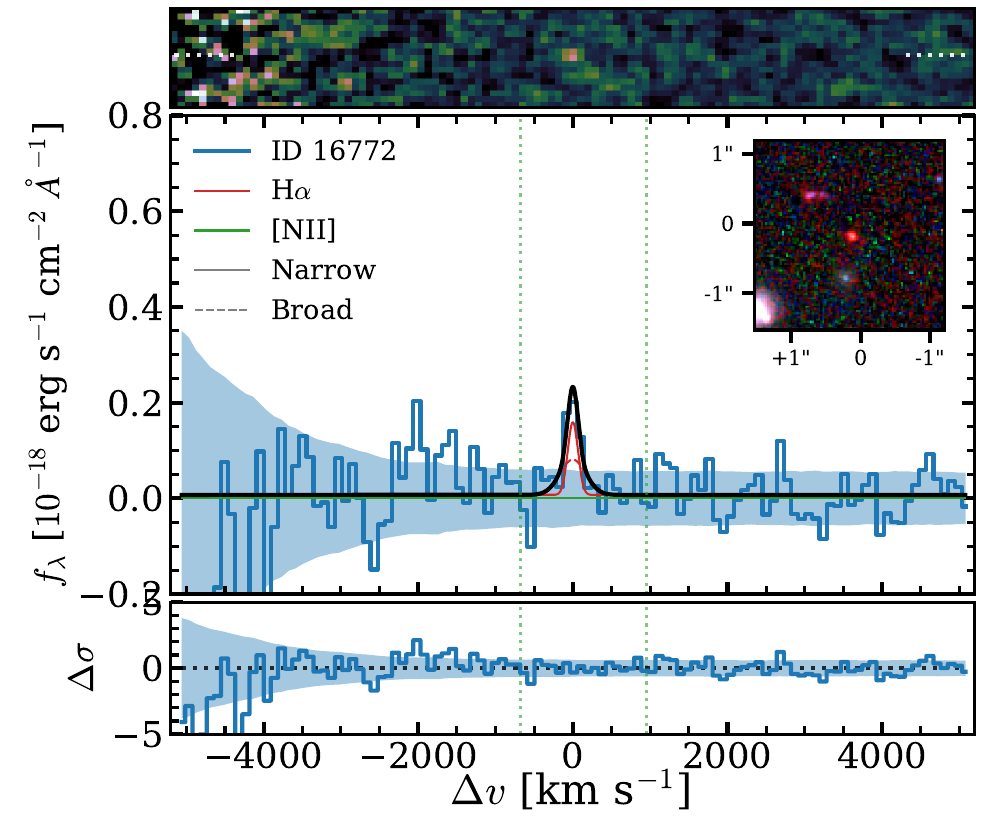} & \hspace{-0.28cm}
    \includegraphics[width=6cm]{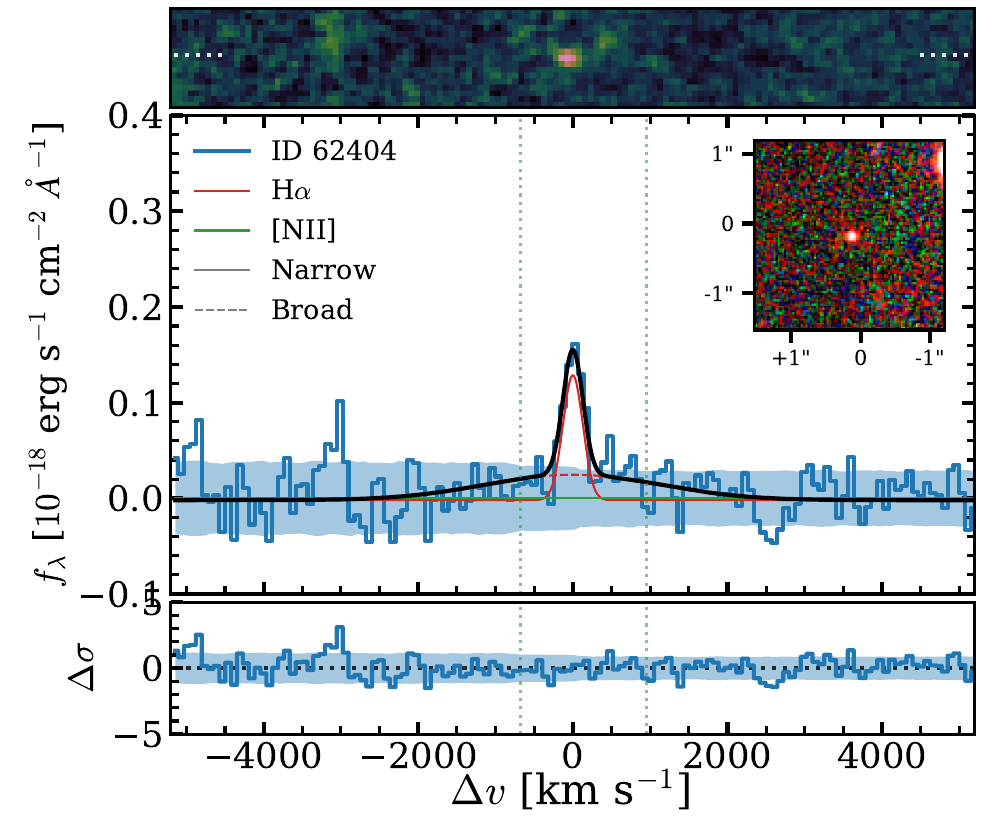} & \hspace{-0.28cm}    
    \includegraphics[width=6cm]{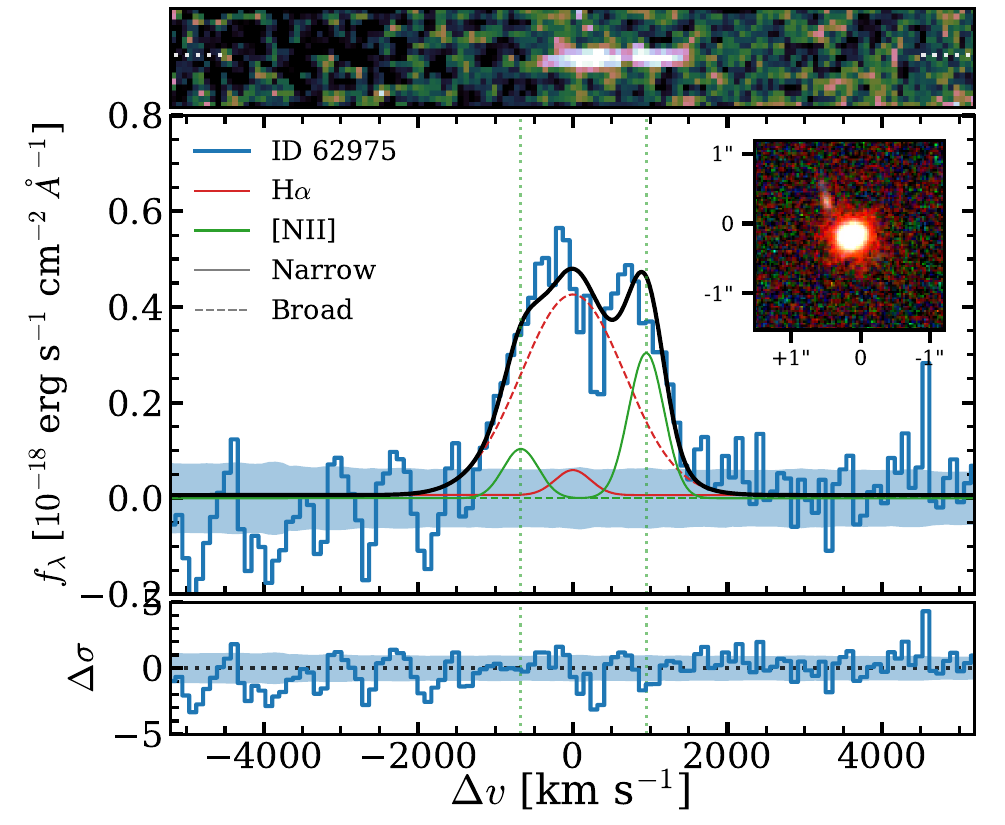} \\
    \end{tabular}
    \caption{The H$\alpha$ profiles of the three suspected broad line emitters (as Fig. $\ref{fig:Lineprofiles}$). Top panels show the 2D continuum-subtracted grism spectra. The middle panels show the optimally extracted 1D spectra. Blue lines show the data, where shaded regions show the errors. The black line shows the combined fit that is composed of a narrow and a broad H$\alpha$ line and narrow [NII]. The red dashed and solid components show the broad and narrow H$\alpha$ component and green shows the best-fit [NII] line, whose wavelength we highlight with dotted green lines.  Bottom shows the residuals of the spectral fit. Inset panels show pseudo-RGB images constructed from NIRCam F115W/F200W/F356W images, highlighting the point-source morphology of the objects. }
    \label{fig:Extra_Lineprofiles}
\end{figure*}

\begin{figure*}
    \centering
    \begin{tabular}{ccc} \hspace{-0.3cm}
    \includegraphics[width=5.8cm]{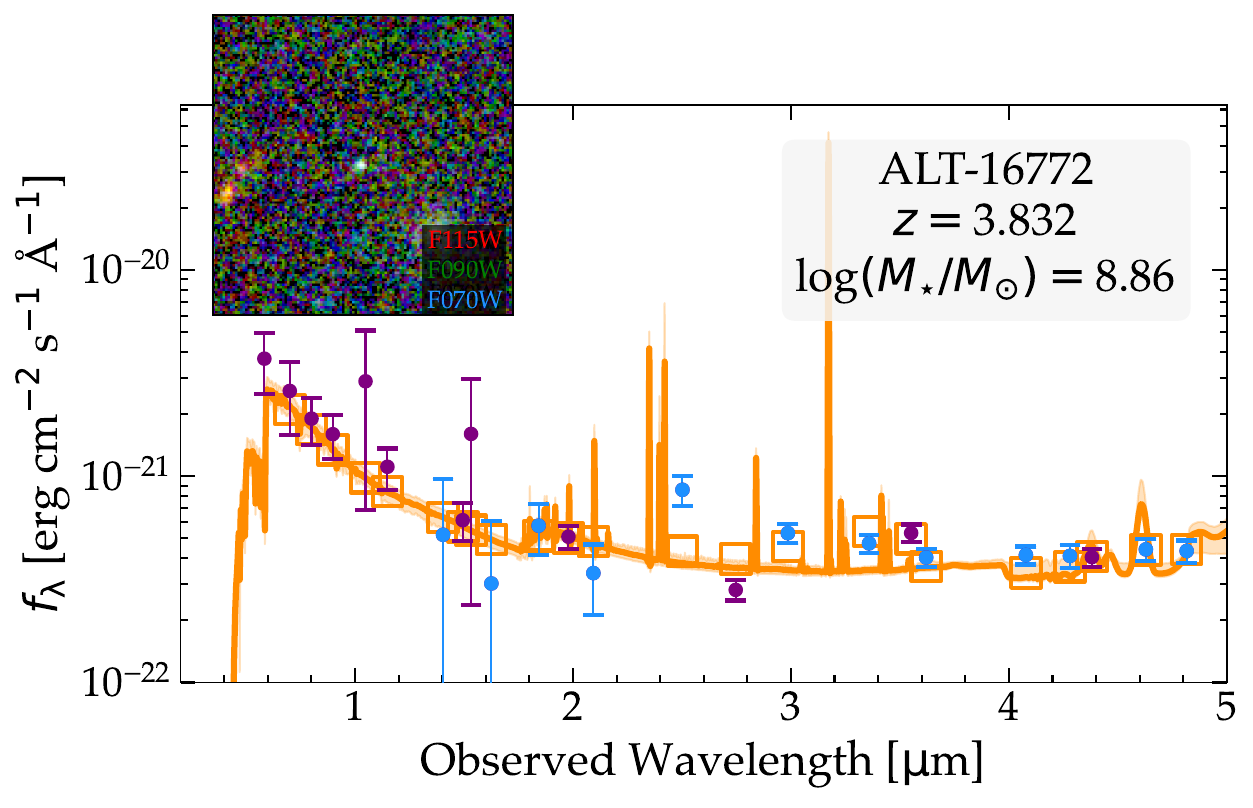}  & \hspace{-0.3cm}
    \includegraphics[width=5.8cm]{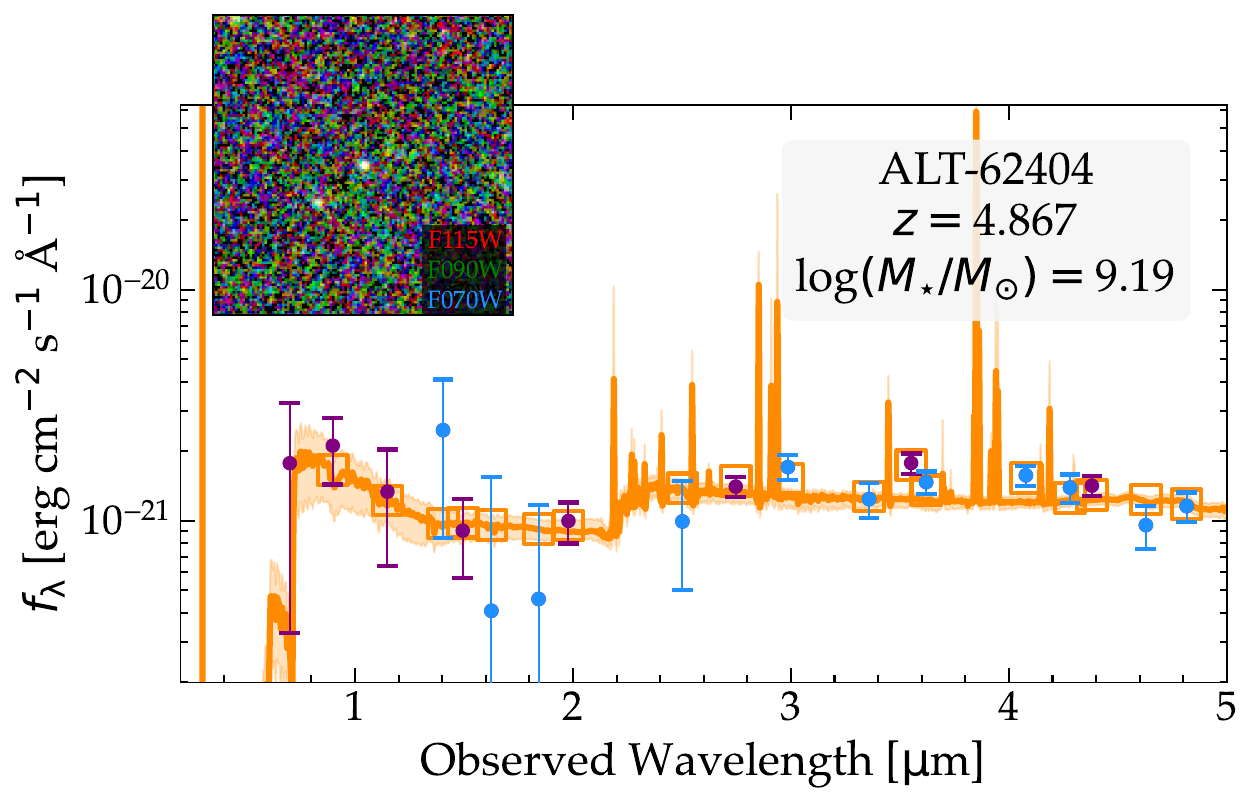}  &\hspace{-0.3cm}
    \includegraphics[width=5.8cm]{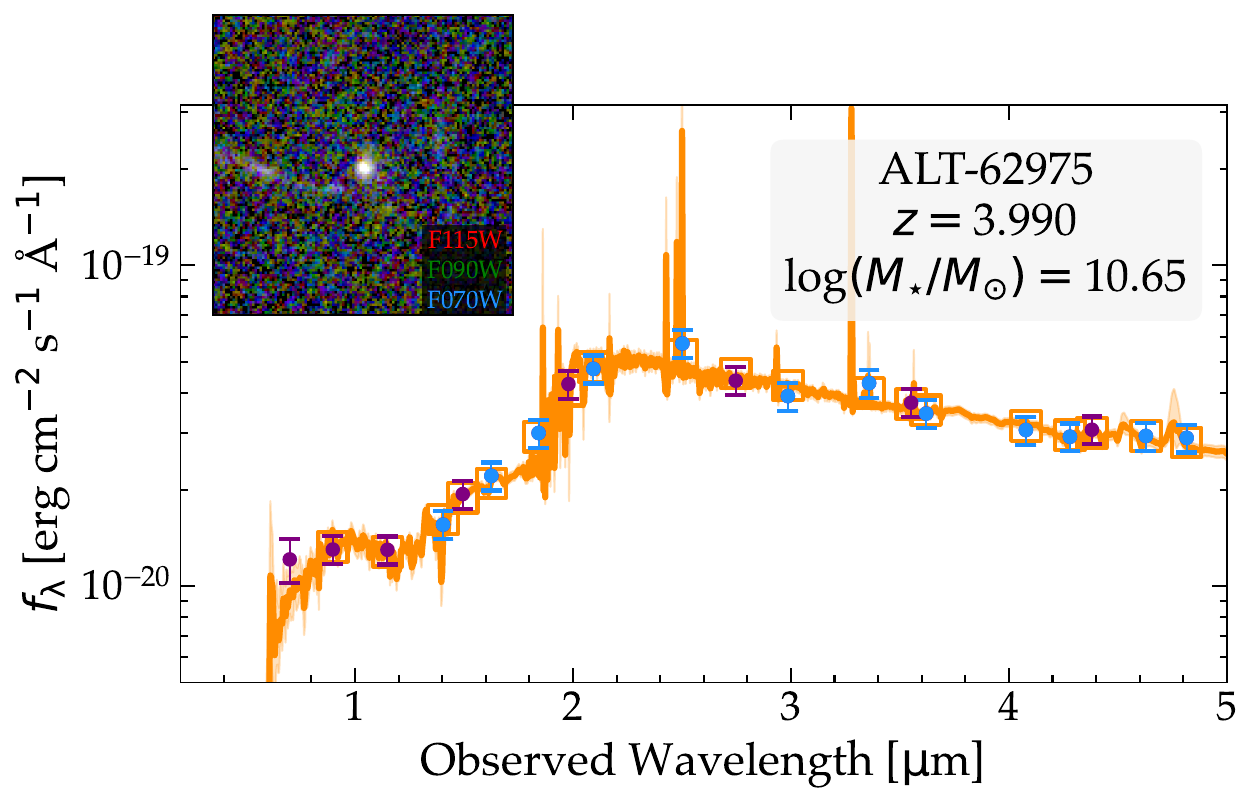}  \\ 
    \end{tabular}
    \caption{SED fits of the three suspected  BL-H$\alpha$ emitters (as Fig. $\ref{fig:LRDs_SEDs}$). Orange curves and shaded regions show the best fit SEDs {\it assuming pure stellar and nebular emission} and their uncertainties that ignore an AGN contribution. Purple data points are measurements in broad-band filters, while blue data-points are medium-band filters. Inset stamps are false-color RGB images of $2.4''\times2.4''$ based on the F070W, F090W and F115W NIRCam imaging data.} 
    \label{fig:Extra_LRDs_SEDs}
\end{figure*}

For completeness, in Figures $\ref{fig:Extra_Lineprofiles}$ and $\ref{fig:Extra_LRDs_SEDs}$, we show the H$\alpha$ spectra and the SEDs of ALT-ID 16772 (RA, DEC = 3.57599, -30.41903, $z=3.832$), 62404 (RA, DEC = 3.54772, -30.3337, $z=4.867$) and 62975 (RA, DEC = 3.56574, -30.33597, $z=3.990$), which we flagged as suspected/possible broad lines in Fig. $\ref{fig:sample}$. 

ALT-16772 and 62404 are very faint galaxies with $M_{\rm UV}\approx-16$ to $-17$ with a compact appearance and a typical UV-blue, optical-red SED. They are particularly red given their faint continuum magnitude. Our sensitivity to broad H$\alpha$ emission is probably preventing the significant detection of a broad component. Both these galaxies are in small over-densities, similar to the typical BL-H$\alpha$ emitters in the sample.

ALT-62975 is among the most luminous systems in the ALT catalog at $z=4-5$, has very red colors, and its SED suggests that it is a very massive galaxy with a stellar mass $4\times10^{10}$ M$_{\odot}$. While the galaxy appears compact, it appears slightly extended in the rest-frame optical, suggesting that a significant fraction of the rest-frame optical light is stellar. The H$\alpha$ profile is complex with two components that are spatially slightly offset. The profile can not be explained by a single narrow line with strong [NII] emission (for example as the massive quenched galaxy at $z\sim5$ identified by \citealt{degraaff24}). The two components are suspected to originate from two components, which could be disentangled with NIRspec spectroscopy or grism spectroscopy at another position angle. The galaxy resides in a large over-density, lending further support that stellar light is the dominant contribution to the SED.

\end{CJK*}
\end{document}